\documentclass[final,2p,times,twocolumn,compress]{elsarticle}
\usepackage{amssymb}
\usepackage{amsmath}
\usepackage{booktabs}
\usepackage{graphicx}
\usepackage{color}
\usepackage{rotating}
\usepackage{subcaption}
\usepackage{setspace}
\usepackage{overpic}
\usepackage{url}

\usepackage{hyperref}
\usepackage{multirow}

\usepackage{mathptmx} % For text and math
\usepackage[T1]{fontenc} % For proper font encoding

\usepackage[utf8]{inputenc}

\begin{document}

\setstretch{0.75} % Reduces line spacing throughout the document
\fontdimen2\font=2.5pt

%-----------------------------------------------------------------------
% \setmainfont{Times New Roman}
%\setsansfont{Arial}
%\setmonofont[Color={0019D4}]{Courier New}
% Use Times-like fonts

%-----------------------------------------------------------------------

\begin{frontmatter}



%%\title{DataScribe Nexus: Developing Deep Encoder-Decoder Models for Complex Tabular Data in Materials Science}
\title{Decoding Non-Linearity and Complexity: Deep Tabular Learning Approaches for Materials Science}


\author[label1]{Vahid Attari} %% Author name
\author[label1]{Raymundo Arroyave} %% Author name

\affiliation[label1]{organization={Department of Materials Science \& Engineering, Texas A\&M University},%Department and Organization
            addressline={}, 
            city={College Station,},
            postcode={77840}, 
            state={TX},
            country={USA}}

%% Abstract
\begin{abstract}
Materials data, especially those related to high-temperature properties, pose significant challenges for machine learning models due to extreme skewness, wide feature ranges, modality, and complex relationships. While traditional models like tree-based ensembles (e.g., XGBoost, LightGBM) are commonly used for tabular data, they often struggle to fully capture the subtle interactions inherent in materials science data. In this study, we leverage deep learning techniques based on encoder-decoder architectures and attention-based models to handle these complexities. Our results demonstrate that XGBoost achieves the best loss value and the fastest trial duration, but deep encoder-decoder learning like Disjunctive Normal Form architecture (DNF-nets) offer competitive performance in capturing non-linear relationships, especially for highly skewed data distributions. However, convergence rates and trial durations for deep model such as CNN is slower, indicating areas for further optimization. The models introduced in this study offer robust and hybrid solutions for enhancing predictive accuracy in complex materials datasets.

%Materials data, particularly those related to high-temperature properties, often present significant challenges for machine learning models due to complexities such as extreme skewness, wide feature ranges, and multi-modal distributions. Traditional models like tree-based ensembles (e.g., XGBoost, LightGBM) are typically favored for tabular data but may struggle to fully capture the intricate relationships inherent in materials science data. In this study, we introduce the DataScribe Nexus framework, specifically designed to handle these challenges by leveraging deep learning models such as encoder-decoder architectures and attention-based approaches. Our results indicate that while XGBoost consistently outperforms deep models on standard features, deep learning techniques offer unique advantages in capturing latent feature representations and complex, non-linear relationships, especially in cases of highly skewed or multi-modal data distributions. Leveraging hybrid approaches that combine both methods could offer a more balanced solution, enhancing predictive performance for complex datasets. The \textbf{``DataScribe Nexus''} framework, introduced as part of this study, provides a robust solution for integrating deep learning into traditional tabular data workflows, offering significant potential for advancing predictive accuracy in materials science and other complex domains.
\end{abstract}



%% Keywords

\end{frontmatter}

\section{Introduction}

%% paragraph 1
Predicting materials behavior is inherently challenging due to the non-linear and interdependent relationships between alloy chemistry, processing conditions, and properties. Unlike datasets in computer vision or natural language processing, tabular materials data can include dense numerical and sparse categorical features with weaker correlations, making pattern recognition more difficult. Deep learning, originally developed for tasks such as sequence-to-sequence modeling and data reconstruction, has shown promise in capturing these complex dependencies for modeling materials properties. However, applying deep learning models to materials data presents unique challenges. In materials science, physics-based relationships require explicit incorporation of domain knowledge, and the “black-box” nature of these models limits interpretability. Their success also hinges on the availability of large, high-quality datasets, which are often costly and scarce. Furthermore, the significant computational demands of deep learning raise questions about its efficiency compared to simpler methods like random forests or physics-informed neural networks. Addressing these challenges could enable deep learning models to offer computationally efficient alternatives to traditional physics-based methods like solving the Schrödinger equation~\cite{kipf2016semi,kalidindi2015materials,raabe2019strategies,butler2018machine,borisov2022deep,xie2018crystal,chen2019atomsets}.

%% paragraph 2
Deep learning has shown remarkable promise across various domains, including materials science, where it has been successfully applied to predict mechanical properties of alloys~\cite{xie2018crystal,jha2018elemnet,agrawal2019deep}, discover new thermoelectric materials~\cite{chen2019atomsets}, and identify phase transitions in complex multicomponent systems~\cite{cubuk2015identifying}. However, while these successes highlight its potential, applying deep learning to tabular data, which is prevalent in materials science, presents unique challenges. Materials science data often spans multiple orders of magnitude, reflecting the diversity of material properties and phenomena. For example, mechanical properties like yield strength range from tens of MPa for polymers to thousands of MPa for metals and ceramics, while electrical conductivity varies from as low as 10$^{-16}$ S/m in insulators to 10$^7$ S/m for conductors like copper. Similarly, thermal conductivity can range from less than 0.1 W/m·K in insulators to over 1000 W/m·K in materials like diamond, and diffusion coefficients vary from 10$^{-25}$ m$^2$/s in solids at low temperatures to 10$^{-8}$ m$^2$/s in liquids. Even creep behavior spans orders of magnitude under high-temperature conditions, influenced by factors such as stress, temperature, and microstructure.

%Paragraph 3  (see Fig.~\ref{fig:MaterialsInsightModel})
Beyond (1) data scarcity, (2) challenges in data preparation, and (3) predictive accuracy, (4) interpretability is vital in materials science applications to ensure models provide actionable insights for experimental validation. To tackle data scarcity and confidentiality concerns, generative approaches like Variational Autoencoders (e.g. Tab-VAE~\cite{tazwar2024tab}) or Tabular Generative Adversarial Networks (~\cite{xu2018synthesizing}) can be employed to create synthetic datasets, enhancing model robustness. Tackling data preparation for largely skewed features can be achieved using robust transformations, such as quantile transformation or log-scaling, to ensure features are more suitable for model training~\cite{das2016brief,shi2023detecting}. When paired with interpretable architectures~\cite{linardatos2020explainable} and efficient inference strategies~\cite{hoefler2021sparsity}, these methods can advance the use of tabular materials data in scientific discovery and design. Recent innovations, such as transformers and hybrid models combining tree-based methods with neural networks, show promise in overcoming these challenges. Transformers leverage self-attention mechanisms to capture complex feature interactions, offering an advantage for the heterogeneous and multiscale nature of materials data. However, inconsistencies in benchmarking practices and unequal levels of optimization hinder direct comparisons with traditional methods like gradient-boosted decision trees (GBDT)~\cite{popov2020neural,ke2017lightgbm,fiedler2021tabular,borisov2021deep,somepalli2021saint}. To fully realize their potential, deep learning models in materials science must balance predictive accuracy with alignment to domain-specific physical principles and interpretability. Achieving this will require improved benchmarking frameworks, access to diverse and high-quality datasets, and the integration of domain knowledge into model development. These advancements could close the performance gap with traditional methods and enable deep learning to address the unique demands of materials science and unlock its potential for solving complex problems.

\begin{figure*}[!ht]
    \centering
    \includegraphics[width=0.95\linewidth]{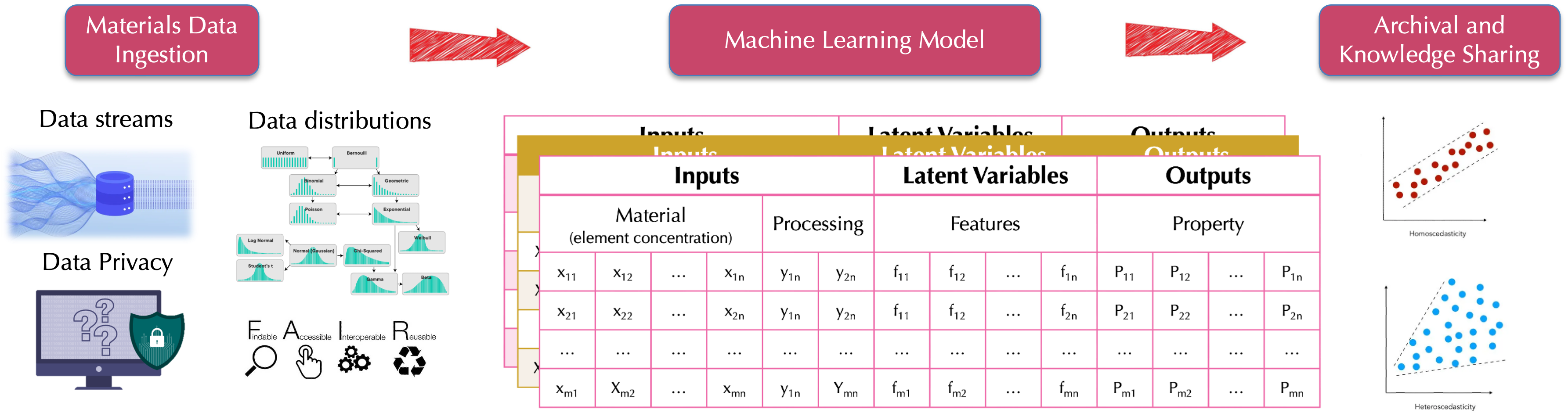}
    \caption{From Materials Data Ingestion to Insightful Models: A streamlined workflow for processing data streams, ensuring privacy, building machine learning models, and driving innovation through archival and knowledge sharing in materials science.}
    \label{fig:MaterialsInsightModel}
\end{figure*}

%Paragraph 5
Several emerging neural architectures, inspired by concepts from the encoder-decoder and other hybrid methods, have shown promise in addressing the limitations of traditional models, particularly in handling tabular data workflows. For example, TabNet enhances interpretability and performance by dynamically selecting relevant features through attention mechanisms\cite{arik2021tabnet}. Neural Oblivious Decision Ensembles (NODE) combine decision rules with neural networks to efficiently model feature interactions~\cite{popov2020neural}. FT-Transformers and TabTransformer employ self-attention to capture complex relationships between numerical and categorical features~\cite{gorishniy2021revisiting,huang2020tabtransformer}. Hybrid approaches like DeepGBM integrate gradient-boosting machines with neural networks to combine feature transformation and prediction~\cite{ke2020deepgbm}. These advancements, along with models like Wide and Deep Networks and DNF-Net (Disjunctive Normal Form Networks), improve scalability, interpretability, and predictive performance, positioning them as robust alternatives to traditional methods.

%% Last paragraph of the introduction

%	Summary of the Study:
In this work, we evaluate the potential of several deep learning architectures to outperform traditional GBDTs in accuracy and efficiency on materials science tabular dataset. Our findings demonstrate that neural architectures tailored for tabular data can effectively handle the complexity and wide range of property scales often encountered in materials science. Beyond static datasets, these models exhibit promise in processing streams of data generated from high-throughput experiments or autonomous systems, making them particularly relevant for real-time analysis. Additionally, their ability to capture intricate dependencies and handle complex, multimodal distributions positions them as critical tools for addressing the inherent variability in materials data. These advancements also align with the principles of FAIR (Findable, Accessible, Interoperable, and Reusable) data, ensuring that the insights derived from these models can be leveraged across diverse research and industrial contexts. In section~\ref{sec:method}, we detail the architectures and approaches used in this study, while section~\ref{sec:results} compares the performance of deep learning models and GBDTs, highlighting their respective strengths in handling complex materials datasets.

%%%%
%%%%

\section{Methods}\label{sec:method}

\subsection{\textbf{Baseline Encoder-Decoder Approach}}

The encoder-decoder model is widely recognized for its ability to reconstruct data and capture complex relationships between features by learning a latent representation. For regression tasks, it can be adapted to learn complex input space and predict continuous output values. The encoder maps the input data $\mathbf{X}$ into a latent space $\mathbf{Z}$, which can either have a lower-dimensional representation (undercomplete model) or a higher-dimensional representation (overcomplete model). The decoder then maps $\mathbf{Z}$ to the predicted output $\hat{\mathbf{Y}}$. These transformations are defined as:
\begin{equation}
\mathbf{Z} = f_{\theta}(\mathbf{X}), \quad \hat{\mathbf{Y}} = g_{\phi}(\mathbf{Z}),
\end{equation}
\noindent
where the model is trained by minimizing a loss function $\mathcal{L}$, which measures the difference between the predicted output $\hat{\mathbf{Y}}$ and the target data $\mathbf{Y}$. The objective is to find the parameters \(\theta\) and \(\phi\) that minimize the following:

\begin{equation}
\min_{\theta, \phi} \mathcal{L}(\mathbf{Y}, \hat{\mathbf{Y}}) = \mathcal{L}(\mathbf{Y}, g_{\phi}(f_{\theta}(\mathbf{X})))
\end{equation}

Training is typically performed using gradient-based methods. After training, the model can be used to transform new data \(\mathbf{X}_{\text{new}}\) into predictions \(\hat{\mathbf{Y}}_{\text{new}}\):

\begin{equation}
\hat{\mathbf{Y}}_{\text{new}} = g_{\phi}(f_{\theta}(\mathbf{X}_{\text{new}}))    
\end{equation}

This baseline encoder-decoder approach is not inherently interpretable largely due to lack of transparency in latent representations. In the following section, we will discuss various neural network architectures that can enhance interpretability and prediction accuracy when paired with encoder-decoder concepts. The hierarchy of models, from the most complex to the simplest, starts with Transformer-based encoder-decoder models. These are state-of-the-art for tasks like language translation and text generation, leveraging self-attention mechanisms to handle long-range dependencies. CNN-based encoder-decoder models are specialized for spatial data, excelling in tasks like image segmentation. Simpler architectures include basic encoder-decoder models that map inputs to latent space and back for structured or variable-length data, and at the base level are multi-layer perceptrons (MLPs), suited for fixed-size input-output tasks in structured datasets. This hierarchy reflects increasing model complexity and specialization as one moves from MLPs to Transformers. Table~\ref{tab:nn_arch_comparison} provides a detailed comparison of these neural network architectures, highlighting their key characteristics, advantages, and considerations related to assumptions, multicollinearity, overfitting, interpretability, and sensitivity to outliers.

%%%%%%%%

\begin{table*}[!ht]
\centering
\scriptsize
\renewcommand{\arraystretch}{0.9} % Increases vertical spacing between rows
\caption{Comparison of neural network architectures across key characteristics and challenges}
\begin{tabular}{l p{2.2cm} p{2.75cm} p{8.25cm}}
\toprule
\textbf{Architecture} & \textbf{Key Characteristics} & \textbf{Advantages} & \textbf{Assumptions, Multicollinearity, Overfitting, Interpretability, Outliers} \\ 
\midrule
\textbf{Fully Dense} & Fully connected layers between neurons. & Captures complex, non-linear relationships; versatile across domains; easy to implement. & \textbf{Assumptions}: No strict assumptions about data. \newline
\textbf{Multicollinearity}: Can occur with highly correlated features. \newline
\textbf{Overfitting}: Prone to overfitting with large models. Regularization techniques (dropout, L2) are often used. \newline
\textbf{Interpretability}: Low, hard to explain weights. \newline
\textbf{Outliers}: Sensitive to outliers.\\
\midrule
\textbf{DNNF} & Combines logical AND/OR operations. & Interpretable; effective for tabular data; good at modeling logical rules. & \textbf{Assumptions}: No strong assumptions. \newline
\textbf{Multicollinearity}: Handles multicollinearity well by logical splitting. \newline
\textbf{Overfitting}: Prone to overfitting in small datasets, mitigated by regularization.\newline
\textbf{Interpretability}: High, uses logical rules that are easy to interpret. \newline
\textbf{Outliers}: Somewhat robust to outliers due to logical decision-making. \\
\midrule
\textbf{TabNet} & Attention-based feature selection for tabular data. & Interpretable; dynamic feature selection; handles both numerical and categorical features. & \textbf{Assumptions}: No strong assumptions. \newline
\textbf{Multicollinearity}: Manages it via attention-based feature selection. \newline
\textbf{Overfitting}: May overfit without proper regularization, especially with small datasets. \newline
\textbf{Interpretability}: Moderate, offers feature importance via attention scores. \newline
\textbf{Outliers}: Robust to outliers, but large outliers may affect performance. \\
\midrule
\textbf{Variational} & Encodes data into latent distributions. & Good for generative tasks; handles uncertainty in data; provides probabilistic outputs. & \textbf{Assumptions}: Assumes that the latent space follows a distribution (e.g., Gaussian). \newline
\textbf{Multicollinearity}: No direct handling of multicollinearity.\newline
\textbf{Overfitting}: Prone to overfitting if the latent space is poorly regularized. \newline
\textbf{Interpretability}: Low, latent space is abstract. \newline
\textbf{Outliers}: Sensitive to outliers; unusual data points can skew distribution. \\
\midrule
%\textbf{LSTM} & Handles sequential data with memory gates. & Good for time series, speech, and text; captures long-term dependencies; effective for sequential prediction. & \textbf{Assumptions}: No specific assumptions for data distribution.\newline
%\textbf{Multicollinearity}: Can manage sequential dependencies, but not designed to handle multicollinearity explicitly. \newline
%\textbf{Overfitting}: Prone to overfitting, often mitigated by dropout or regularization. \newline
%\textbf{Interpretability}: Low, hard to interpret long-range dependencies. \newline
%\textbf{Outliers}: Sensitive to outliers in sequence data.\\
%\midrule
\textbf{1D-CNN} & Convolutions for structured data, e.g., time series. & Efficient feature extraction; can capture local patterns; faster than recurrent networks. & \textbf{Assumptions}: Assumes some local structure in data. \newline
\textbf{Multicollinearity}: Not explicitly handled. \newline
\textbf{Overfitting}: Can overfit if too deep or wide; regularization or pooling helps. \newline
\textbf{Interpretability}: Low to moderate, depending on depth and feature extraction. \newline
\textbf{Outliers}: Can be robust to small outliers but affected by extreme ones. \\
\midrule
\textbf{XGBoost} & Boosting trees with gradient descent. & Feature selection; strong performance on tabular data; built-in regularization; handles missing data well. & \textbf{Assumptions}: Does not require strong assumptions. \newline
\textbf{Multicollinearity}: Tolerates multicollinearity; feature importance still useful. \newline
\textbf{Overfitting}: Handles overfitting well via regularization and early stopping. \newline
\textbf{Interpretability}: Moderate, with feature importance analysis. \newline
\textbf{Outliers}: Robust to moderate outliers, depending on tree depth and splits. \\
\bottomrule
\end{tabular}
\label{tab:nn_arch_comparison}
\end{table*}

%%%%%
\subsubsection{\textbf{Fully Dense Neural Network Block}}

A fully dense neural network, also known as a fully connected or feedforward network, is a foundational deep learning architecture that remains widely used for solving diverse machine learning problems across various dataset types~\cite{balderas2024optimizing}. In this architecture, each neuron in one layer connects to every neuron in the next, allowing the model to capture complex, non-linear relationships between features. For a given input vector \(\mathbf{x}\), the output of the \(i\)-th layer is computed as:

\begin{equation}
\mathbf{y}_i = f(\mathbf{W}_i \mathbf{x} + \mathbf{b}_i)    
\end{equation}

\noindent
where \(\mathbf{W}_i\) is the weight matrix, \(\mathbf{b}_i\) the bias term, and \(f\) a non-linear activation function such as ReLU, sigmoid, or tanh. Multiple layers refine the data representation, improving model predictions. The parameters \(\theta\) include the weight matrices, biases, and, if applicable, trainable activation parameters like the slope \(\alpha\) in LeakyReLU. Regularization terms, such as the L2 coefficient \(\lambda\), help control overfitting by penalizing large weights. These parameters collectively determine how the dense block processes and learns from data. Fully connected layers are not transparent, offering little insight into their decision-making process~\cite{barnes2024architectural}.

%%%%%

\subsubsection{\textbf{A Disjunctive Normal Neural Form (DNNF) Block}}

A Disjunctive Normal Neural Form (DNNF) block is a neural network architecture designed to model logical structures, similar to decision trees, making it interpretable and efficient for capturing complex relationships in tabular data. The DNNF block operates by combining multiple logical conjunctions (AND conditions), which are then aggregated using disjunctions (OR conditions). Each conjunction represents feature interactions modeled through linear transformations and activation functions, mimicking the behavior of a logical AND operation. The outputs of these conjunctions are then aggregated to form a disjunction, effectively modeling a logical OR operation. This hierarchical structure enables the DNNF block to identify and represent complex feature relationships in a manner that is both interpretable and computationally efficient. Key parameters of a DNNF block include the number of clauses (representing distinct logical interactions), the number of literals per clause (defining how many features contribute to each interaction), the choice of activation function (e.g., ReLU or Sigmoid), and regularization techniques such as dropout to prevent overfitting. This structure makes DNNF particularly suitable for applications requiring interpretability and robust modeling of feature interactions.

\subsubsection{\textbf{TabNet Block}}

TabNet employs a sequential attention mechanism to dynamically select the most relevant features at each decision step, enabling it to efficiently capture complex relationships in tabular datasets~\cite{arik2021tabnet}. Unlike traditional fully connected neural networks, TabNet performs feature selection dynamically at each layer, mirroring the interpretability of decision tree models while retaining the flexibility and power of neural networks. At each decision step, TabNet applies a transformation that includes an attention mechanism to focus on the most relevant features. Attention mechanisms have emerged as a pivotal architectural element in deep neural
networks (DNNs), significantly enhancing their interpretability~\cite{barnes2024architectural}. This mechanism ensures that the outputs from one step guide the feature selection process for subsequent steps.  By iteratively refining feature selection and transformation across multiple steps, TabNet effectively models intricate feature interactions and relationships, culminating in a final output that leverages the entire feature space. The model’s parameter set encompasses weight matrices, bias terms, and parameters for the attention mechanism, which collectively determine the feature selection and transformation process. To prevent overfitting, regularization techniques such as L2 regularization and sparsity-inducing penalties are commonly applied.

%%%%%

\subsubsection{\textbf{1D-Convolutional Neural Network Block}}

A 1D-Convolutional Neural Network (1D-CNN) block is specifically designed to extract local patterns from sequential or ordered data, such as time series or audio signals~\cite{kiranyaz20211d}. The architecture employs convolutional filters that slide across the input data, capturing localized features through a combination of linear transformations and non-linear activation functions. These filters enable the model to detect patterns such as trends or periodicities, which are essential for understanding structured data. Typically, the convolutional layers are followed by pooling layers that down-sample the data, reducing computational complexity while preserving critical information. This hierarchical structure allows 1D-CNNs to progressively capture more abstract patterns as the data flows through deeper layers. Parameters of the model include the weights of the convolutional filters, biases, and the choice of activation functions, such as ReLU, which introduces non-linearity to the model. 1D-CNNs can be combined with fully connected layers to refine feature representations, making them versatile for a variety of tasks. Recent advancements have demonstrated the efficacy of these neural network blocks in competitions and real-world applications by leveraging shortcut connections and innovative architectures to improve feature extraction and accuracy, particularly in domains where spatial locality is less pronounced, such as tabular data.

\subsection{\textbf{Deep generative tabular learning use VAE}}

The deterministic design of classical encoder-decoder models limits their ability to generate diverse outputs, constraining their effectiveness in capturing and navigating complex data distributions. A variational autoencoder (VAE) shares an encoder-decoder structure, but extends it by incorporating probabilistic modeling in the latent space, making it ideal for generative tasks that require both dimensionality reduction and data generation. The encoder transforms the input \(\mathbf{x}\) into a distribution over the latent space, rather than a single point. Specifically, the encoder outputs the mean \(\mu_i\) and variance \(\sigma_i^2\) for the latent variable \(\mathbf{z}\):

\[
\mathbf{z}_i = \mu_i + \sigma_i \odot \mathbf{\epsilon}, \quad \mathbf{\epsilon} \sim \mathcal{N}(0, 1)
\]

\noindent
where \(\mathbf{z}_i\) represents a latent variable sampled from the distribution, \(\mu_i\) and \(\sigma_i\) are learned parameters that describe the mean and variance, and \(\mathbf{\epsilon}\) is a random noise sampled from a standard normal distribution, enabling stochasticity in the latent space. 

The decoder then reconstructs the data by mapping the latent variable \(\mathbf{z}_i\) back to the output space:

\[
\hat{\mathbf{y}}_i = f_{\text{decoder}}(\mathbf{W}_i \mathbf{z}_i + \mathbf{b}_i)
\]

\noindent
where \(f_{\text{decoder}}\) is a non-linear transformation (e.g., a fully connected layer with ReLU), \(\mathbf{W}_i\) represents the weight matrix for the decoder, and \(\mathbf{b}_i\) is the bias term.

The objective of the variational encoder-decoder is to maximize the evidence lower bound (ELBO), which consists of two terms: the reconstruction loss (e.g., mean squared error between the input \(\mathbf{x}\) and the reconstruction \(\hat{\mathbf{y}}\)) and the Kullback-Leibler (KL) divergence between the learned latent distribution and a standard normal distribution. The full loss function can be written as:

\[
\mathcal{L} = \mathbb{E}_{q(\mathbf{z}|\mathbf{x})}[\log p(\mathbf{y}|\mathbf{z})] - D_{\text{KL}}(q(\mathbf{z}|\mathbf{x}) \| p(\mathbf{z}))
\]

The parameter set \(\theta\) in a variational encoder-decoder block includes the weights and biases of the encoder and decoder, the mean and variance parameters \(\mu_i\) and \(\sigma_i\), and any regularization terms used to control overfitting. This architecture is particularly useful for generating new samples, data imputation, and learning robust latent representations in tasks involving high-dimensional data.

\subsubsection{Extreme Gradient Boosting}

Extreme Gradient Boosting (XGBoost) is a powerful ML algorithm based on decision trees, optimized for speed and performance. It belongs to the family of boosting algorithms, where multiple weak learners (typically decision trees) are combined sequentially to form a strong predictive model. XGBoost works by fitting a new tree to correct the errors of the previous trees, iteratively improving the model's predictions. Given an input vector $\mathbf{x}$, the output of the model at step $t$ is computed as:

\[
\hat{y}_t = \sum_{i=1}^{t} f_i(\mathbf{x})
\]

\noindent
where \(f_i(\mathbf{x})\) represents the prediction from the \(i\)-th tree, and $\hat{y}_t$ is the cumulative prediction at step $t$. The trees are added one by one to minimize the residual errors of the previous trees. XGBoost optimizes a regularized objective function that consists of a loss term (e.g., mean squared error for regression or log loss for classification) and a regularization term to prevent overfitting:

\[
\mathcal{L} = \sum_{i=1}^{n} \ell(y_i, \hat{y}_i) + \sum_{k=1}^{T} \Omega(f_k)
\]

\noindent
where $\ell(y_i, \hat{y}_i)$ is the loss function, which measures the difference between the true label $y_i$ and the predicted label $\hat{y}_i$, and $\Omega(f_k)$ is a regularization term that controls the complexity of each tree, penalizing large trees to avoid overfitting. XGBoost incorporates several advanced features, such as regularization, missing values by automatically learning the best imputation strategy during training. The parameter set \(\theta\) for XGBoost includes the weights of the decision trees, the regularization coefficients (L$_1$, L$_2$), and the learning rate, which controls how quickly the model adapts to errors during training. 

%%%%
%%%%
%%%%
\subsection{\textbf{Bayesian Optimization of Hyperparameters}}

While deep learning has achieved various breakthrough successes, its performance highly depends on the proper set- tings of its hyperparameters~\cite{chen2018bayesian}. In this work, we use a widely-used versatile Bayesian optimization method, tree-structured Parzen estimator (TPE), to handle hyperparameter optimization~\cite{bergstra2011algorithms,bergstra2013making,optuna_2019}. The goal is to minimize the objective function $f(x)$ as follows:
\begin{equation}
    x_{\text{opt}} \in \operatorname*{arg\,min}_{x\in \mathcal{X}} f(x)
\end{equation}
\noindent
where $x_{\text{opt}}$ is a list of model parameters (e.g., number of layers, learning rate, regularization, and dropout rate) that exhibits the best performance. The algorithm efficiently narrows down the hyperparameters, reducing computational cost compared to grid or random searches. We performed 130 trials per dataset, tuning 5–10 key hyperparameters depending on the model. Model performance was evaluated using MSE (RMSE for XGBoost). For datasets spanning multiple orders of magnitude, we used logarithmic-based metrics like Mean Squared Logarithmic Error (MSLE) and Symmetric Mean Absolute Percentage Error (SMAPE) to normalize errors across different scales. Instead of early stopping, we evaluated all models at fixed epoch number of 50 (number of estimators for XGBoost). Statistical significance in model performance differences was evaluated using Friedman's test at a 95\% confidence level.

Before optimization, we also evaluated both first- and higher-order statistics in the dataset, including measures such as mean, standard deviation, skewness, and kurtosis. Skewness, a third-order statistic, quantifies the asymmetry of a distribution, indicating whether the data is skewed to the left (negative skewness) or to the right (positive skewness). A distribution with skewness close to zero is considered symmetrical. Kurtosis, a fourth-order statistic, measures the “tailedness” or the presence of extreme values in the distribution. High kurtosis suggests heavy tails and a greater occurrence of outliers, while low kurtosis indicates lighter tails and fewer extreme values. Building on this statistical understanding of the dataset, we then analyzed the performance of neural encoder-decoder models and XGBoost, focusing on accuracy, training efficiency, and optimization time. The goal was to assess whether the neural encoder-decoder models offer advantages over classical models in tabular data tasks.

\begin{table*}[!ht]
    \centering
    \caption{Statistical Summary of the Dataset including Skewness and Kurtosis of main features}
    \scriptsize
    \label{tab:extended_data_stats_skew_kurt}
    \begin{tabular}{lccccccccc}
        \toprule
        \textbf{Feature} & \textbf{Mean} & \textbf{Std. Dev.} & \textbf{Min} & \textbf{Max} & \textbf{Median} & \textbf{25\%} & \textbf{75\%} & \textbf{Skewness} & \textbf{Kurtosis} \\
        \midrule
        Nb & 0.22 & 0.19 & 0 & 0.95 & 0.2 & 0.05 & 0.35 & 0.91 & 0.29 \\
        Cr & 0.21 & 0.18 & 0 & 0.95 & 0.15 & 0.05 & 0.3 & 0.96 & 0.46 \\
        V & 0.22 & 0.19 & 0 & 0.95 & 0.2 & 0.05 & 0.35 & 0.94 & 0.4 \\
        W & 0.14 & 0.11 & 0 & 0.9 & 0.1 & 0.05 & 0.2 & 0.7 & 0.14 \\
        Zr & 0.21 & 0.18 & 0 & 0.9 & 0.15 & 0.05 & 0.3 & 0.91 & 0.3 \\
        YS 1000$^\circ$C & 1.2$\times10^{3}$ & 6.4$\times10^{2}$ & 3.8$\times10^{-6}$ & 3.4$\times10^{3}$ & 1.1$\times10^{3}$ & 6.7$\times10^{2}$ & 1.6$\times10^{3}$ & 0.48 & -0.26 \\
        EQ 1273K THCD (W/mK) & 14 & 13 & 0.024 & 63 & 9.3 & 3.4 & 23 & 0.97 & 0.0098 \\
        EQ 1273K Density (g/cc) & 9.6 & 1.8 & 6.1 & 17 & 9.3 & 8.3 & 11 & 0.77 & 0.56 \\
        \textbf{1300 Min Creep CB [1/s]} & 0.0025 & 0.098 & 2.5$\times10^{-22}$ & 4 & 5.5$\times10^{-15}$ & 3$\times10^{-16}$ & 1.7$\times10^{-12}$ & \textbf{39} & \textbf{1500} \\
        PROP 1500C CTE (1/K) & 1.3$\times10^{-5}$ & 4.1$\times10^{-6}$ & 5.9$\times10^{-6}$ & 2.9$\times10^{-5}$ & 1.1$\times10^{-5}$ & 9.8$\times10^{-6}$ & 1.5$\times10^{-5}$ & 1.1 & 0.35 \\
        YS 1500C PRIOR & 8.1$\times10^{2}$ & 5$\times10^{2}$ & 9.4$\times10^{-9}$ & 2.7$\times10^{3}$ & 7.4$\times10^{2}$ & 4.2$\times10^{2}$ & 1.1$\times10^{3}$ & 0.66 & -0.0047 \\
        Pugh Ratio PRIOR & 2.6 & 0.51 & 1.4 & 4.5 & 2.6 & 2.2 & 2.9 & 0.39 & -0.085 \\
        Scheil LT & 1.6$\times10^{3}$ & 1.6$\times10^{2}$ & 1.5$\times10^{3}$ & 2.4$\times10^{3}$ & 1.5$\times10^{3}$ & 1.5$\times10^{3}$ & 1.6$\times10^{3}$ & 2 & 2.9 \\
        Kou Criteria & 3.5$\times10^{2}$ & 4$\times10^{2}$ & 0.13 & 4.7$\times10^{3}$ & 2.4$\times10^{2}$ & 63 & 4.6$\times10^{2}$ & 2.2 & 6.8 \\
        Creep Merit & 7.8e$\times10^{4}$ & 4.3$\times10^{5}$ & -8.3$\times10^{2}$ & 1$\times10^{7}$ & 5.1$\times10^{2}$ & 41 & 6.2$\times10^{3}$ & 12 & 200 \\
        \bottomrule
    \end{tabular}
\end{table*}

%%%%%%
%%%%%%
%%%%%%
\section{Results and Discussion}\label{sec:results}

The statistical summary of data presented in Table~\ref{tab:extended_data_stats_skew_kurt} highlights key distribution characteristics, such as skewness and kurtosis, which have significant implications for model performance. Features like “1300 Min Creep CB” exhibit extreme skewness (39) and kurtosis (1500), indicating highly non-normal distributions with substantial outliers. These characteristics point to the need for targeted preprocessing techniques. For example, log or quantile transformations can help normalize such skewed features, while more robust models may be required to handle data with extreme outliers. In contrast, features with relatively low skewness and kurtosis, such as “Yield Strength (YS) 1000°C,” follow more normal distributions, making them suitable for direct training. Applying these transformations is crucial for avoiding biased predictions and improving model stability and accuracy during training.

On the other hand, more normally distributed features, such as Yield Strength (YS) 1000$^\circ$C and Pugh Ratio, show relatively low skewness and kurtosis values, indicating symmetrical distributions with fewer outliers. These features are less likely to cause instability in the trained models and typically contribute to more reliable and consistent predictions. However, features with moderate skewness and kurtosis, such as Scheil LT and Kou Criteria, suggest some degree of asymmetry and outliers, which could still impact model performance if not accounted for during preprocessing. 

Selecting Kou Criteria for hyperparameter optimization offers several advantages, particularly because of its high variability and significance in determining material behavior. The feature’s relatively high skewness (2.2) and kurtosis (6.8) indicate a wide range of values, capturing both common and extreme cases, which can be beneficial for exploring a broader parameter space during optimization. This variability allows the model to learn from diverse scenarios, potentially enhancing its ability to capture complex relationships. Additionally, optimizing hyperparameters based on Kou Criteria could lead to a model that is more sensitive to critical outliers and boundary conditions, making it well-suited for applications where extreme performance metrics are of interest.

%%%%%%
%%%%%%

\subsection{\textbf{Effect of Hyperparameter Optimization}}

%%% Explaining the hyper-parameter optimization

\begin{figure}[!ht]
    \centering
    \scriptsize
    \begin{overpic}[width=0.98\linewidth]{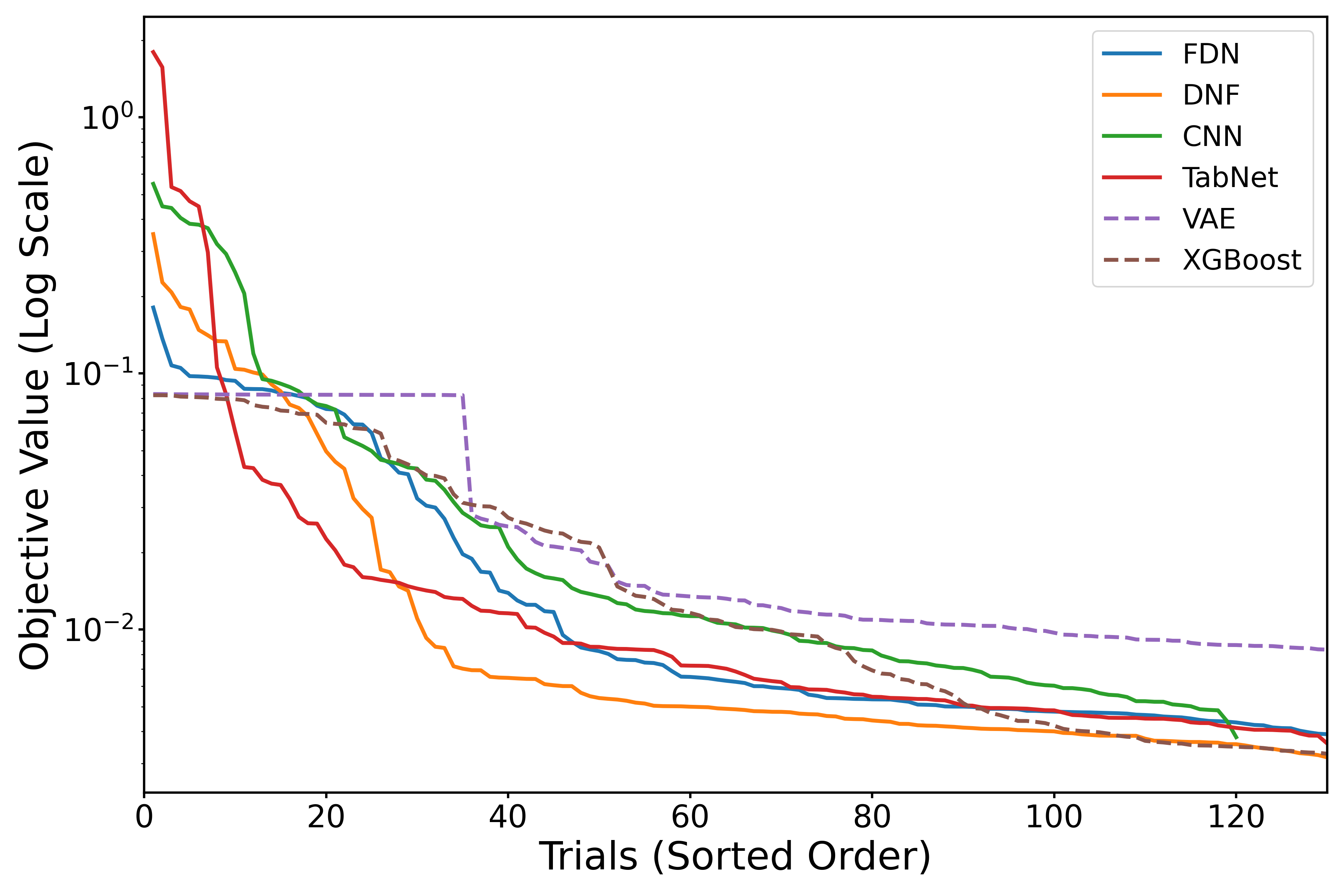}%
    \put(1,65){(a)}%
    \end{overpic}%
    \\ \vspace{0.3cm}
    \begin{overpic}[width=0.32\linewidth]{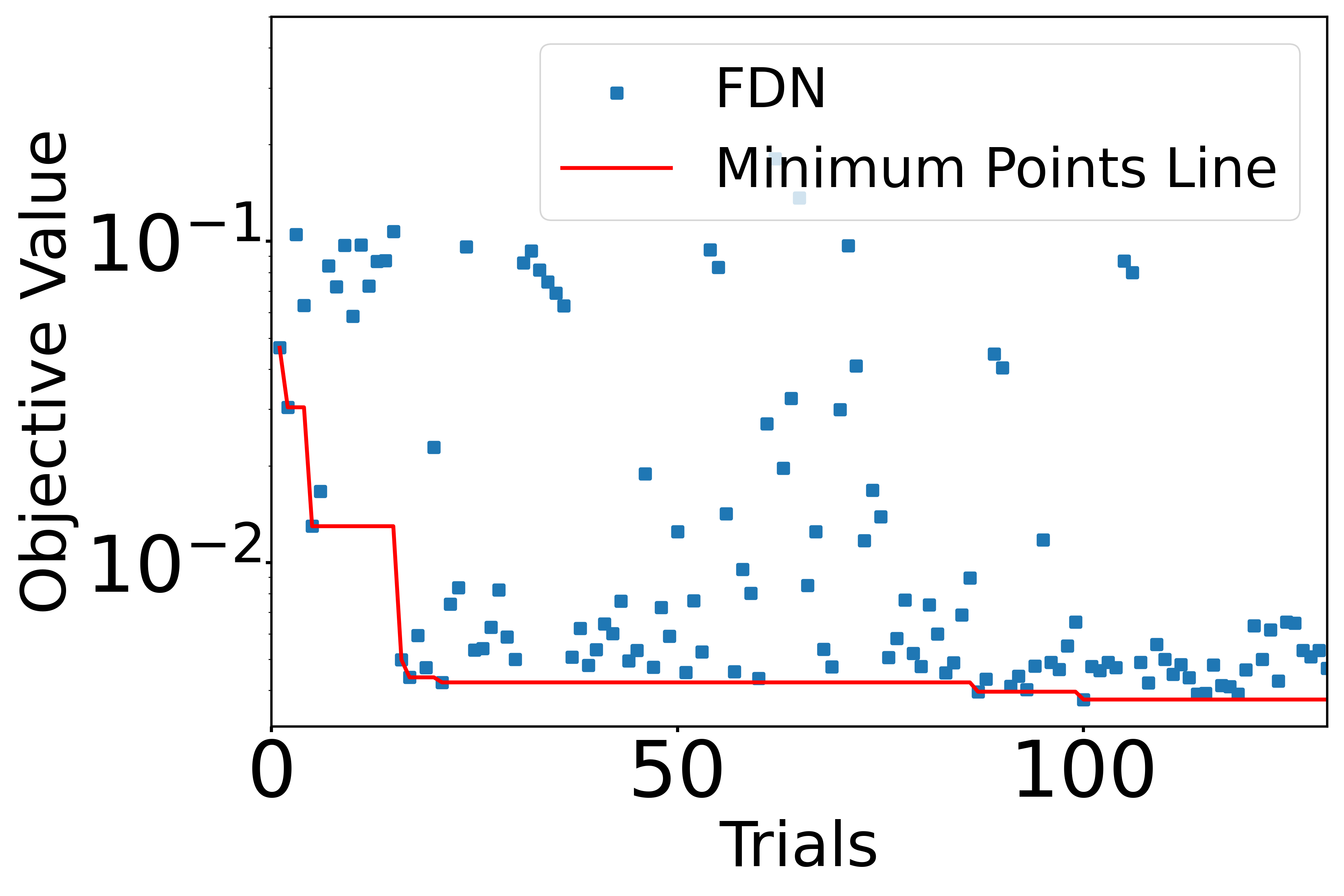}%
    \put(1,70){(b)}%
    \end{overpic}%
    \begin{overpic}[width=0.32\linewidth]{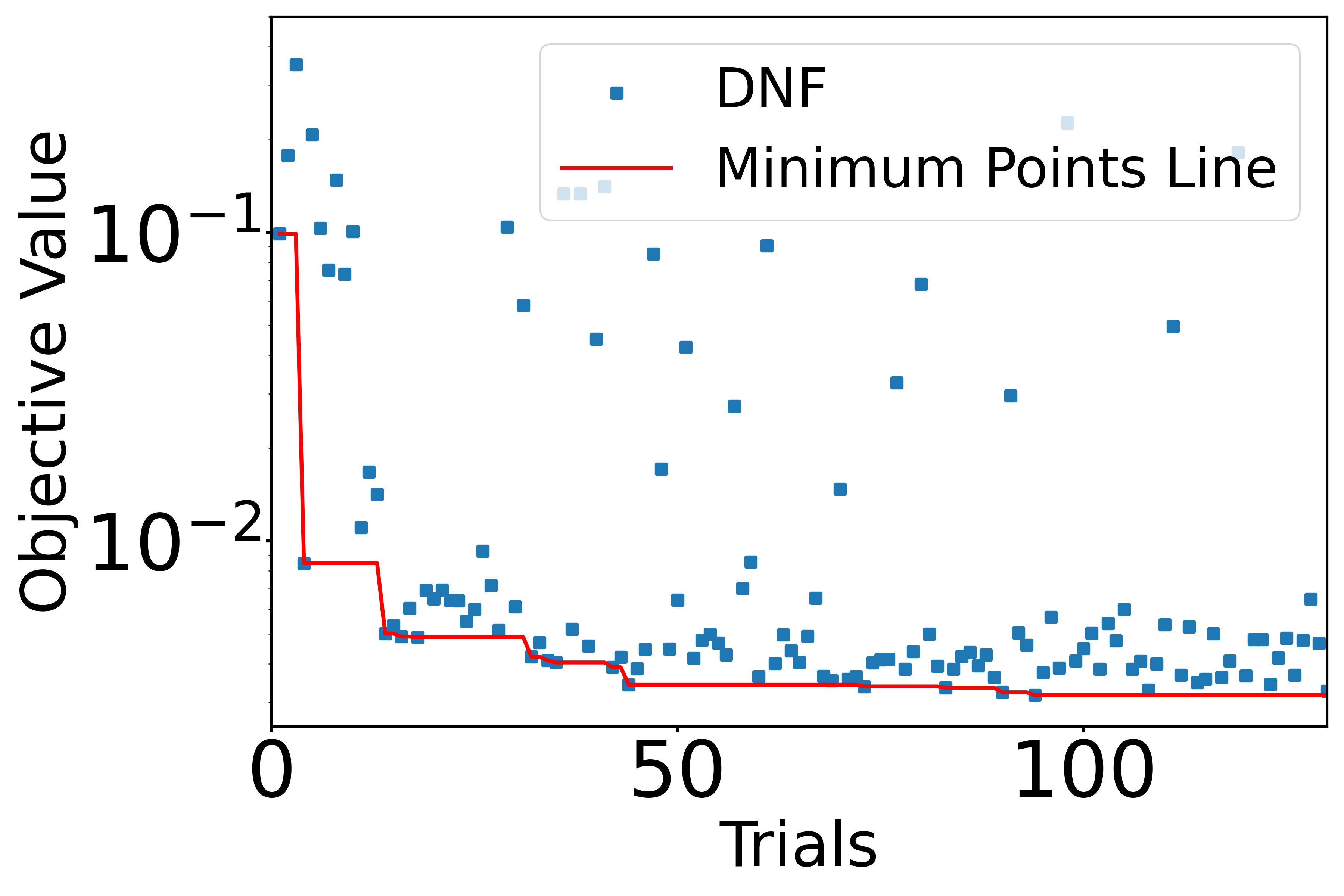}%
    \put(1,70){(c)}%
    \end{overpic}%
    \begin{overpic}[width=0.32\linewidth]{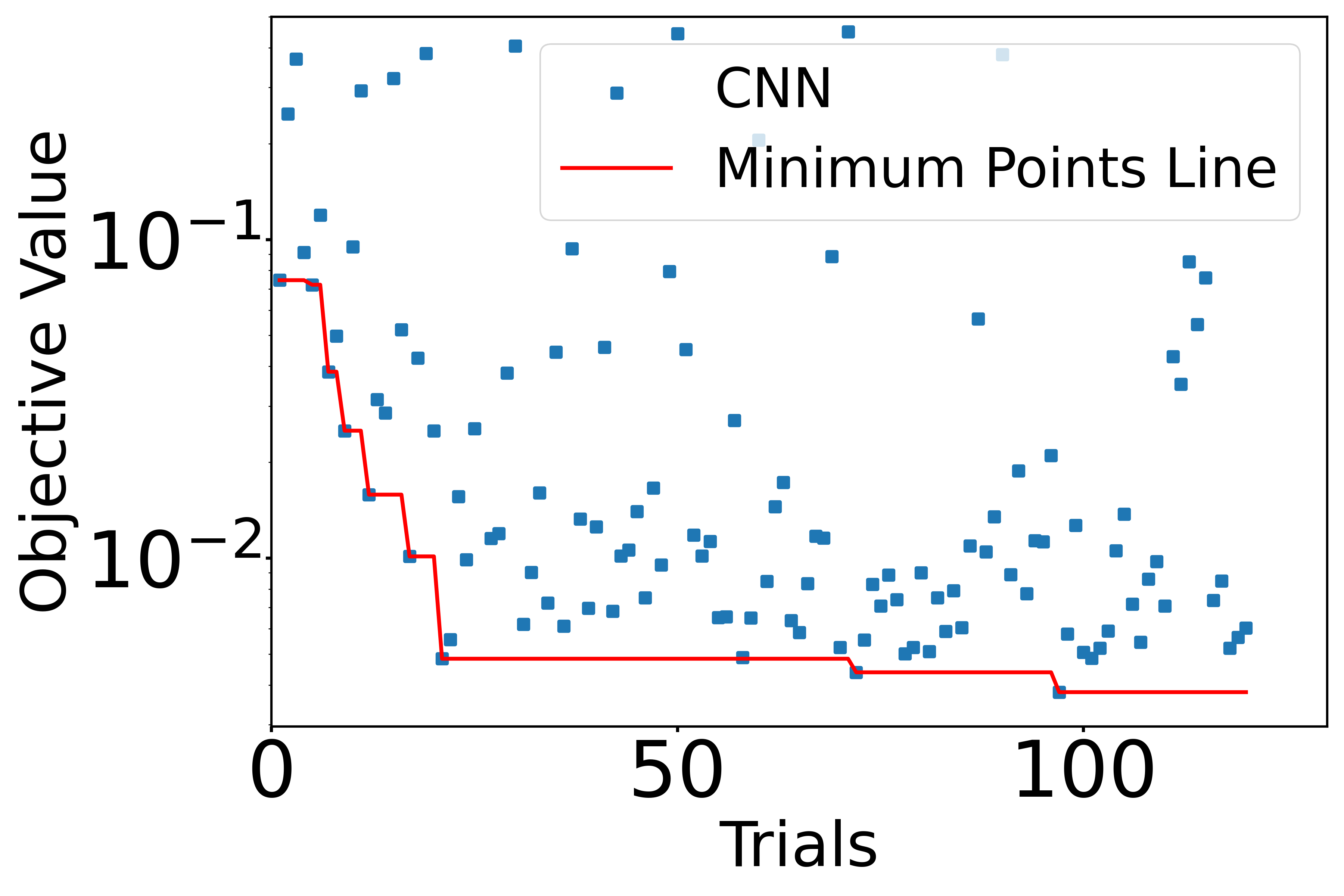}%
    \put(1,70){(d)}%
    \end{overpic}%
    \\
    \begin{overpic}[width=0.32\linewidth]{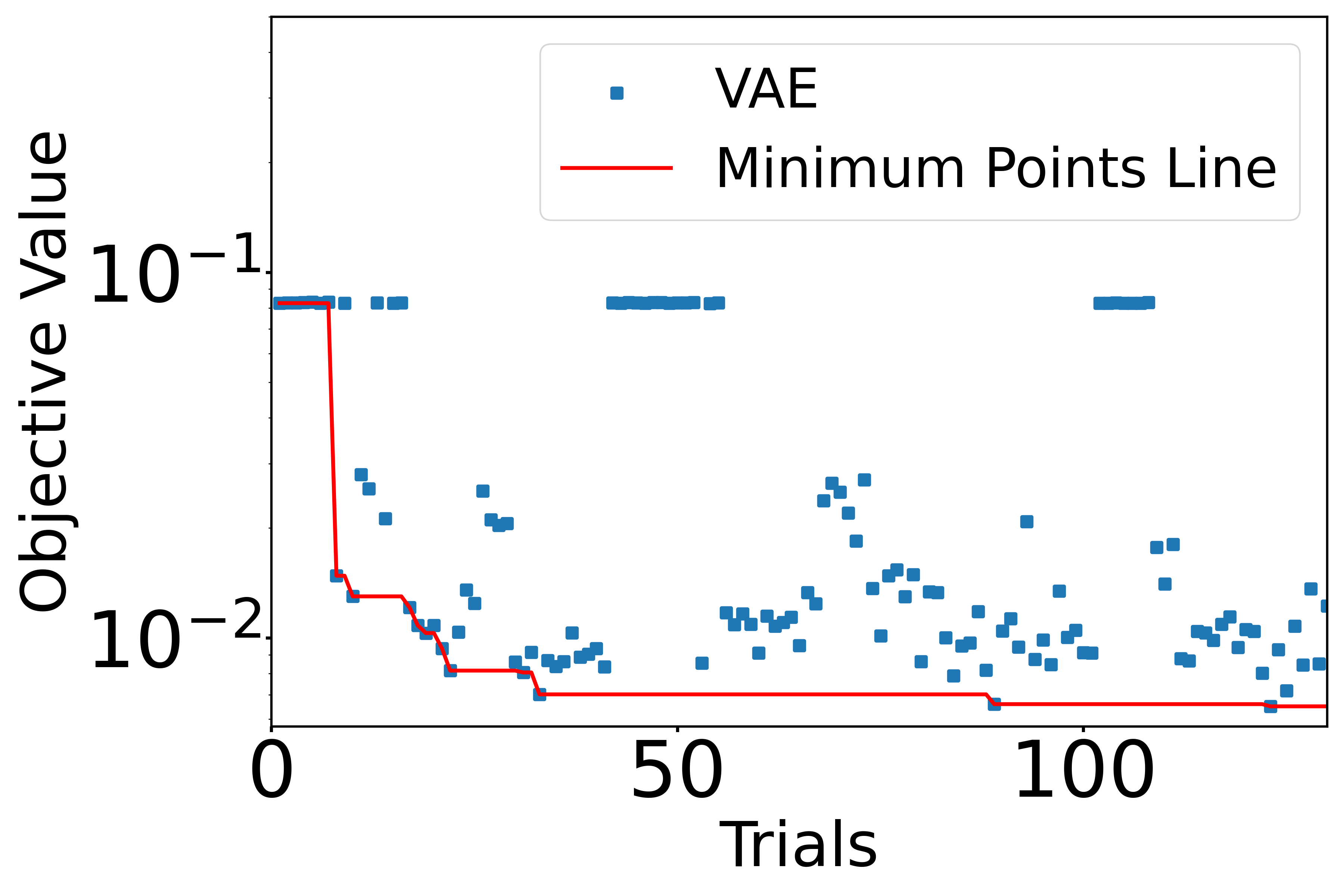}%
    \put(1,70){(e)}%
    \end{overpic}%
    \begin{overpic}[width=0.32\linewidth]{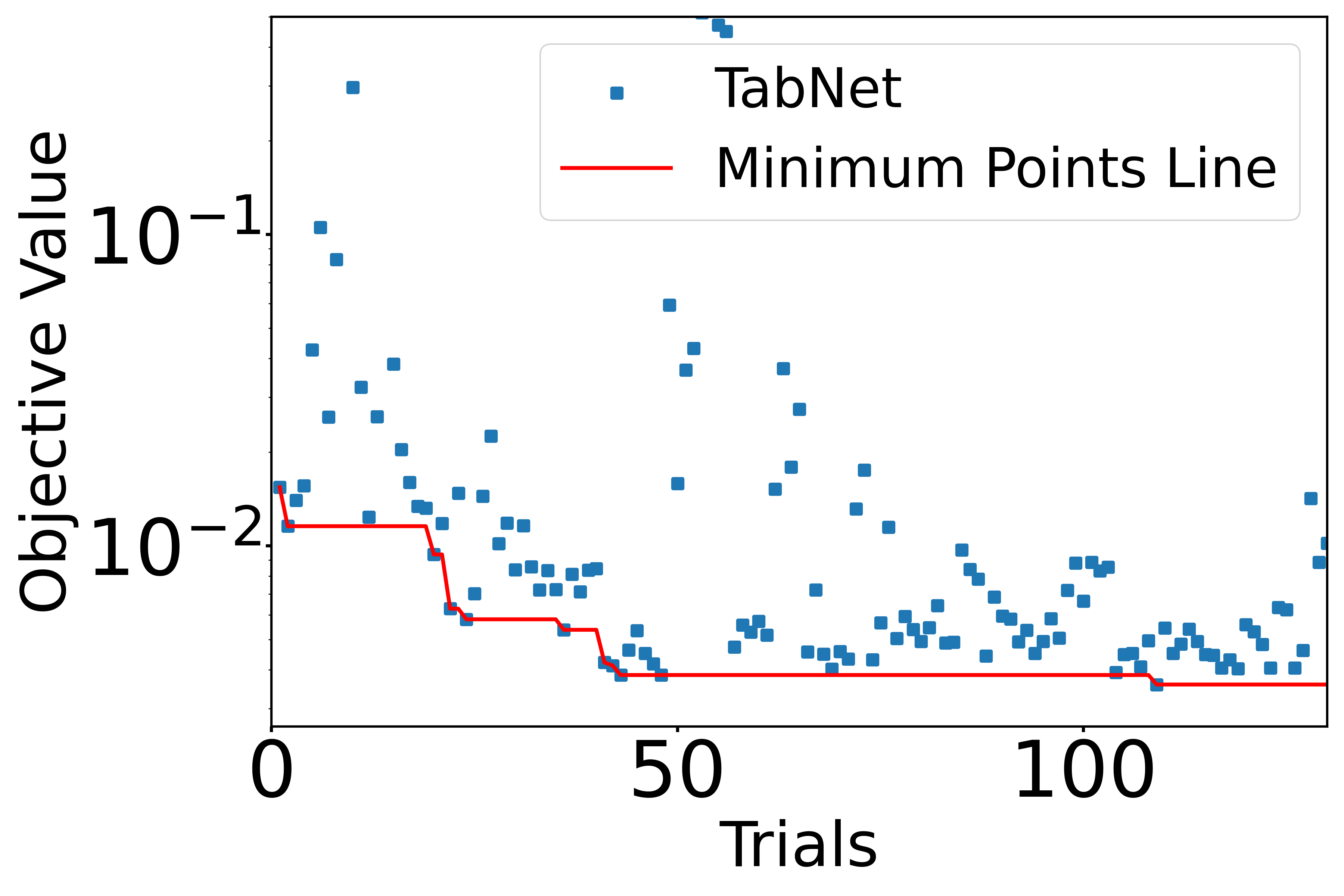}%
    \put(1,70){(f)}%
    \end{overpic}%
    \begin{overpic}[width=0.32\linewidth]{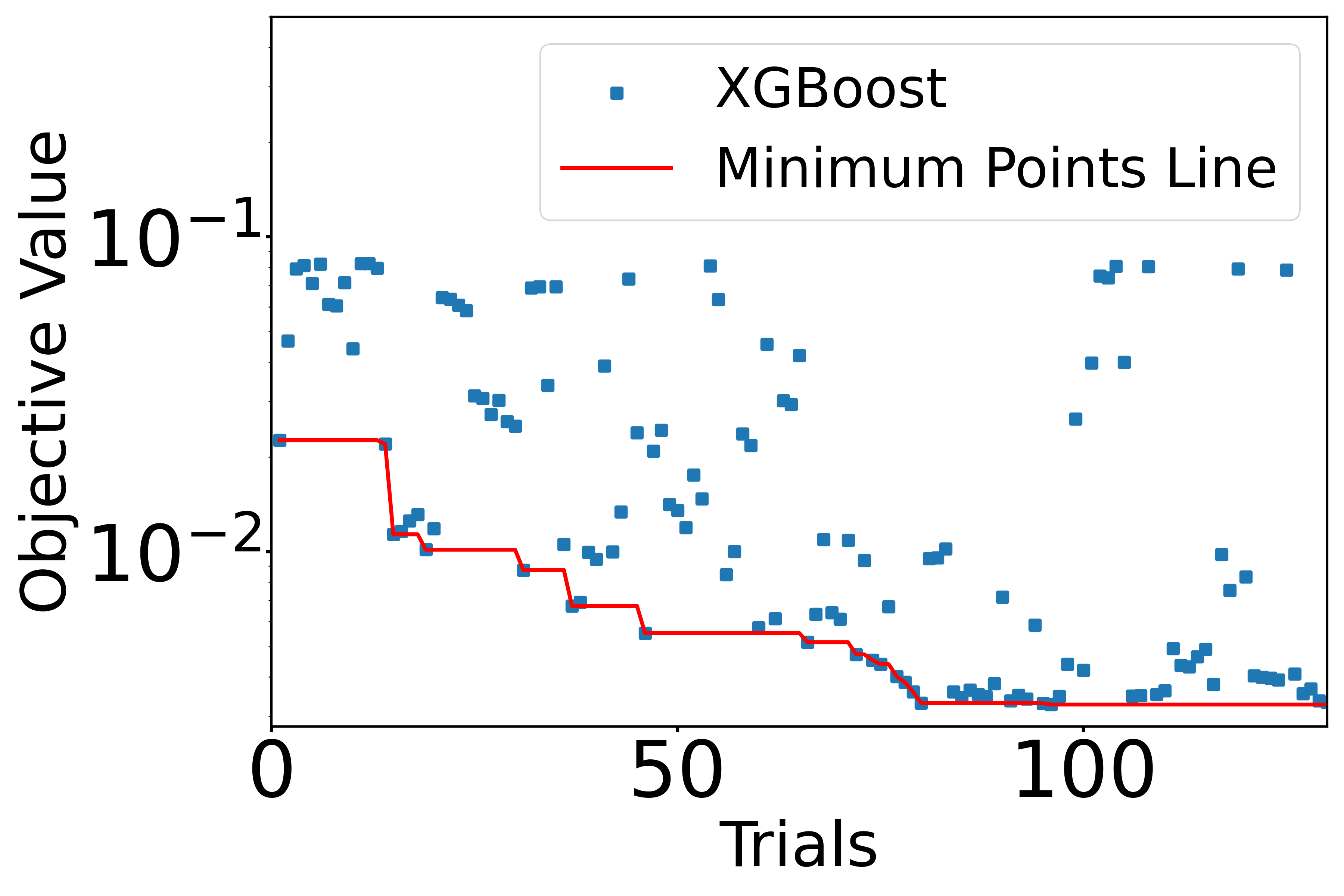}%
    \put(1,70){(g)}%
    \end{overpic}%
    \caption{Optimization history for different models, (a) sorted trials, origial sequence of Bayesian optimization for (b) FDN, (c) DNF, (c) CNN, (d) VAE, (e) TabNet, (f) XGBoost. }
    \label{fig:optuna-history}
\end{figure}

Figure~\ref{fig:optuna-history} shows the variance in model performance during hyperparameter optimization, comparing multiple encoder-decoder architectures with VAE and XGBoost conducted using Bayesian optimization. The y-axis represents the objective value on a log scale, while the x-axis shows the number of trials, sorted by their objective performance. The goal is to minimize the loss, with lower curves indicating better optimization results. The individual convergence plots provide a clearer understanding of how each model responds to Bayesian optimization, offering insights beyond the sorted order view of Figure~\ref{fig:optuna-history}. The optimization speed and behavior vary significantly across the models, as illustrated in Fig.~\ref{fig:optuna-history}. Both the FDN and DNF models exhibit rapid initial improvements, with sharp drops in objective values within the first 20-30 trials, quickly converging to stable solutions. In contrast, models like VAE and XGBoost show slower, more gradual improvements across trials, requiring more extensive exploration to identify optimal configurations. This variation in convergence speed has practical implications: while FDN and DNF can quickly find strong solutions with fewer trials, models like VAE and XGBoost benefit from longer optimization runs to thoroughly explore their parameter spaces. Understanding these differences is crucial for efficiently allocating computational resources during hyperparameter tuning, particularly when scaling to larger datasets.

%%%%%%%%

Table~\ref{tab:best_hyperparameters} highlights the tailored hyperparameter configurations across models. The FDN and DNF employ smaller learning rates (10$^{-4}$) compared to the higher rates used by TabNet and XGBoost (10$^{-2}$). Batch sizes also vary, with DNF using smaller batches (32), while VAE benefits from larger batches (128). Regularization terms ($\lambda$) are applied in FDN, DNF, and VAE to prevent overfitting. Adam is the preferred optimizer for most models, except for TabNet, which benefited using RMSprop. Additionally, architecture-specific parameters like conjunction units (112) are critical for DNF performance. 

\begin{table*}[!ht]
    \centering
    \caption{Best Hyperparameters for Different Models in the Study}
    \scriptsize
    \label{tab:best_hyperparameters}
    \begin{tabular}{llllllllc}
        \toprule
        \textbf{Study Name} & \textbf{Latent dim} & \textbf{Drop-out rate} & \textbf{Learning rate} & \textbf{Optimizer} & \textbf{Batch size} & \textbf{Epochs} & \textbf{Additional Hyperparameters} \\
        \midrule
        FDN & 192 & 0.1 & 1.26$\times10^{-4}$ & adam & 96 & 50 & $\lambda$=$3.69\times10^{-6}$, $\alpha$=0.0164 \\
        DNF & 64 & 0.1 & 5.39$\times10^{-4}$ & adam & 32 & 50 & num\_conjunctions=10, conjunction\_units=112 \\
        CNN & 64 & 0.1 & 3.77$\times10^{-4}$ & adam & 32 & 50 & Kernel size=4 \\
        %LSTM & 32 & 0.2 & 4.09$\times10^{-4}$ & adam & 8 & 50 & LSTM Units=224 \\
        TabNet & - & - & 4.24$\times10^{-3}$ & RMSprop & 72 & 50 & $n_d$=24, $n_a$=32, $\lambda$=7.98$\times10^{-4}$ \\
        VAE & 16 & 0.2 & 3.58$\times10^{-4}$ & adam & 128 & 50 & $\lambda$=2.87$\times10^{-5}$, $\alpha$=0.045 \\
        XGBoost & - & - & 9.58$\times10^{-2}$ & - & - & - & n\_estimators=50, Max depth=10 \\
        \bottomrule
    \end{tabular}
\end{table*}

\begin{figure*}[!ht]
    %\centering
    \tiny
    % First row of images
    \begin{overpic}[width=0.32\textwidth]{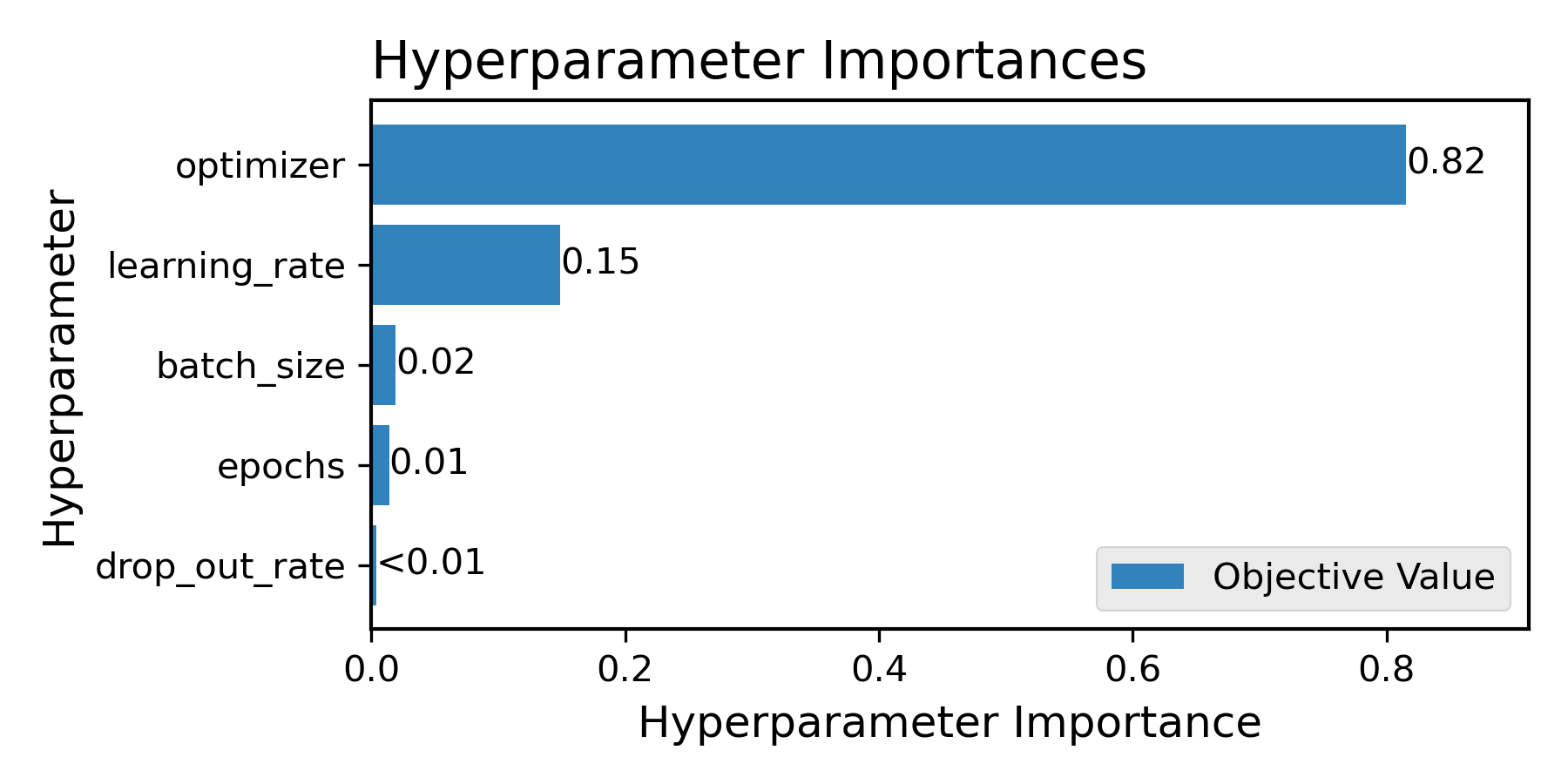}%
    \put(10,47){(a)}%
    \end{overpic}%
    \begin{overpic}[width=0.32\textwidth]{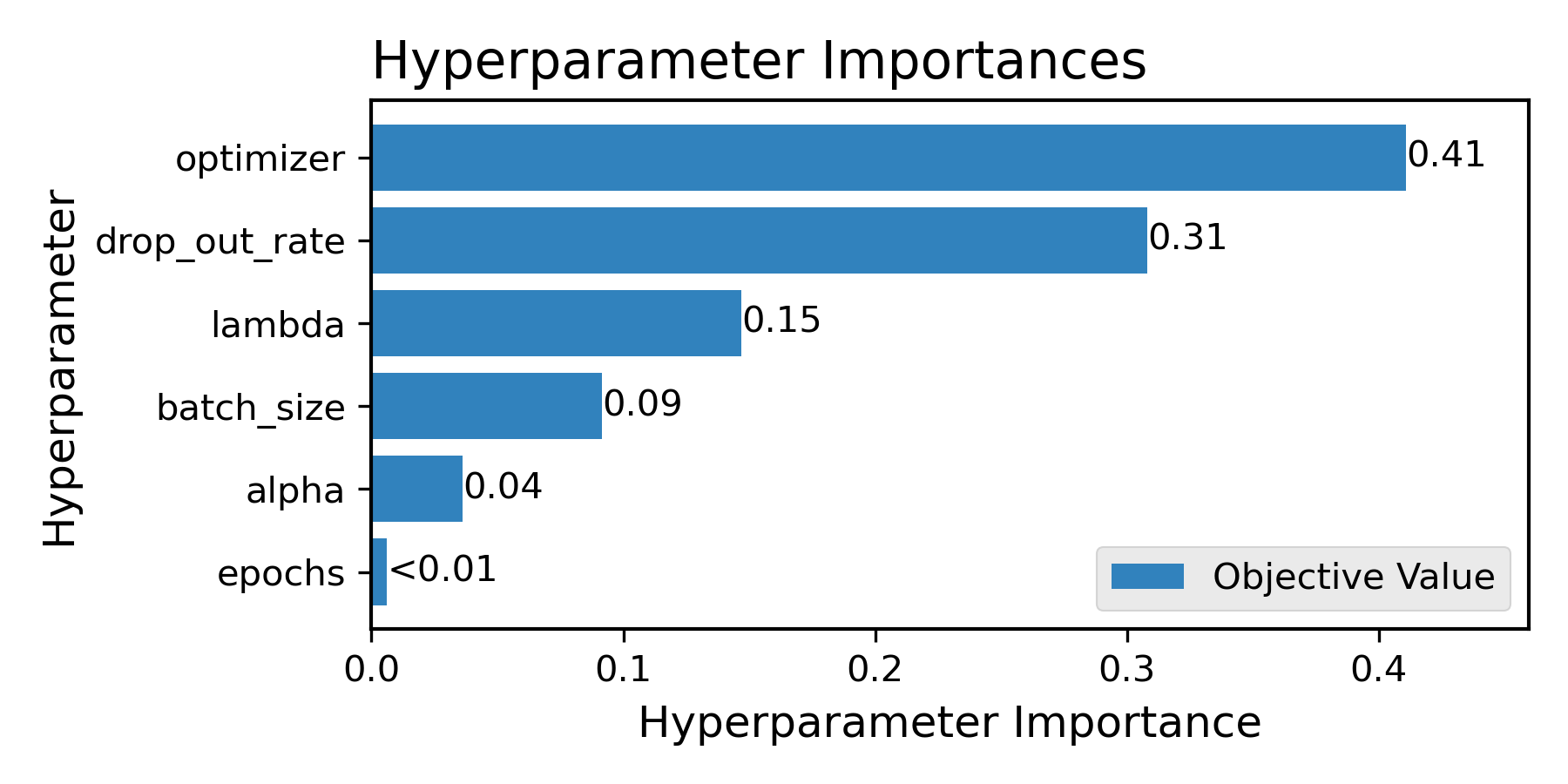}%
    \put(10,47){(b)}%
    \end{overpic}%
    \begin{overpic}[width=0.32\textwidth]{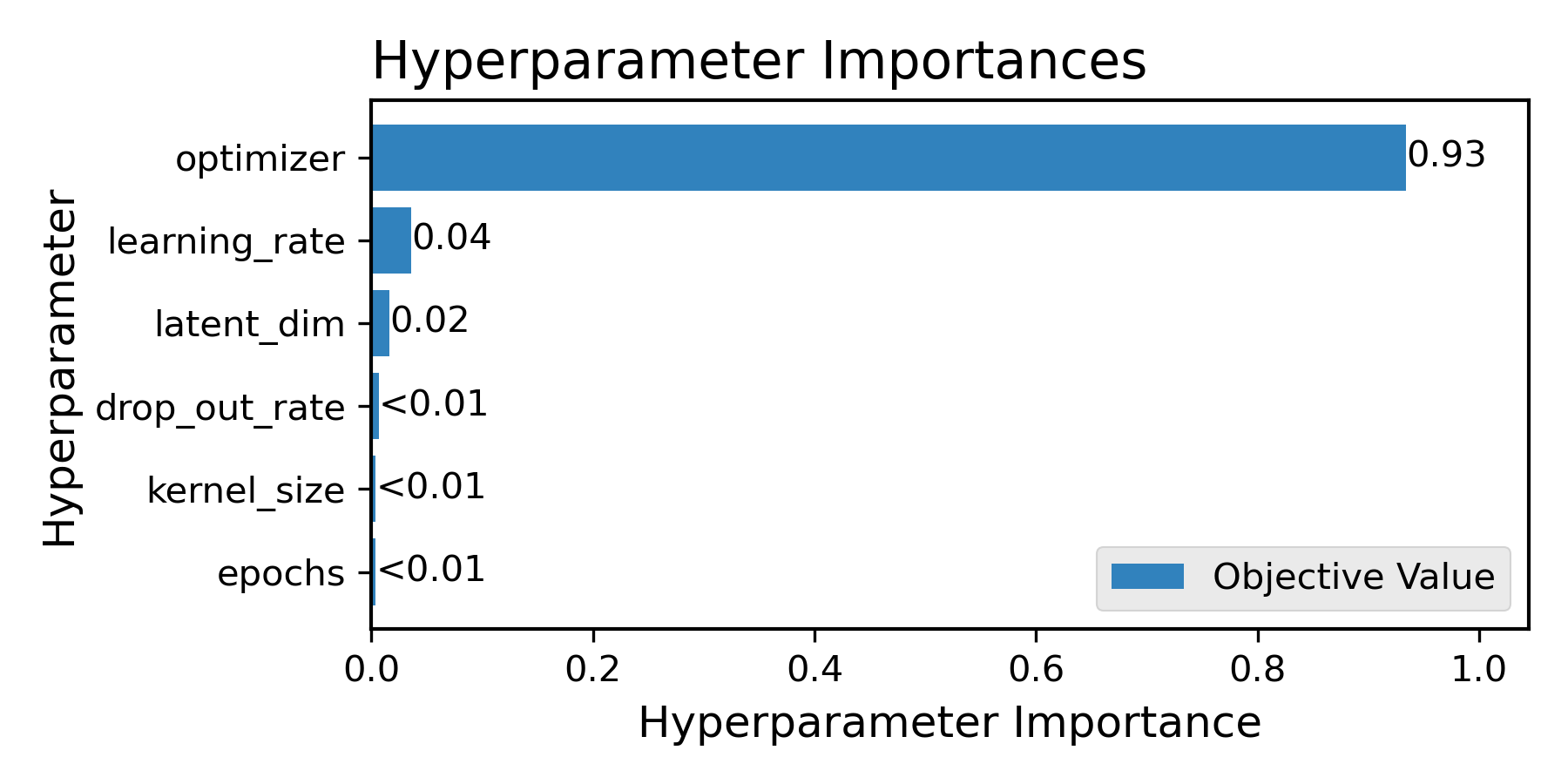}%
    \put(10,47){(c)}%
    \end{overpic}%
    \\
    % Second row of images
    \begin{overpic}[width=0.32\textwidth]{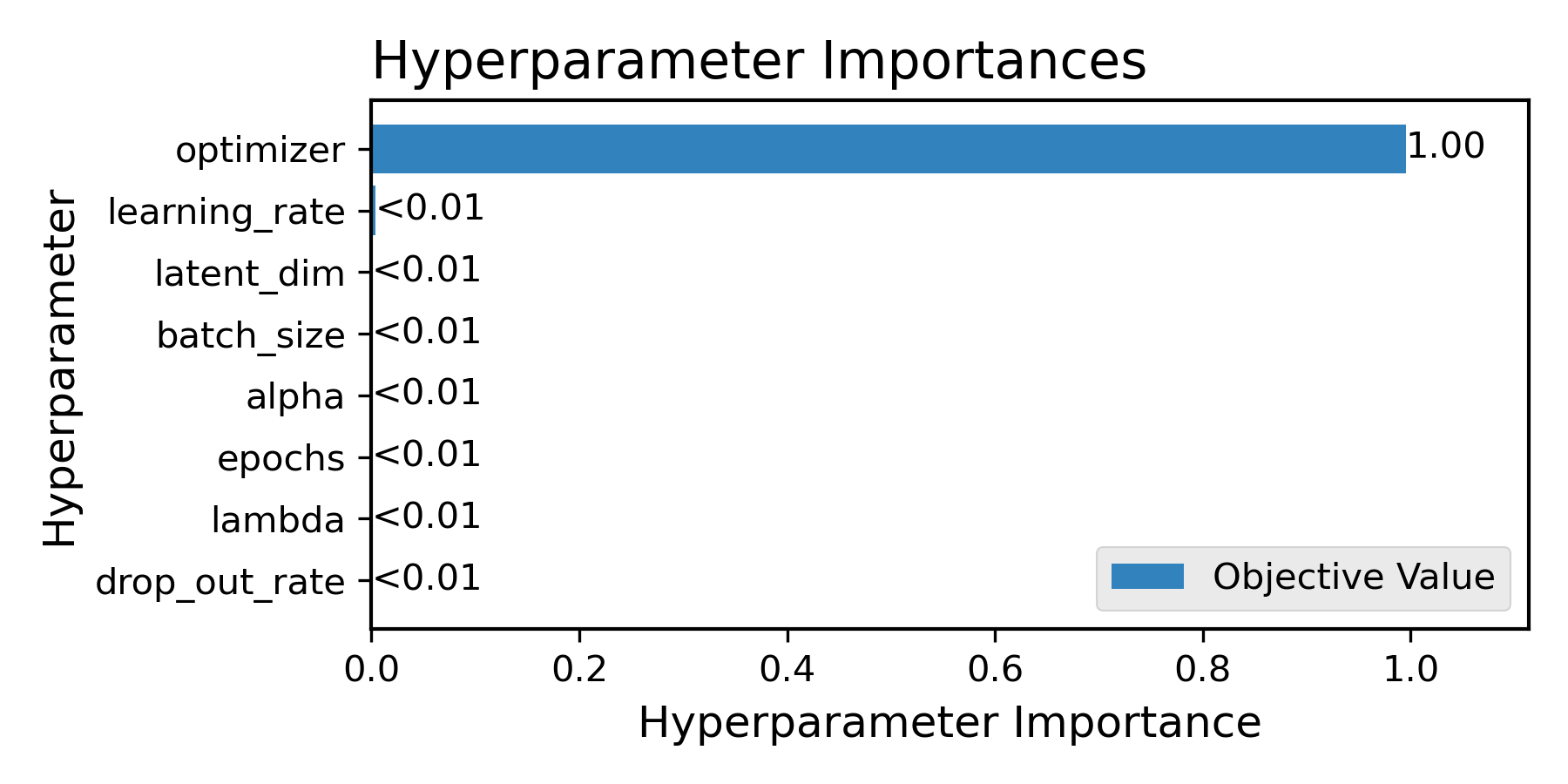}%
    \put(10,47){(d)}%
    \put(44,11){\includegraphics[width=0.1683\textwidth]{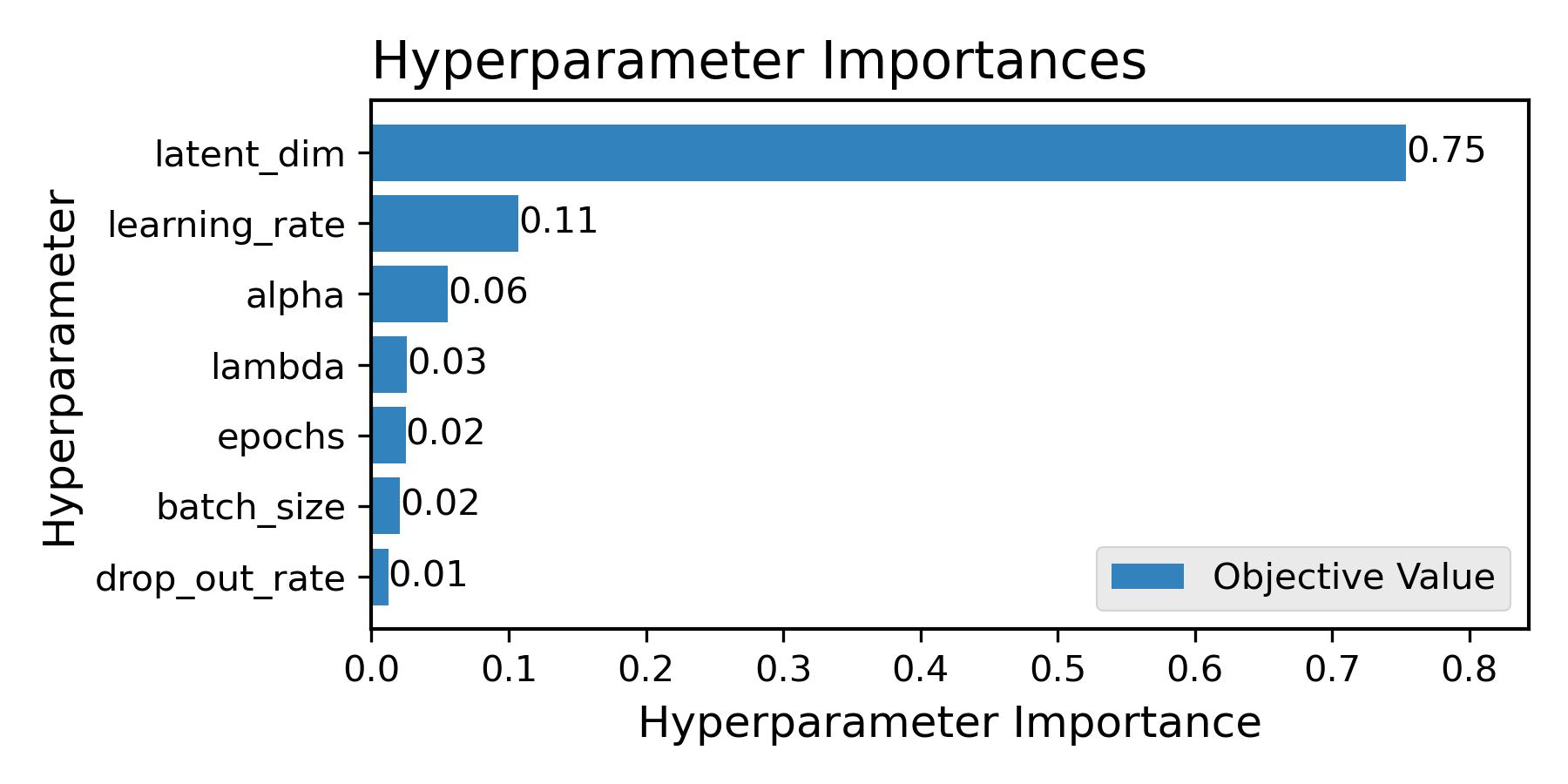}}%
    \end{overpic}%
    \begin{overpic}[width=0.32\textwidth]{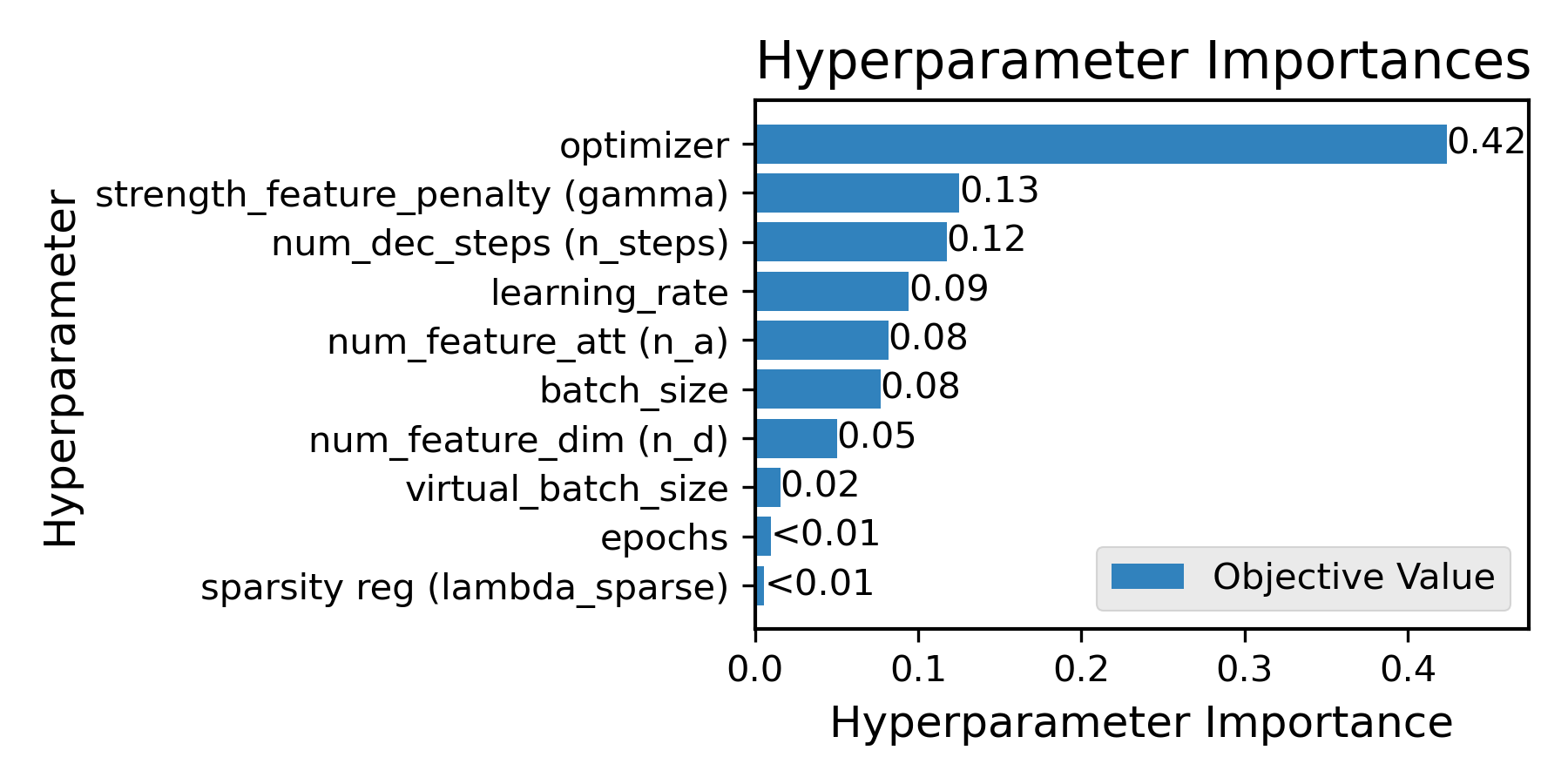}%
    \put(10,47){(e)}%
    \end{overpic}%
    %\begin{overpic}[width=0.32\textwidth]{hyperparam-importance_distributed-LSTM-Optuna.png}%
    %\put(10,47){(f)}%
    %\end{overpic}%
    %\\
    % Third row
    % Embedded table as part of the figure
    \begin{overpic}[width=0.32\textwidth]{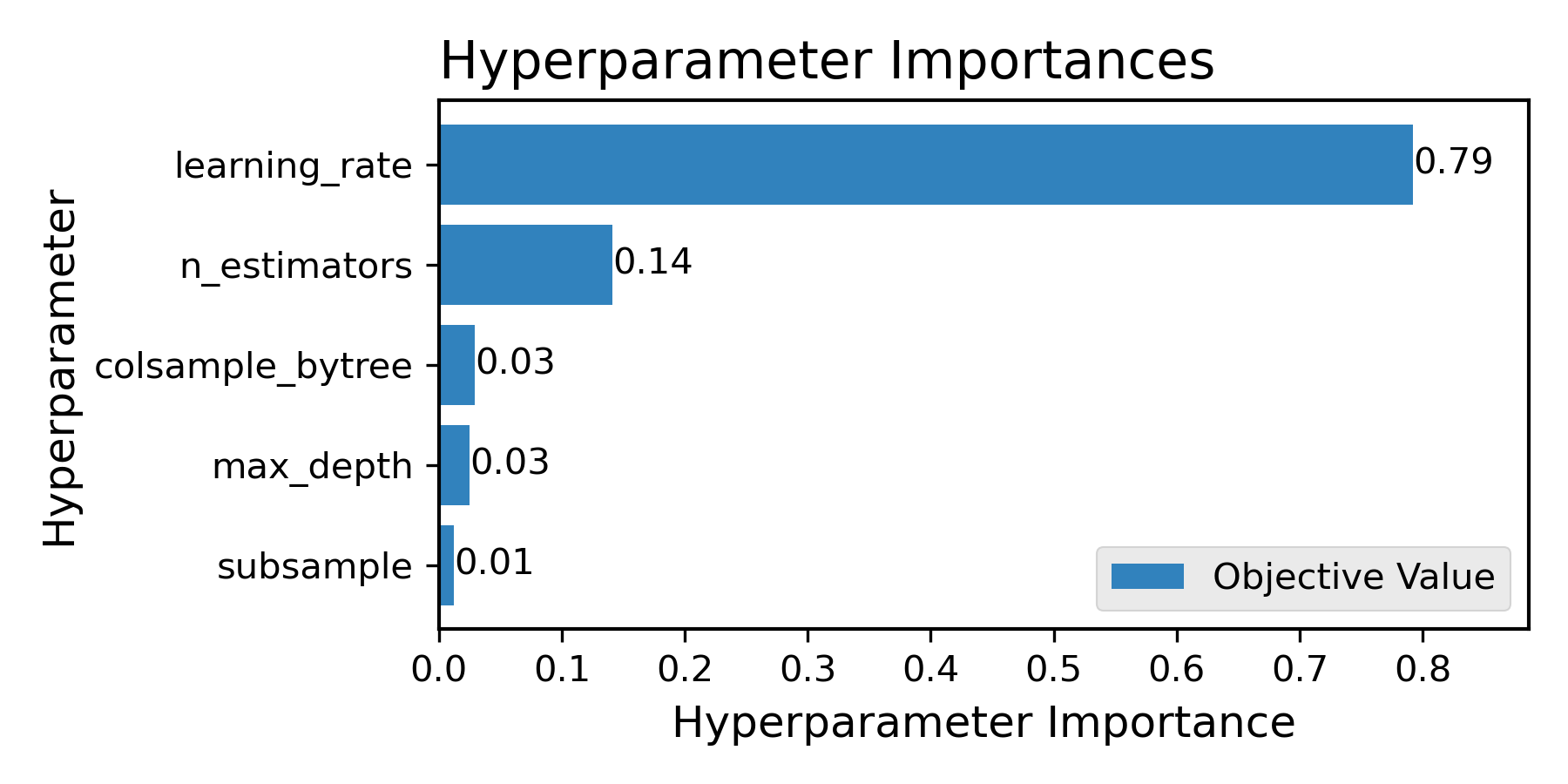}%
    \put(10,47){(f)}%
    \put(-140,-8){(g)}%
    \end{overpic}%

    %    \put(115,25){

    \vspace{0.25cm}
    \centering
        \begin{tabular}{lccccc}
            \toprule
            \textbf{Model} & \textbf{Best Loss Value} & \textbf{Optimizer} & \textbf{Best Trial Duration} & \textbf{Test R$^2$} & \textbf{Avg. Convergence Rate (\%)} \\
            \midrule
DNF & \textbf{0.003166} & adam & 29.39 & \textbf{0.962} & \textbf{-232.7} \\
XGBoost & 0.003270 & -- & \textbf{0.63} & 0.960 & -75.2 \\
TabNet & 0.003586 & RMSprop & 152.17 & -- & -95.0 \\
FDN & 0.003748 & adam & 330.67 & -- & -74.6 \\
CNN & 0.003796 & adam & 324.58 & 0.954 & -212.1 \\
%LSTM & 0.004317 & adam & 512.07 & 0.948 & -129.9 \\
VAE & 0.006506 & adam & 89.35 & 0.921 & -32.5 \\
            \bottomrule
        \end{tabular}
        %}
    %\end{overpic}%

    \caption{The hyper-parameters optimization process for different models. (a) DNF (b) FDN (c) CNN, (d) VAE, (e) TabNet, (f) XGBoost. (g) The enclosed table provides best Loss Values, best optimizer, best trial duration, best test set R$^2$, and average optimization convergence rate   for each model trained on ``Kou Criteria" feature. The convergence rate is negative because the model’s loss function value is decreasing}
    \label{fig:hyper-parameter-importance}
\end{figure*}

Figure~\ref{fig:hyper-parameter-importance} shows the hyperparameter importance for all investigated models. Each subplot shows the relative importance of various hyperparameters in determining model performance, providing insights into which parameters have the greatest impact on the objective value during optimization. For example, the DNF and FDN models (Figures \ref{fig:hyper-parameter-importance}a and \ref{fig:hyper-parameter-importance}b) highlight distinct hyperparameters as the most influential, whereas the CNN and VAE models (Figures \ref{fig:hyper-parameter-importance}c and \ref{fig:hyper-parameter-importance}d) demonstrate different sensitivity patterns. The variation in hyperparameter importance across models underscores the unique tuning needs of each architecture.

The results show that the optimizer is often more important than other hyperparameters because it directly controls how the model updates its weights during training. Optimizers like Adam and RMSprop can significantly influence convergence speed, training stability, and the quality of the final solution by adapting learning rates and managing gradients effectively. This adaptability often makes the optimizer more impactful than other hyperparameters like dropout rate or batch size, which primarily regulate model complexity or regularize training without fundamentally altering the learning process.

For FDN architecture, the type of optimizer (Adam, SGE, Adadelta) has significant impact on the model’s performance, contributing 100\% to the objective value’s reduction. If we fix the optimizer to adam, the most important hyperparameters are the latent dimension contributing 75\%, learning rate contributing 11\%, and negative slope of activation function contributing 6\%, and the rest of the parameters show minor importance. The FDN model shows consistent improvement across multiple runs with the best hyper-parameters, with progressively lower validation losses. Friedman’s test reveals a significant difference in performance between the runs, with a test statistic of 191.176 and a p-value of $1.11\times10^{-18}$, indicating that the observed performance variations are statistically meaningful. This suggests that factors such as random initialization or data splits may influence the model’s performance, warranting further investigation to enhance consistency. Alternatively, for VAE, the results of Friedman’s test, with a test statistic of 28.70 and a p-value of 0.052, suggest that there is no statistically significant difference in the performance of the model across multiple training runs at a 95\% confidence level. Although the p-value is close to the threshold of 0.05, it is slightly higher, indicating that we cannot reject the null hypothesis of no significant difference. This implies that the variations observed in the validation loss across the runs are likely due to random chance rather than meaningful differences in the model’s performance.

The enclosed table in Fig.~\ref{fig:hyper-parameter-importance} offers key insights into model performance, emphasizing variations in optimization efficiency, test R$^2$ values, and convergence rates. The DNF model stands out with the lowest loss value (0.003166) and the highest test R$^2$ (0.962), demonstrating its strong predictive capabilities. However, its negative convergence rate (-232.7\%) indicates challenges in achieving optimization stability, which may complicate its application in high-efficiency contexts. XGBoost, with a slightly higher loss (0.003270), exhibits exceptional computational efficiency, achieving the shortest trial duration (0.63 seconds), making it the most suitable for large-scale datasets where time is a critical factor. Despite its fast performance, XGBoost’s convergence rate (-75.2\%) signals potential limitations in optimization stability over extended trials. TabNet, while competitive in loss value (0.003586), exhibits a significantly longer trial duration (152.17 seconds) and a moderately negative convergence rate (-95.0\%), suggesting that its attention-based mechanism, although useful for interpretability, introduces additional computational complexity. The VAE, while showing reasonable test R$^2$ scores, face challenges with prolonged trial durations and suboptimal convergence rates, indicating inefficiencies in both computational cost and optimization robustness. These results indicate that DNF and XGBoost are well-suited for scenarios requiring both predictive accuracy and computational efficiency. However, models such as VAE, and TabNet, with their longer trial durations and convergence challenges, may be less suited for time-sensitive or large-scale applications. These insights can help guide model selection based on the trade-off between performance and optimization speed.

%%%%%%
\subsection{\textbf{Generalization of Models to Other Features}}
%%%%%%

\begin{figure*}[!ht]
    \centering
    \includegraphics[width=0.18\linewidth]{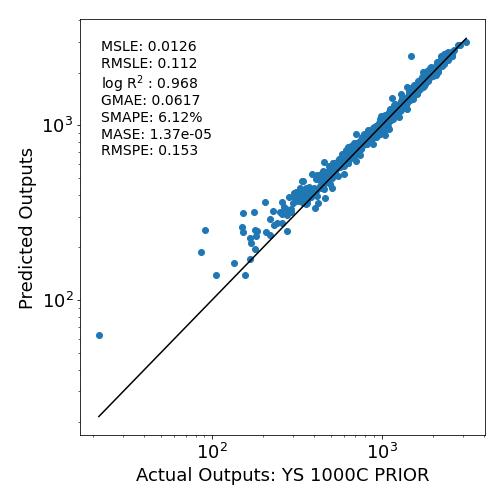}
    \includegraphics[width=0.18\linewidth]{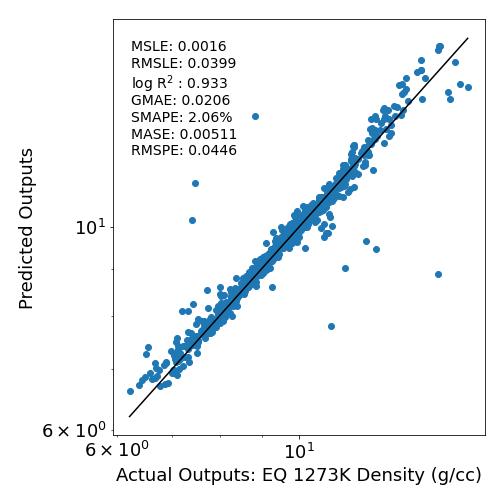}
    \includegraphics[width=0.18\linewidth]{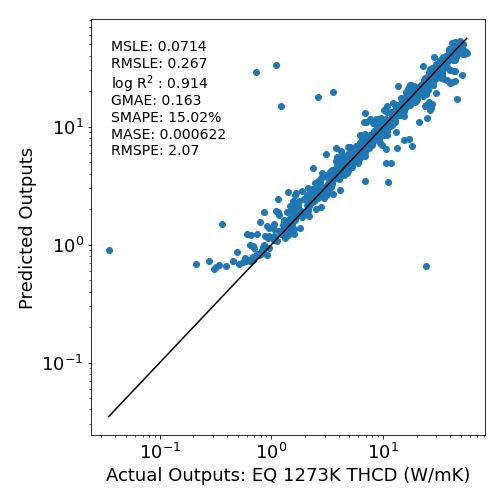}
    \includegraphics[width=0.18\linewidth]{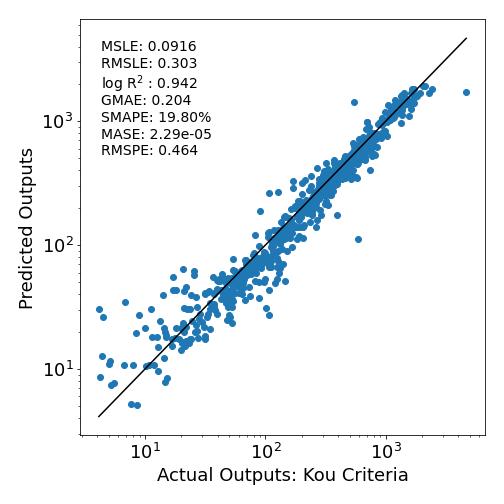}
    \includegraphics[width=0.18\linewidth]{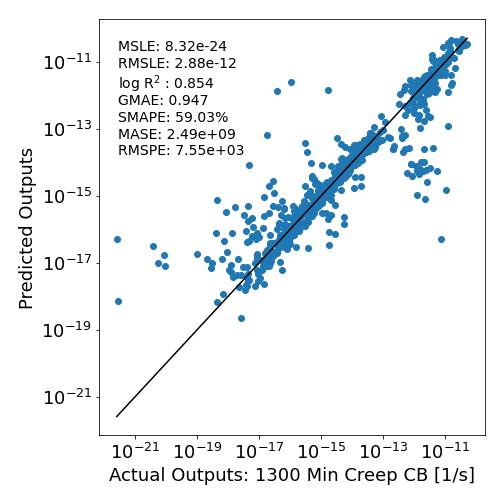}
    \\
    \includegraphics[width=0.18\linewidth]{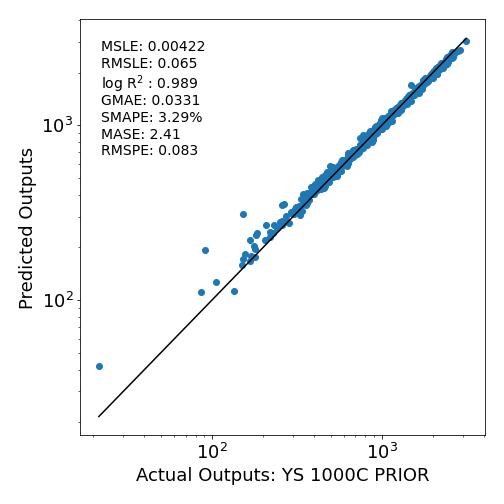}
    \includegraphics[width=0.18\linewidth]{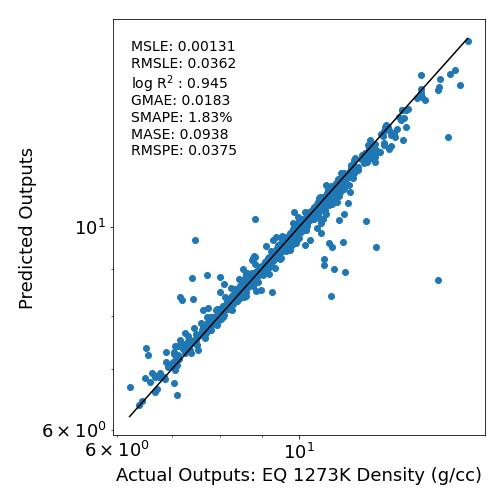}
    \includegraphics[width=0.18\linewidth]{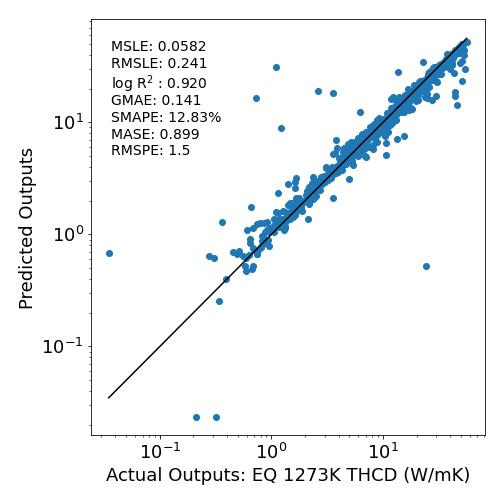}
    \includegraphics[width=0.18\linewidth]{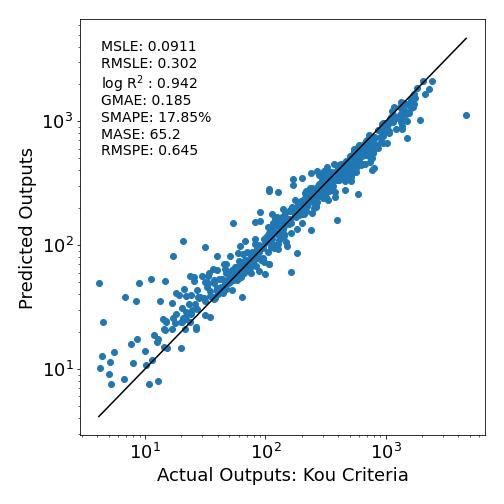}
    \includegraphics[width=0.18\linewidth]{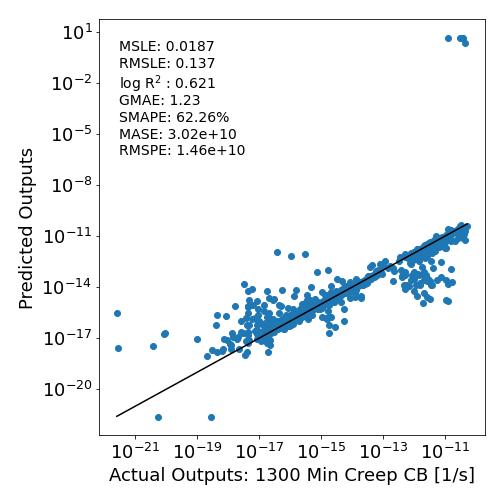}
    \\
    \includegraphics[width=0.18\linewidth]{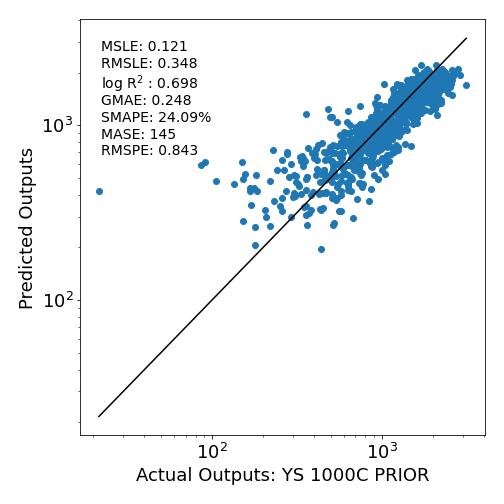}
    \includegraphics[width=0.18\linewidth]{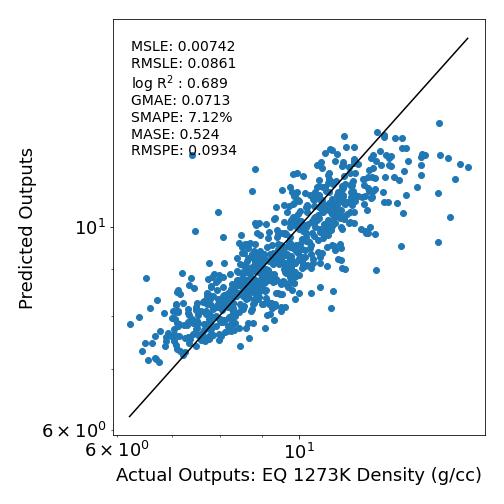}
    \includegraphics[width=0.18\linewidth]{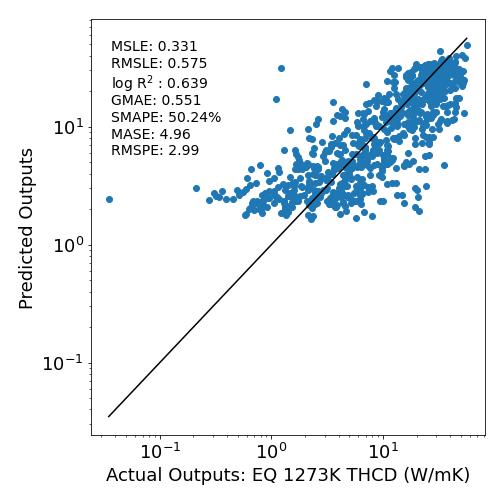}
    \includegraphics[width=0.18\linewidth]{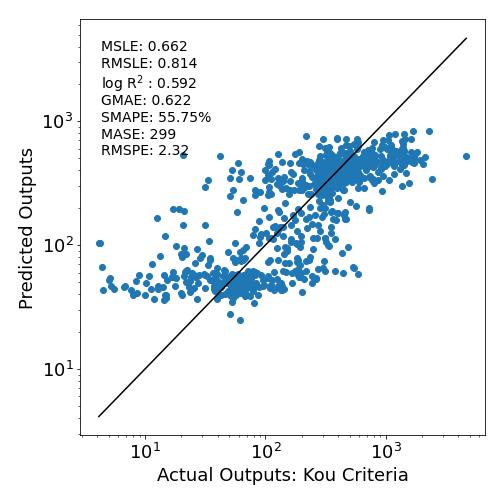}
    \includegraphics[width=0.18\linewidth]{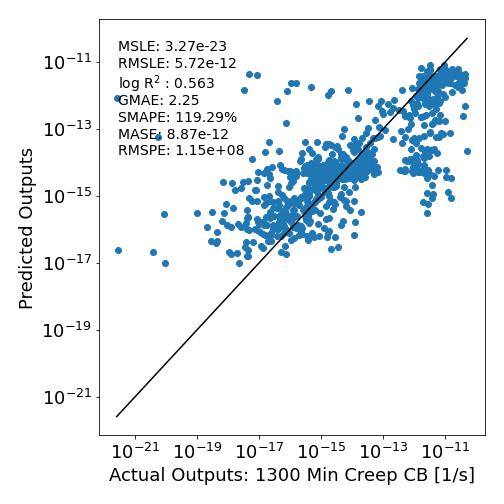}
    \\
    \includegraphics[width=0.18\linewidth]{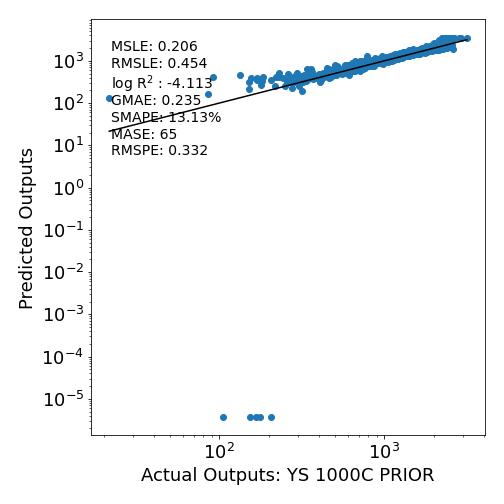}
    \includegraphics[width=0.18\linewidth]{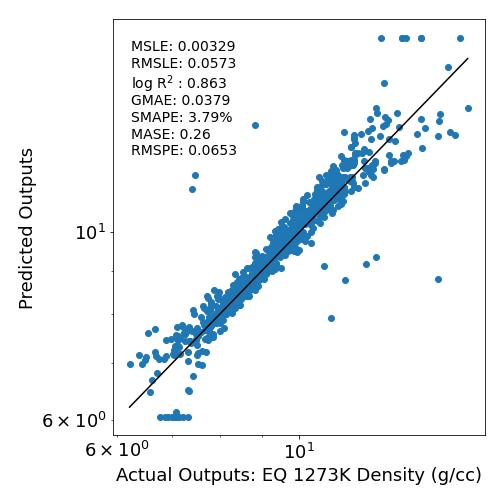}
    \includegraphics[width=0.18\linewidth]{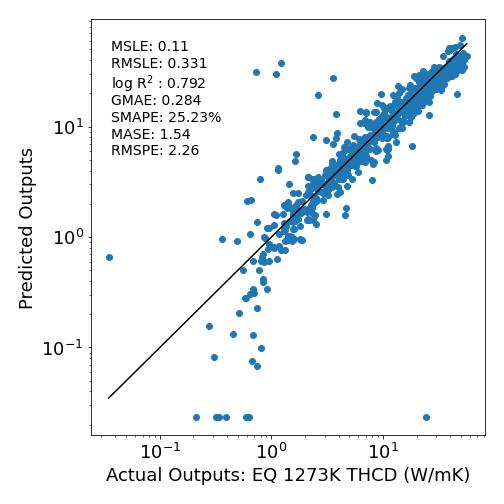}
    \includegraphics[width=0.18\linewidth]{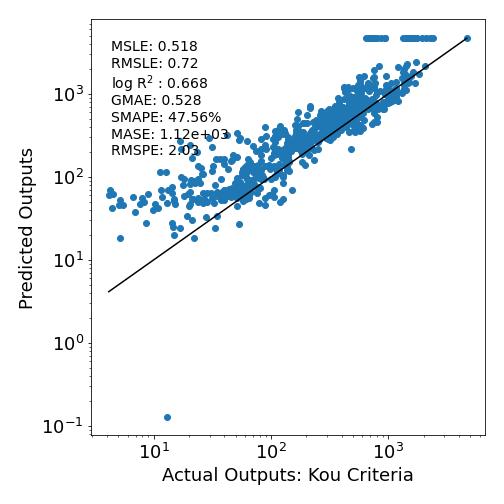}
    \includegraphics[width=0.18\linewidth]{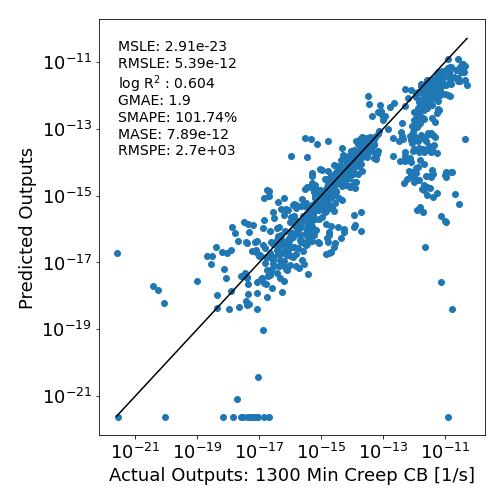}

    \caption{Parity plots illustrating the predictive performance of the FDN, XGBoost, VAE, and CNN models on five different output features from the dataset using optimzied parameters.}
    \label{fig:parity-plots-others}
\end{figure*}

The parity plots in Fig.~\ref{fig:parity-plots-others} compare the predictive performance of the FDN, XGBoost, VAE, and CNN models across five features: YS 1000C PRIOR, EQ 1273K Density, EQ 1273K THCD (W/mK), Kou Criteria, and 1300 Min Creep CB [1/s]. The metrics displayed in each plot—MSLE, RMSLE, and log R$^2$—show how well each model generalizes across a variety of feature distributions, ranging from relatively symmetric to highly skewed.

For features like YS 1000C PRIOR and EQ 1273K Density, the FDN and XGBoost models exhibit strong predictive performance, with low MSLE values (e.g., 0.00825 for FDN) and high log R$^2$ values (around 0.97 for both models). These results, along with low SMAPE and RMSPE values, suggest that these models can handle well-behaved, symmetric data distributions effectively. In contrast, the VAE and CNN models show a slight degradation in performance on these simpler features, especially for CNN on more complex features like 1300 Min Creep CB.

As feature complexity increases with Kou Criteria and 1300 Min Creep CB [1/s], both FDN and XGBoost maintain strong, though gradually declining, performance. However, VAE and CNN struggle significantly with these more skewed features, as shown by higher MSLE and RMSPE values. Particularly for 1300 Min Creep CB [1/s], the VAE model records an RMSPE as high as 1.12e+03, highlighting the model’s inability to handle extreme values and outliers. This comparison suggests that while FDN and XGBoost are more robust in handling complex, skewed features, VAE and CNN models show clear limitations in generalizing to these types of data.

%% generalization for other features - FDN Model

The FDN model excels with symmetric features such as YS 1000C PRIOR and EQ 1273K Density, showing high log R$^2$ values (around 0.97 and 0.93) and low MSLE and GMAE metrics. This reflects the model’s capability to capture underlying relationships effectively in standard material properties with well-behaved distributions. However, for highly skewed features like 1300 Min Creep CB [1/s], the FDN model’s performance degrades, with log R$^2$ dropping to 0.85 and SMAPE spiking to 59\%. This indicates difficulties in predicting extreme values, and suggests that specialized techniques, such as tailored loss functions or more feature engineering, may be required to enhance its performance on skewed data.

%% generalization for other features - VAE Model

Comparing this with the VAE model, the latter exhibits a similar trend, performing well on more balanced features but struggling with complex ones like 1300 Min Creep CB [1/s]. Despite lower MSLE values for symmetric features, the VAE model shows significant errors in skewed distributions, with an RMSPE over 1.12e+03 for 1300 Min Creep CB [1/s]. This highlights the challenge both models face with skewed data, though FDN generally shows more resilience than VAE, suggesting that advanced methods are necessary to improve their generalization on more complex features.

%% generalization for other features - XGBoost Model

The XGBoost model demonstrates strong generalization across all features, outperforming the other models in handling both symmetric and skewed distributions. For relatively well-behaved features like YS 1000C PRIOR and EQ 1273K Density, XGBoost shows very low MSLE values (around 0.0083 and 0.0102, respectively) and high log R$^2$ values (close to 0.97), indicating excellent predictive accuracy. The model’s robustness is further highlighted by low SMAPE and RMSPE values, confirming its ability to capture underlying patterns in the data effectively. However, what sets XGBoost apart is its resilience when faced with more challenging, skewed features like Kou Criteria and 1300 Min Creep CB [1/s]. While other models such as FDN and VAE show a decline in performance on these features, XGBoost manages to maintain relatively high accuracy, with a log R$^2$ of 0.92 for Kou Criteria and 0.88 for 1300 Min Creep CB [1/s], alongside lower error metrics than the other models. This suggests that XGBoost’s tree-based structure and inherent ability to handle non-linear relationships and extreme values make it particularly well-suited for complex, skewed data distributions.

Severe skewness, whether left or right, can significantly impair the performance of regression models, particularly when assumptions like normality of residuals are violated. Many models, including linear regressions, rely on the assumption that residuals are normally distributed, and skewed data disrupts this, leading to biased and inefficient estimates of the model’s coefficients. This distortion impacts the accuracy and reliability of the model’s predictions. Moreover, skewness often introduces outliers, particularly in the tails of the distribution, which can act as leverage points, disproportionately influencing the model and exacerbating issues like heteroscedasticity—where the variance of the residuals is inconsistent. As a result, skewed data can lead to poor model fits and inaccurate predictions, underscoring the need for advanced techniques or feature transformations to mitigate these effects.

%%\begin{table*}[]
%    \centering
%    \scriptsize
%    \begin{tabular}{l|ccccc}
%        \toprule
%        \textbf{Model Name} & Density 1000C & THCD 1000C (W/mK) & Yield Strength 1000C [] & Nabarro Herring Creep 1000C [] & Kou Criterion [] \\ \midrule
%        XGBoost    & \textcolor{blue}{0.003 (0.968)} & \textcolor{blue}{0.004 (0.954)} & \textcolor{blue}{0.000 (0.999)} & \textcolor{blue}{0.073 (0.145)} & \textcolor{blue}{0.016 (0.810)} \\
%        Encoder-Decoder (DNF-Net)    & 0.005 (0.943) & 0.020 (0.773) & 0.001 (0.992) & 0.085 (-0.001) & 0.006 (0.926) \\
%        Encoder-Decoder (FDN) & 0.004 (0.947) & 0.009 (0.890) & 0.001 (0.983) & 0.101 (-0.192) & 0.006 (0.928) \\
%        Encoder-Decoder (TabNet)      & 0.005 (0.935) & 0.007 (0.916) & 0.002 (0.978) & 0.083 (0.022) & 0.007 (0.910) \\
%        Encoder-Decoder (1D CNN) & 0.006 (0.930) & 0.008 (0.912) & 0.001 (0.983) & 0.084 (0.011) & 0.007 (0.915) \\ 
%        Encoder-Decoder (1D LSTM) & 0.005 (0.943) & 0.010 (0.889) & 0.002 (0.973) & 0.000 (-0.141) & 0.007 (0.918) \\ 
%        Encoder-Decoder (Variational) & 0.008 (0.909) & 0.009 (0.890) & 0.007 (0.997) & 0.085 (-0.002) & 0.005 (0.934) \\ 
%        \bottomrule
%    \end{tabular}
%    \caption{Test results on tabular dataset. Presenting the performance for each model. MSE (R$^2$) are presented for five features.}
%    \label{tab:model_performance1}
%\end{table*}

\begin{table*}[!ht]
    \centering
    \scriptsize
    \resizebox{\textwidth}{!}{%
    \begin{tabular}{l|ccccc}
        \toprule
        \textbf{Model Name} & YS 1000C PRIOR & 1273K Density (g/cc) & 1273K THCD (W/mK) & Kou Criteria & 1300C Creep CB [1/s] \\ 
        Skewness    & 0.48 & 0.77 & 0.97 & 2.2 & 39 \\ \midrule
XGBoost & 1613 (R$^2$: 0.995) & 0.1684 (R$^2$: 0.939) & 13.27 (R$^2$: 0.925) & 2.61e+04 (R$^2$: 0.854) & 0.111 (R$^2$: -3.06e+21) \\        
Encoder-Decoder (DNNF) & 8949 (R$^2$: 0.976) & 0.2755 (R$^2$: 0.902) & 15.43 (R$^2$: 0.913) & 2.59e+04 (R$^2$: 0.855) & 1.19$\times10^{-23}$ (R$^2$: 0.671) \\
Encoder-Decoder (FDN) & 5443 (R$^2$: 0.985) & 0.2106 (R$^2$: 0.925) & 14.56 (R$^2$: 0.918) & 1.93e+04 (R$^2$: 0.892) & 8.32$\times10^{-24}$ (R$^2$: 0.771) \\
TabNet & 1210 (R$^2$: 0.969) & 0.2501 (R$^2$: 0.919) & 13.73 (R$^2$: 0.917) & 1.361e+05 (R$^2$: 0.161) & 2.58$\times10^{-23}$ (R$^2$: 0.403) \\
Encoder-Decoder (1D CNN) & 4348 (R$^2$: 0.881) & 0.4662 (R$^2$: 0.833) & 22.76 (R$^2$: 0.872) & 4.48e+05 (R$^2$: -1.5) & 2.91$\times10^{-23}$ (R$^2$: 0.201) \\
Encoder-Decoder (VAE) & 9683 (R$^2$: 0.736) & 0.941 (R$^2$: 0.664) & 73.1 (R$^2$: 0.588) & 1.19e+05 (R$^2$: 0.332) & 3.26$\times10^{-23}$ (R$^2$: 0.102) \\
\bottomrule
    \end{tabular}
    } % End resizebox
    \caption{Test results on tabular dataset after hyper-parameter tuning. Presenting the performance for each model. MSE (R$^2$) are presented for five features.}
    \label{tab:model_performance2}
\end{table*}

%\textcolor{red}{To address skewness in regression modeling, several techniques can be applied. One common approach is to transform the skewed variables using logarithmic, square root, or reciprocal transformations for right-skewed data, or exponential transformations for left-skewed data. Power transformations, such as Box-Cox or Yeo-Johnson, can also reduce skewness effectively. In cases where the skewness suggests a non-linear relationship, using non-linear regression models can provide a better fit. Additionally, robust regression techniques like Ridge or Lasso can minimize the influence of outliers caused by skewness. If transformations are not effective, generalized linear models (GLMs) that do not assume normally distributed residuals can be used to handle skewed distributions more appropriately. Each of these methods helps improve the accuracy, efficiency, and reliability of regression models when dealing with severely skewed data.}

%%%% Table 

Table~\ref{tab:model_performance2} presents the performance of various models in predicting five critical material properties: Yield Strength (YS) at 1000°C, Density at 1273K, Thermal Conductivity (THCD) at 1273K, Kou Criteria, and 1300°C Creep CB [1/s]. The performance metrics are represented as Mean Squared Error (MSE) with the corresponding R-squared (R$^2$) values in parentheses, offering insight into how well each model captures the variance in the data for these specific properties.

Among the models, XGBoost stands out as the most consistent performer, achieving the lowest MSE values and highest R$^2$ values across most features. For instance, it delivers excellent predictive accuracy for YS 1000C PRIOR and 1273K THCD, with MSE values of 1613 and 13.27, and R$^2$ values of 0.995 and 0.925, respectively. However, XGBoost shows weaker performance for the highly skewed 1300C Creep CB feature, where it records an extremely high MSE of 0.111 and a highly negative R$^2$, indicating challenges in capturing the behavior of this feature.

In contrast, the Encoder-Decoder models demonstrate mixed results. The FDN variant performs relatively well, especially for THCD and Kou Criteria, with MSE values of 14.56 and 1.93e+04, and R$^2$ values of 0.918 and 0.892, respectively. However, like XGBoost, it struggles with predicting 1300 Min Creep CB, as reflected by a substantial drop in R$^2$. Other models, such as the Variational Autoencoder (VAE) and 1D CNN, show larger errors and lower R$^2$ values across most features, with VAE particularly underperforming for more complex properties like THCD and Creep CB, where its R$^2$ values drop to 0.588 and 0.102, respectively. Overall, while XGBoost excels across the majority of features, the Encoder-Decoder models offer competitive performance on simpler features, though they require further optimization to handle more complex, skewed properties like 1300 Min Creep CB.

\subsection{\textbf{Scaling and Quantile Effects on Model Performance}}

\begin{figure*}[!ht]
    \centering
    % First column
    \begin{overpic}[width=0.25\linewidth]{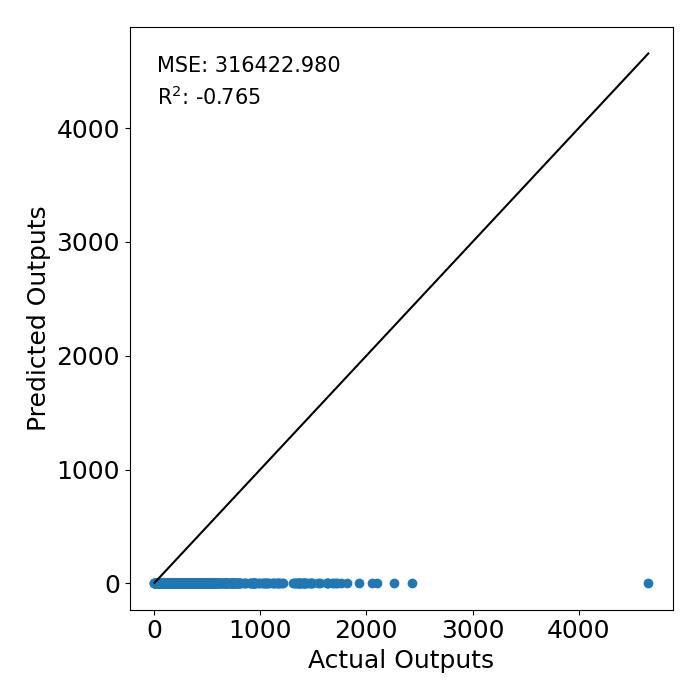}
        \put(60,60){\includegraphics[width=0.10\linewidth]{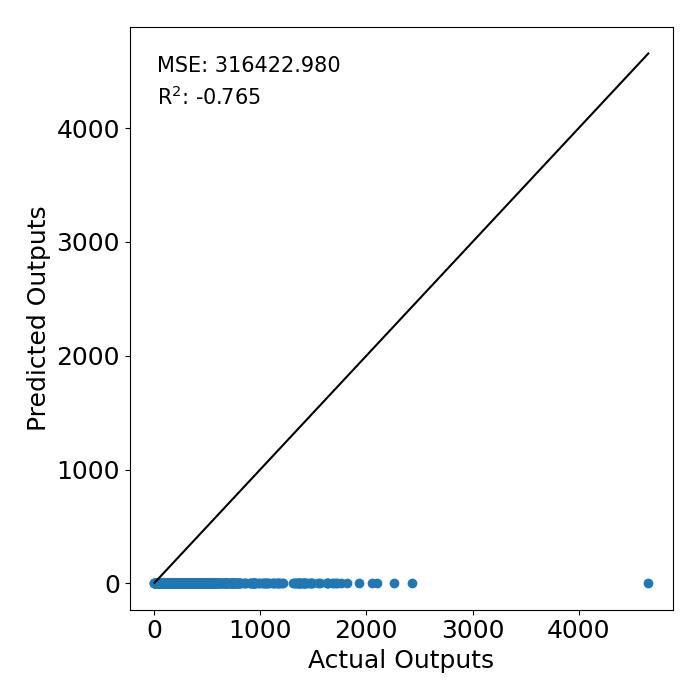}}
    \end{overpic}
    % Second column
    \begin{overpic}[width=0.25\linewidth]{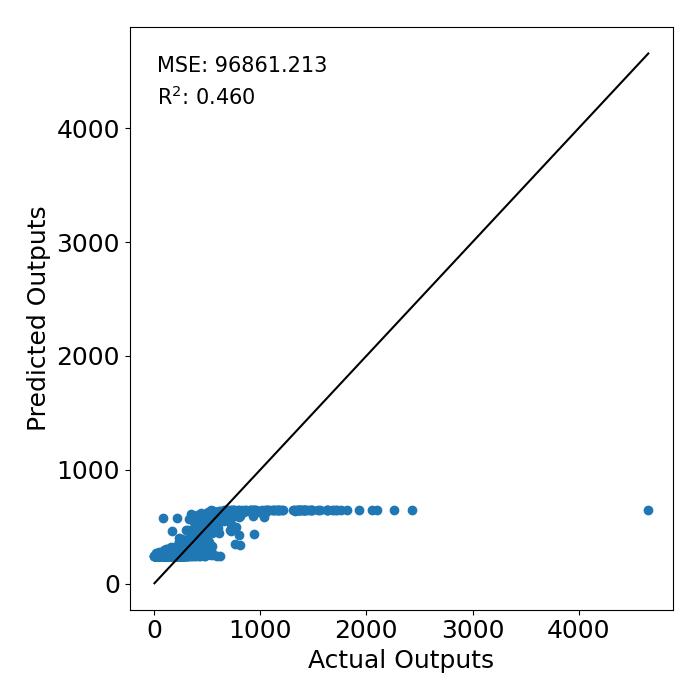}
        \put(60,60){\includegraphics[width=0.10\linewidth]{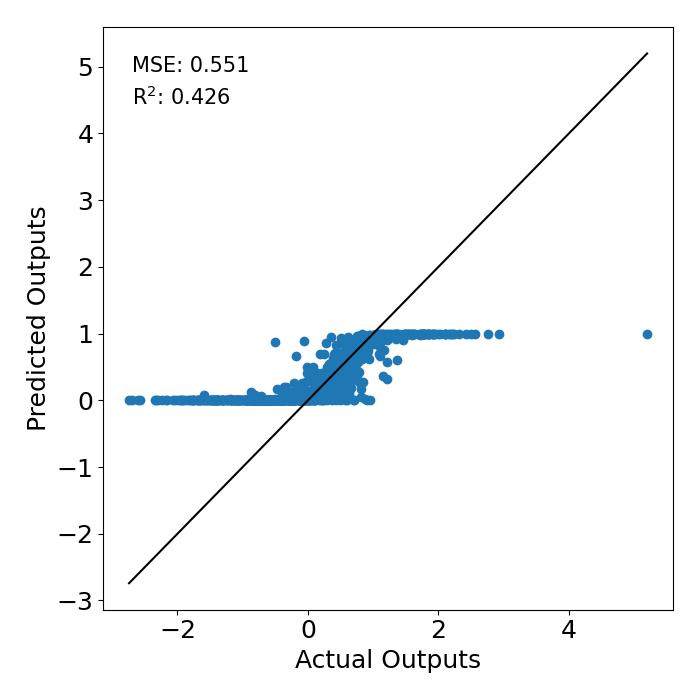}}
    \end{overpic}
    % Third column
    \begin{overpic}[width=0.25\linewidth]{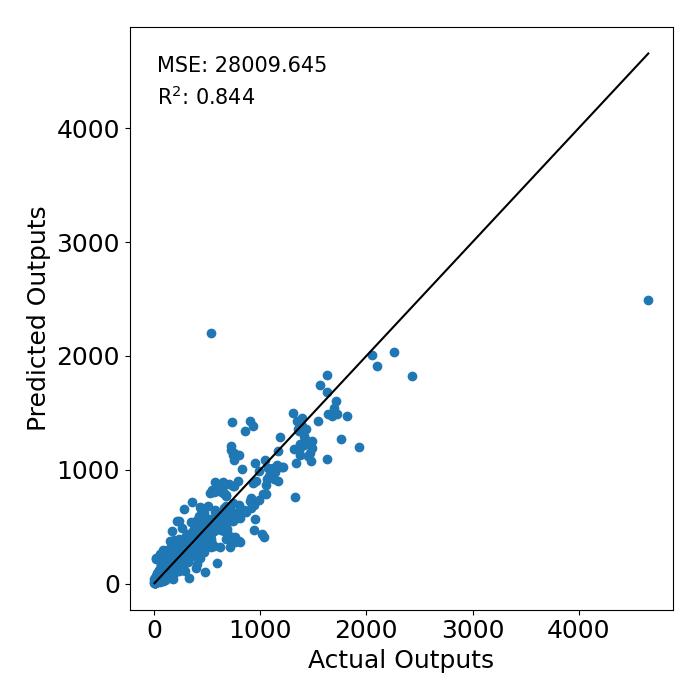}
        \put(60,60){\includegraphics[width=0.10\linewidth]{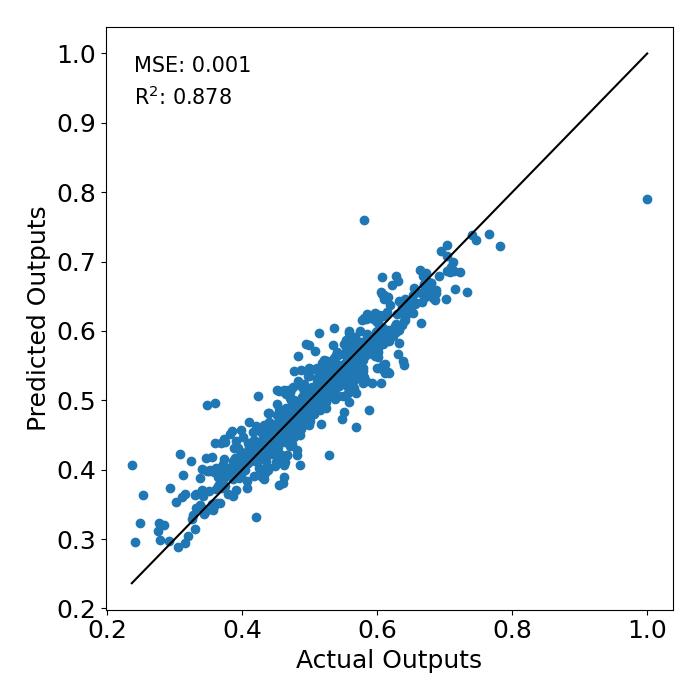}}
    \end{overpic}
    \\
    % Fourth column
    \begin{overpic}[width=0.25\linewidth]{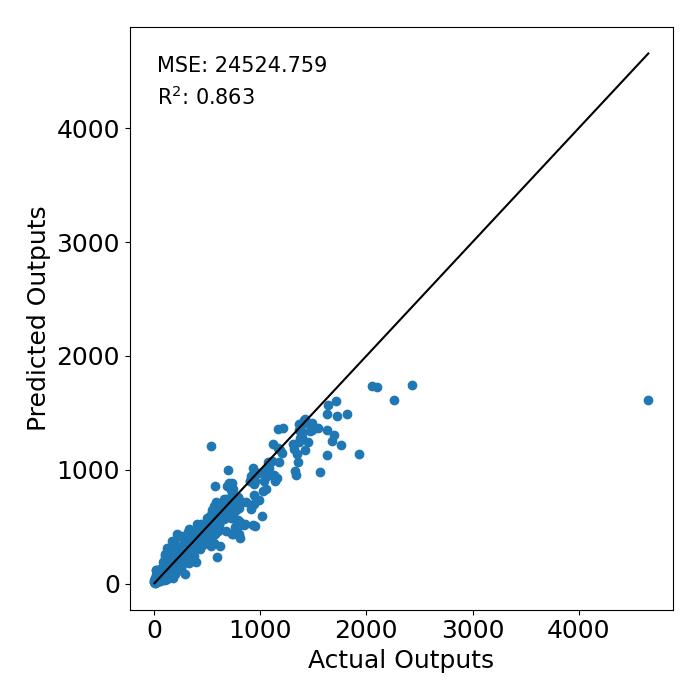}
        \put(60,60){\includegraphics[width=0.10\linewidth]{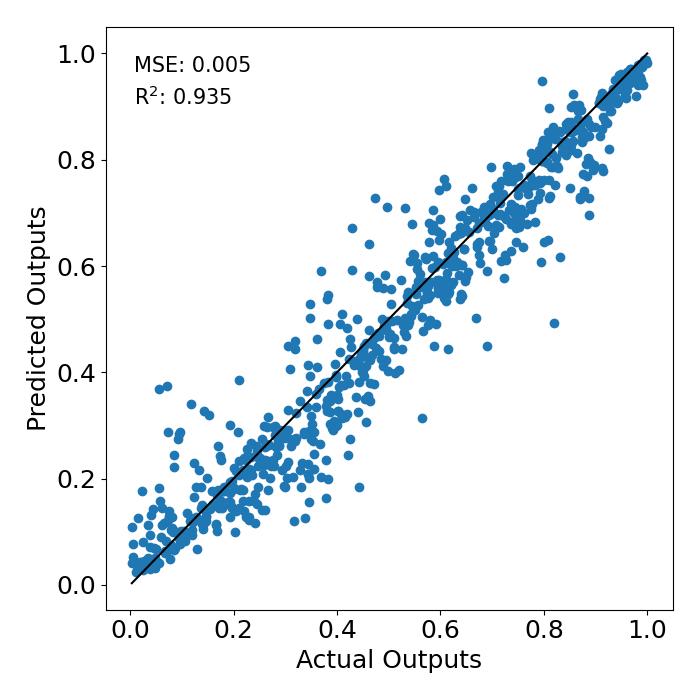}}
    \end{overpic}
    % Fifth column
    \begin{overpic}[width=0.25\linewidth]{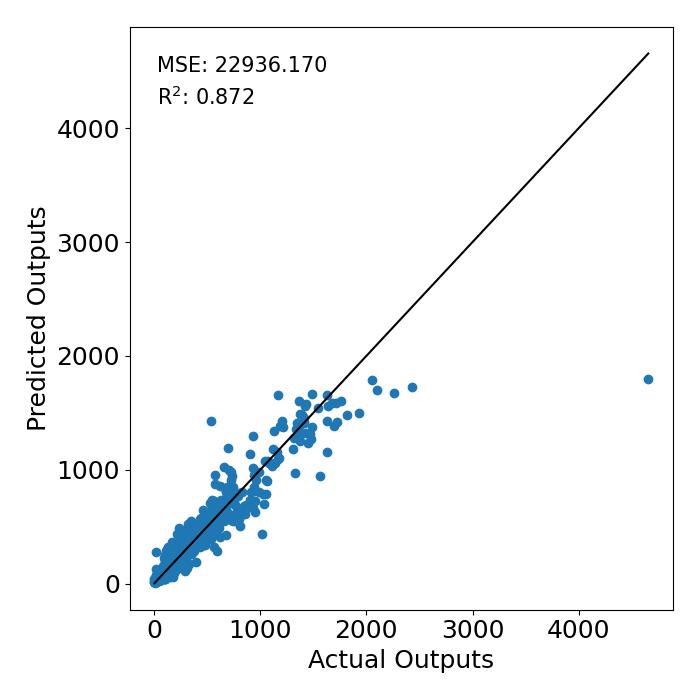}
        \put(60,60){\includegraphics[width=0.10\linewidth]{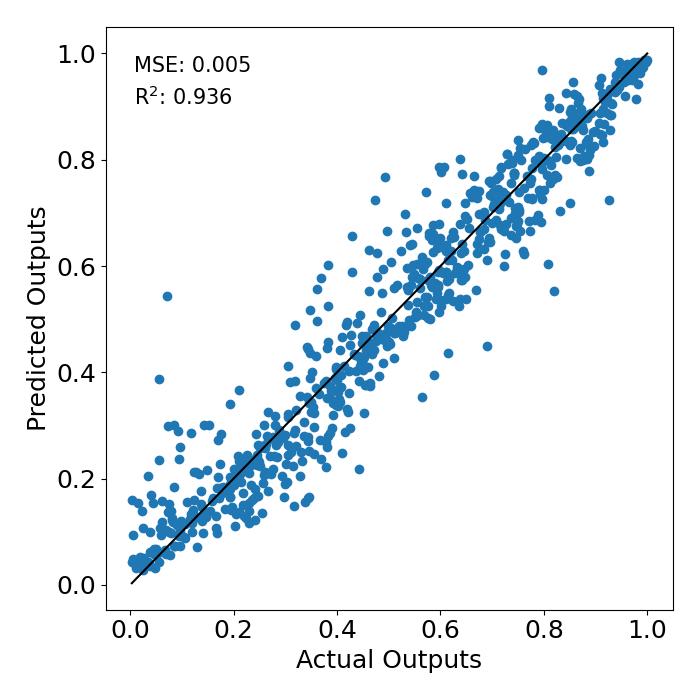}}
    \end{overpic}
    % Six column
    \begin{overpic}[width=0.25\linewidth]{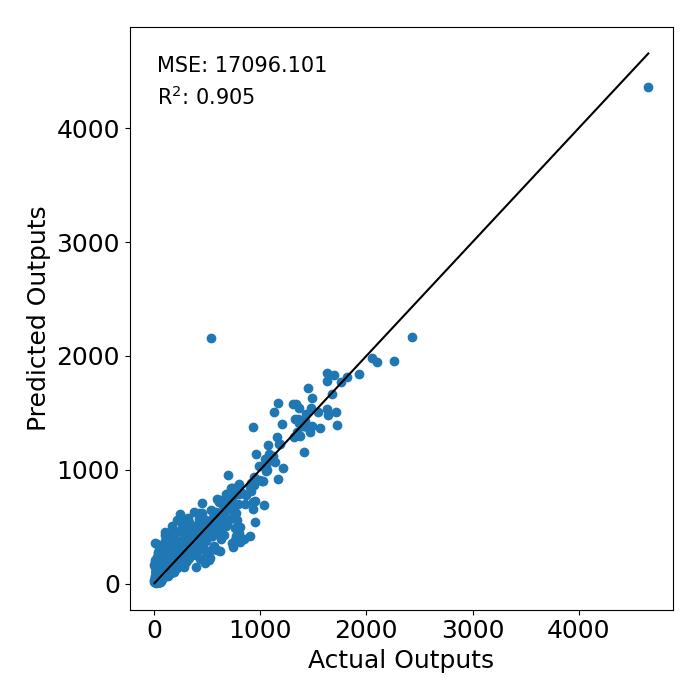}
        \put(60,60){\includegraphics[width=0.10\linewidth]{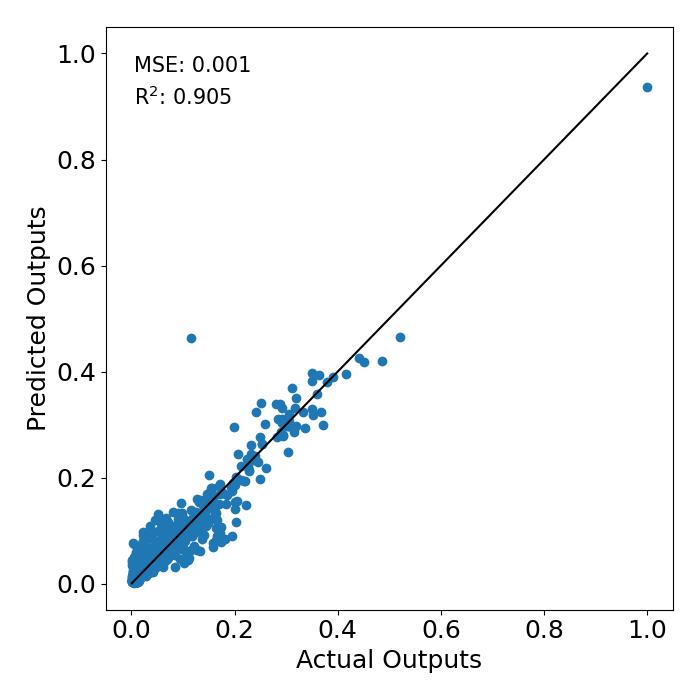}}
    \end{overpic}
    
    \caption{Parity plots showing the effect of different scaling methods on the FDN model’s predictive performance (test data) after 50 epochs. The smaller insets display the training response on scaled data. MSE (Mean Squared Error) and R$^2$ values are provided for quantitative assessment of the fit.}
    \label{fig:scaling-effect}
\end{figure*}

The results shown in Fig.~\ref{fig:scaling-effect} demonstrate the impact of various scaling and quantile transformation techniques on the predictive performance of the FDN model using test data. In the main parity plots, the predicted outputs for Kou criteria are plotted against the existing (actual) outputs for the test dataset, while the smaller inset plots show the training phase results on scaled data. These inset plots allow a direct comparison between training and testing performance. Each plot includes Mean Squared Error (MSE) and R$^2$ values, which provide a quantitative measure of how well the predicted values fit the actual ones. In the first plot, where no scaling or quantile transformation is applied, the model struggles, evidenced by a high MSE and a negative R$^2$, indicating poor predictive ability. As transformations, particularly min-max scaling and quantile transformations, are introduced, the FDN model’s performance improves markedly. The final two plots, which use uniform quantile transformation, achieve the best results with the lowest MSE and the highest R$^2$ values, reflecting more accurate and reliable predictions. 

The last plot shows the FDN model’s performance with min-max scaling applied but without any quantile transformation, resulting in improved predictions compared to using uniform/normal quantile transformation, as reflected by a lower MSE and a relatively high R$^2$ value. While min-max scaling works effectively for the ``Kou Criteria'', as demonstrated by the improved performance in the last image, it may not always be the best option for extremely skewed features. In cases where the data distribution is heavily skewed, min-max scaling can compress the data into a narrow range, leading to a loss of important variation in the features. As a result, other scaling methods, such as quantile transformation, become essential to normalize these distributions and better capture the relationships in the data, ultimately enhancing the model’s predictive capabilities.

The difference between the scaled and re-scaled performance, as seen in Fig.~\ref{fig:scaling-effect}(e), where the R$^{2}$ metric for scaled data is 0.92 but reduces to 0.872 after re-scaling, can be attributed to the use of quantile transformation. Quantile transformation effectively handles skewed distributions by mapping them to a uniform or normal distribution, which helps improve the model’s performance during training. However, when the data is re-scaled back to the original range, some of the nuanced variations captured by the model might be lost. This is because the test data may not perfectly match the distribution seen during training, especially if it contains outliers or different characteristics, leading to a slight reduction in performance. This highlights that while quantile transformation is highly useful for handling skewed data, it can introduce challenges during re-scaling, particularly when generalizing to unseen data.

\section{Discussion: Suitability of Models for Predicting the Creep Feature}

%%%%%%%
We evaluated the performance of various models in predicting high-temperature material properties such as density, thermal conductivity, yield strength, and creep resistance. While the models demonstrate reasonable predictive capabilities for most features, they fail to generalize and accurately predict for highly skewed features such as creep resistance. This section discusses the reasons behind this limitation, focusing on the complexity of creep behavior and the shortcomings in the dataset and models.

This poor performance is primarily due to the complex nature of creep, which is influenced by factors such as microstructural evolution, grain boundary stability, diffusion of alloying elements, and phase stability—factors that are not fully captured by the dataset used in this study. While elemental compositions are included, key microstructural features like grain size, precipitate distribution, and dislocation density, which play a critical role in determining creep behavior, are absent. As a result, the models are unable to generalize effectively to the creep property, as they lack the necessary information to capture its intricate, time-evolving dynamics.

The inclusion of these additional microstructural features would likely enhance the model’s ability to establish a more accurate chemistry-creep relationship. By incorporating such factors, the models could better reflect the underlying physical processes that govern creep, such as diffusion and dislocation movement. This would improve the models’ predictive performance, allowing them to generalize more effectively across both simple and complex material properties. Moreover, the integration of time-dependent features, or the use of advanced approaches like physics-informed neural networks (PINNs), could further strengthen the models’ ability to capture the nuances of creep behavior. These enhancements would not only improve the prediction of creep resistance but also enable more accurate modeling of other complex, nonlinear material properties, ultimately advancing the applicability of machine learning in materials science.

\section{Conclusion}\label{sec:conc}

In this study, we evaluated the performance of several machine learning models, including deep encoder-decoder architectures, VAEs, and XGBoost, for predicting material properties. Our results demonstrate that while these models perform well for normally-distributed features such as yield strength and density, they fail to generalize for more complex and skewed features like creep resistance. %This limitation is attributed to the static treatment of creep and the absence of critical microstructural features in the dataset.

Creep is a highly nonlinear and time-dependent process, influenced by diffusion, microstructural evolution, and dislocation movement, among other factors. The static treatment of creep in these models, combined with the absence of critical microstructural features like grain size, precipitate distribution, and dislocation density in the dataset, limits the ability of the models to make accurate predictions. The results underscore the need for more sophisticated modeling approaches that incorporate time-evolving properties and detailed microstructural information.

In conclusion, predicting creep resistance requires more advanced techniques that account for both the temporal and microstructural complexities of the material. Physics-informed models, which embed domain-specific knowledge into the learning process, and multi-task learning frameworks that can simultaneously learn from multiple material properties, offer promising avenues for improvement. Furthermore, expanding datasets to include a broader range of microstructural features will be crucial in enhancing model performance and generalization.

From a data science perspective, our analysis also highlights the importance of selecting the appropriate model architecture for different types of material properties. While FDN perform well in capturing simple, balanced distributions, they struggle with interpretability and feature selection, particularly when dealing with heterogeneous or skewed data. On the other hand, DNNF architectures, which incorporate logical operators for decision-making, show promise in handling mixed data types and modeling the complex, non-linear interactions between microstructure, temperature, and stress.

%	2.	Broader Implications:
Beyond prediction, encoder-decoder models hold promise for integration into advanced applications such as smart manufacturing, self-driving labs, and autonomous robotics. By combining these models with optimization techniques like Reinforcement Learning or Bayesian Optimization, systems can adaptively explore and optimize processes, accelerating materials discovery. However, challenges such as computational demands, data quality, and the lack of uncertainty quantification in predictions—which is critical for applications like Bayesian Optimization—must be addressed to fully realize their potential. Furthermore, generative variants of these neural architectures can enhance privacy-preserving techniques through synthetic data generation, mitigating risks associated with sharing sensitive or proprietary materials data. These methods, coupled with FAIR data considerations, ensure that the insights and models developed are accessible and reusable while maintaining ethical standards of data governance. By addressing these challenges, novel deep learning architectures pave the way for real-time decision-making, high-speed data processing, and improved precision in applications ranging from materials design to autonomous exploration. 

In summary, while FDN and VAE models demonstrate reasonable performance for features with balanced distributions, they fail to generalize effectively for more skewed and complex features such as 1300 Min Creep CB [1/s]. Further refinements, such as feature-specific adjustments and the integration of more sophisticated modeling techniques, are essential to improve generalization across a wider range of material properties.

\section*{Code Availability}
%The tools developed in this study, named ``DataScribe Nexus'', is implemented within the online platform \url{DataScribe.Cloud}, an online scientific data ingestion and analysis tool developed in Texas A\&M University. This framework is publicly available for applying data analysis to various datasets. Users can access the DataScribe Nexus framework online through \url{DataScribe.Cloud}. a GitHub repository \url{link}.

The data developed as part of this study are publicly available at \url{https://github.com/vahid2364/DataScribe_DeepTabularLearning.git}.

\section*{Data Availability}
The dataset used and analyzed during this study is available upon reasonable request from the corresponding author. %Data used in this study can also be accessed via the \url{DataScribe.Cloud} platform, which supports uploading, transforming, and analyzing custom datasets. For further inquiries regarding access to the data, users are encouraged to contact the authors or utilize the tools provided by \url{DataScribe.Cloud} for seamless data ingestion and analysis.

    \section*{Acknowledgments}
    The authors acknowledge the Grace supercomputing facility at Texas A\&M University for providing essential computational resources that facilitated the research presented in this paper. VA and RA also acknowledge the financial support from the BIRDSHOT Center, provided under Cooperative Agreement Number W911NF-22-2-0106.

%%%%%%
%%%%%%
%%%%%%

\setlength{\bibsep}{0pt}  % Set the bibliography item spacing to zero
\begin{small}
\bibliographystyle{elsarticle-num}
\bibliography{Refs}
\end{small}

%%%%%%
%%%%%%
%%%%%%

\appendix
\renewcommand{\thefigure}{A.\arabic{figure}}  % Changes figure numbering style for the appendix
\setcounter{figure}{0}  % Resets figure counter
\setcounter{table}{0}  % Resets figure counter

%%%%%%
%%%%%%
%%%%%%

\section{Dataset Description and Visualization}

The introduced dataset includes alloy composition and material properties for various temperatures and conditions. The elements Niobium (Nb), Chromium (Cr), Vanadium (V), Tungsten (W), and Zirconium (Zr) are key components of the alloys studied. The primary material properties include thermal conductivity, density, coefficient of thermal expansion (CTE), creep rate at multiple temperature ranges.

\begin{figure*}[!ht]
    \centering
    \vspace{0.2cm}
    % First sub-figure
    \begin{subfigure}[b]{1\textwidth}
    \centering
    \includegraphics[width=0.19\linewidth]{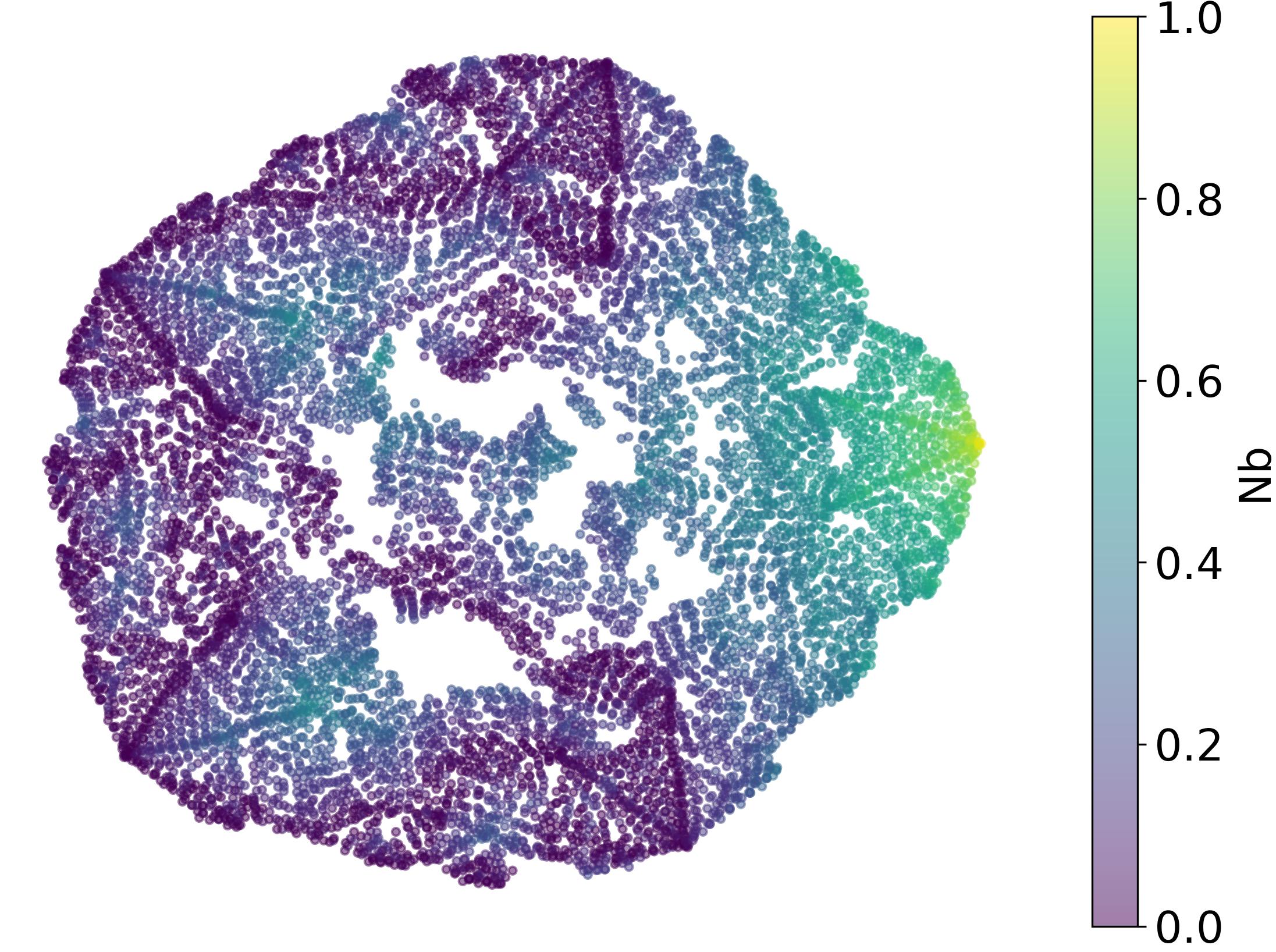}
    \includegraphics[width=0.19\linewidth]{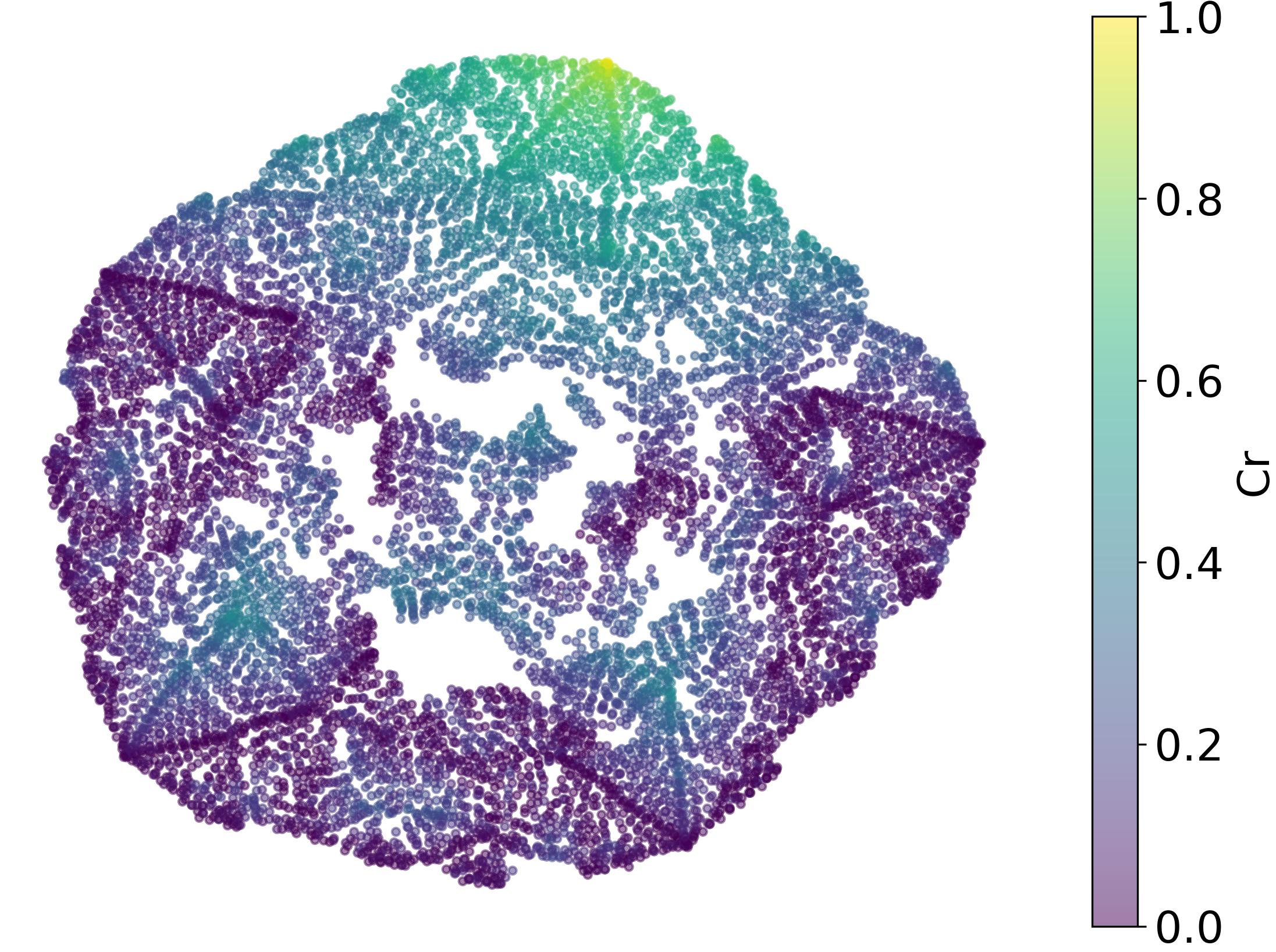}
    \includegraphics[width=0.19\linewidth]{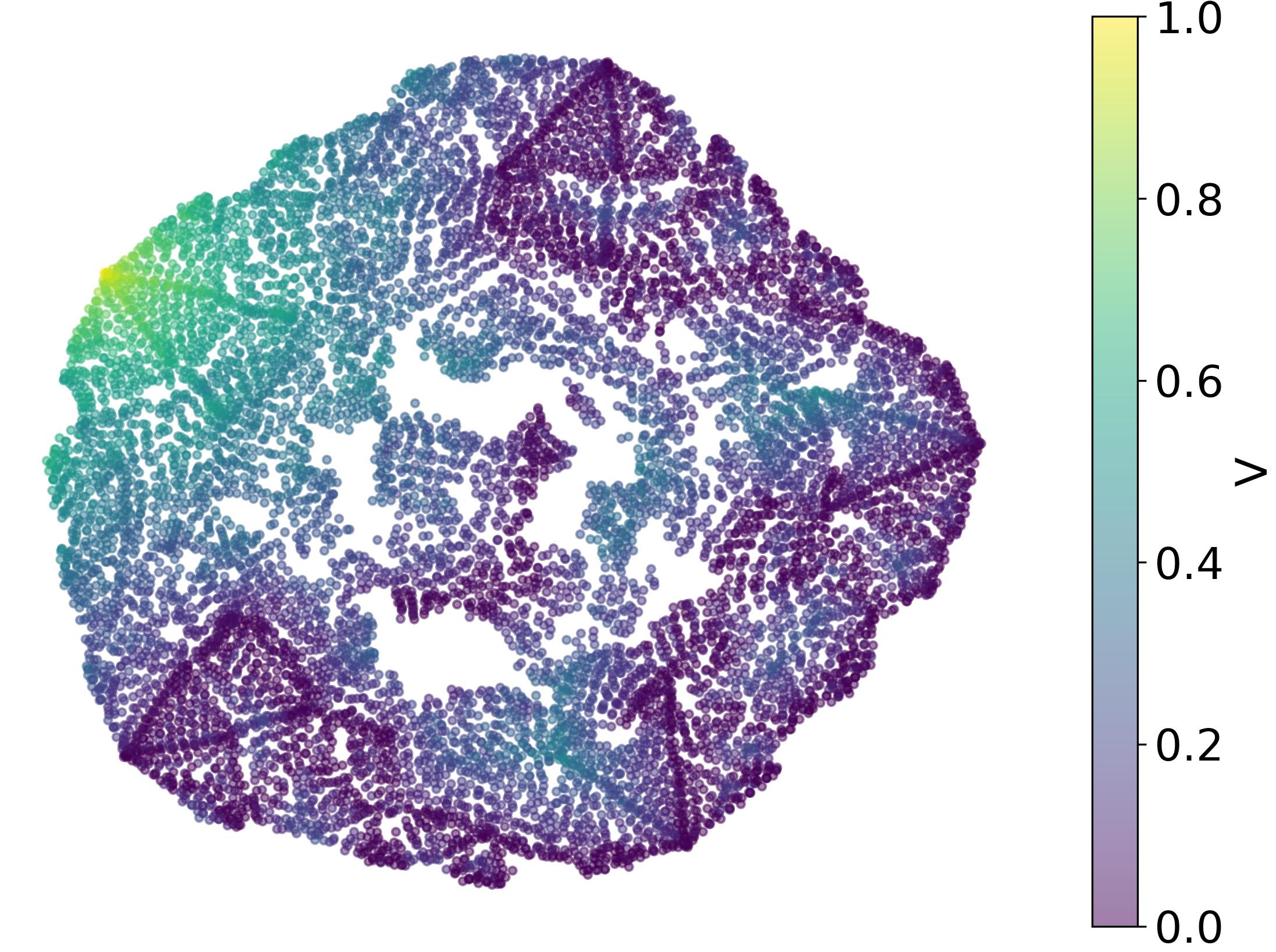}
    \includegraphics[width=0.19\linewidth]{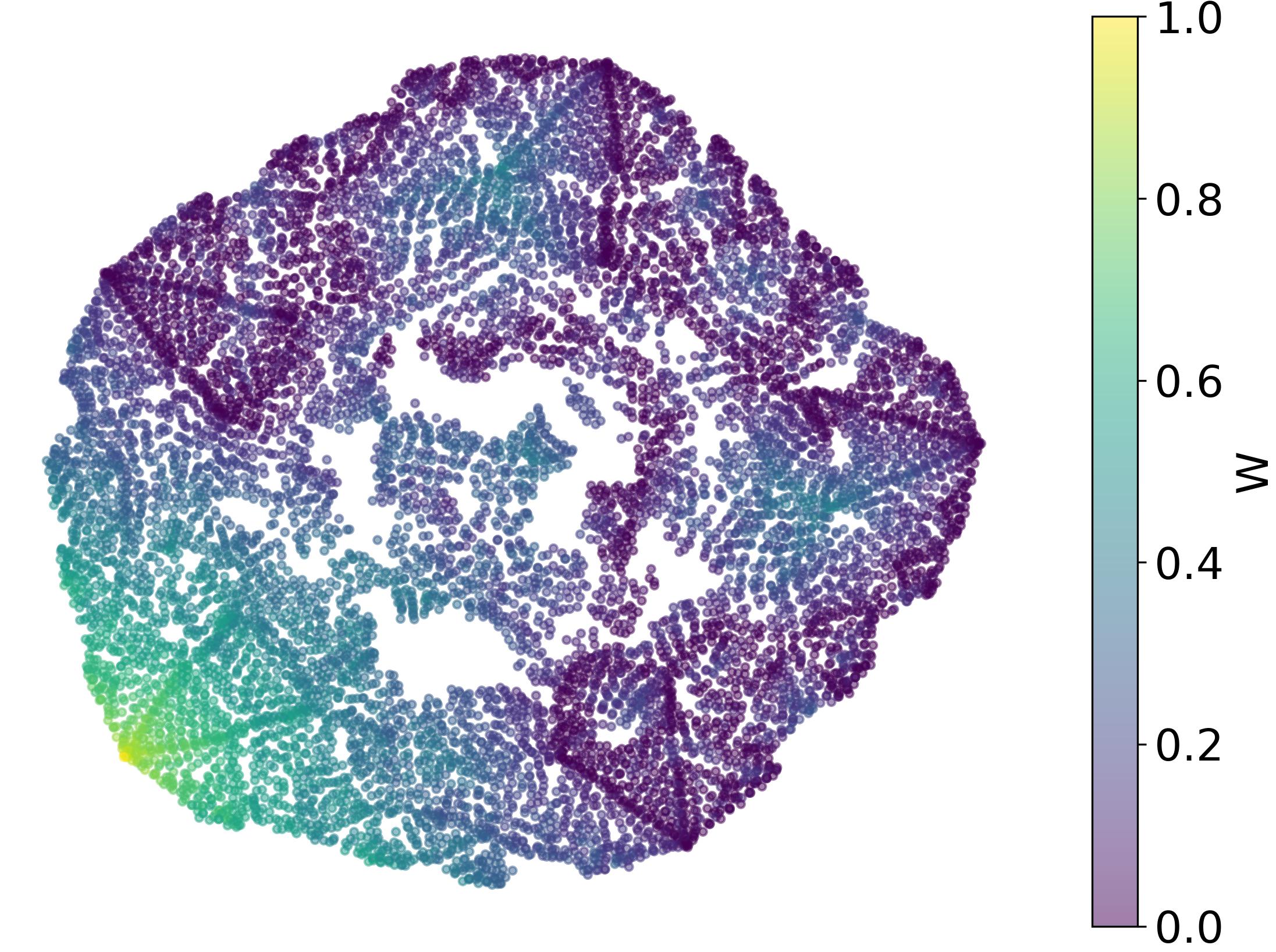}
    \includegraphics[width=0.19\linewidth]{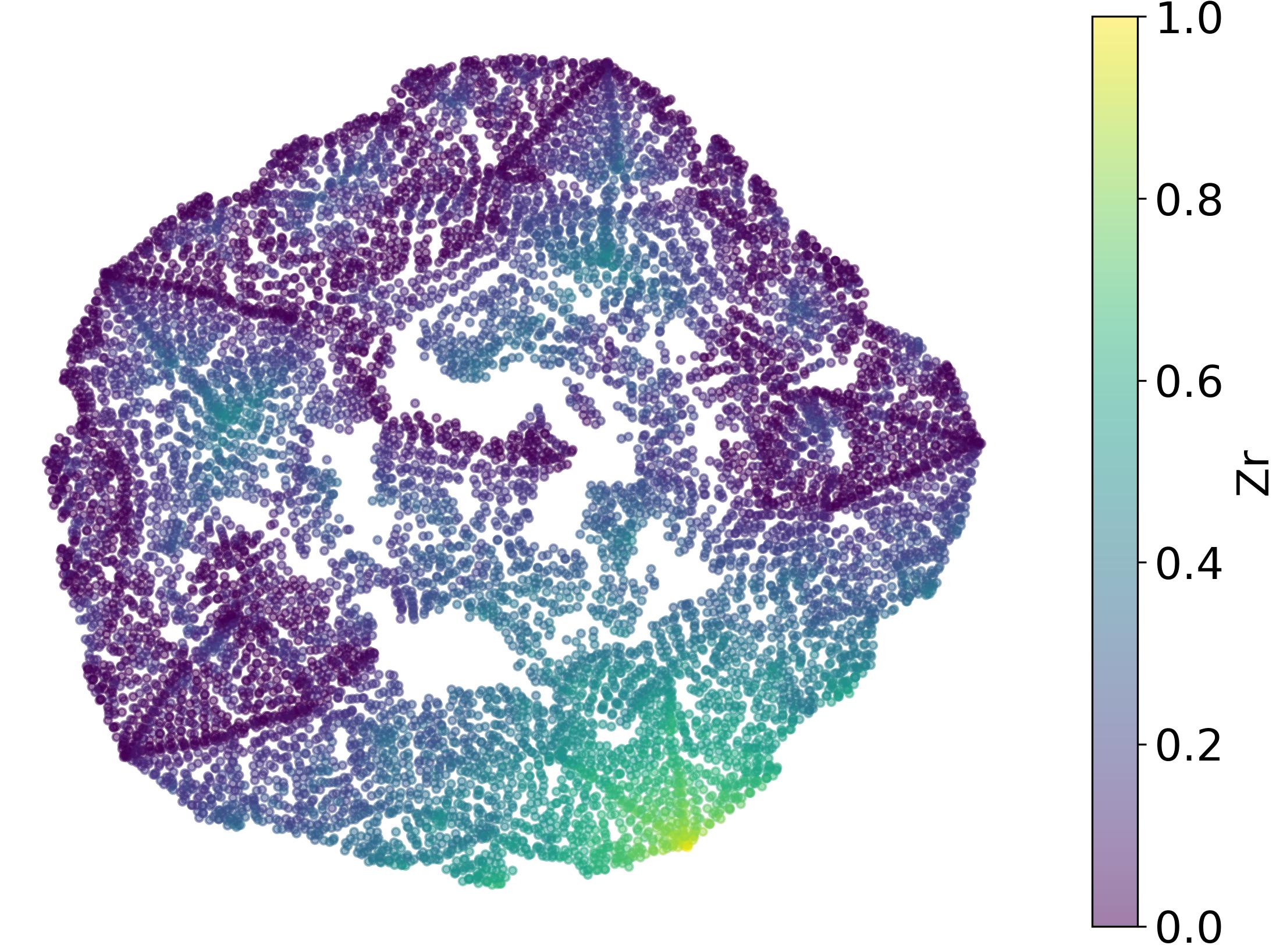}
    \\
    \caption{t-SNE plots for input features: Niobium, Chromium, Vanadium, Tungsten, Zirconium concentrations (wt\%).}
    \label{fig:input_features}
    \end{subfigure}
    \hfill
    % Second sub-figure
    \begin{subfigure}[b]{1\textwidth}
    \centering
    \includegraphics[width=0.19\linewidth]{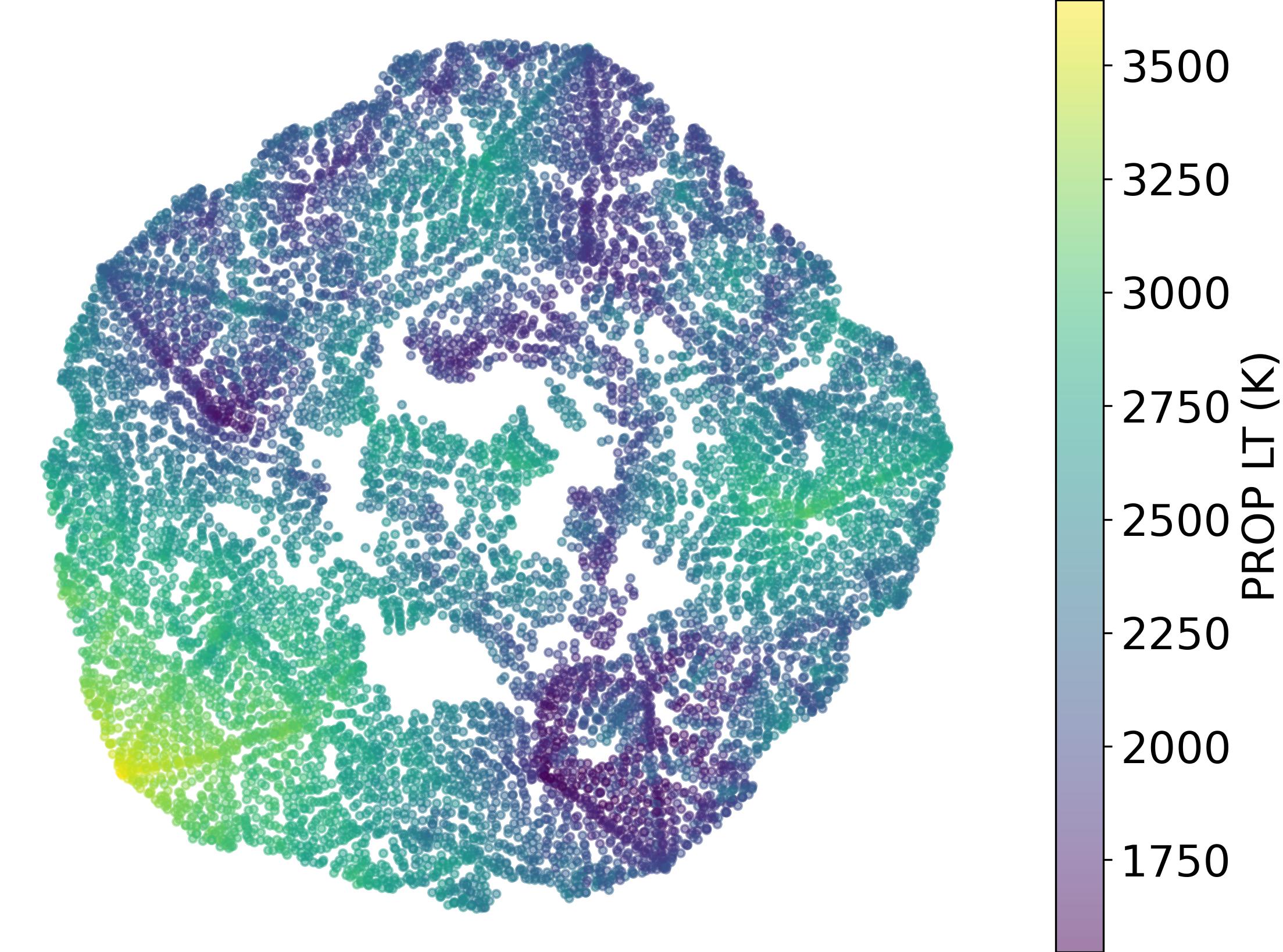}
    \includegraphics[width=0.19\linewidth]{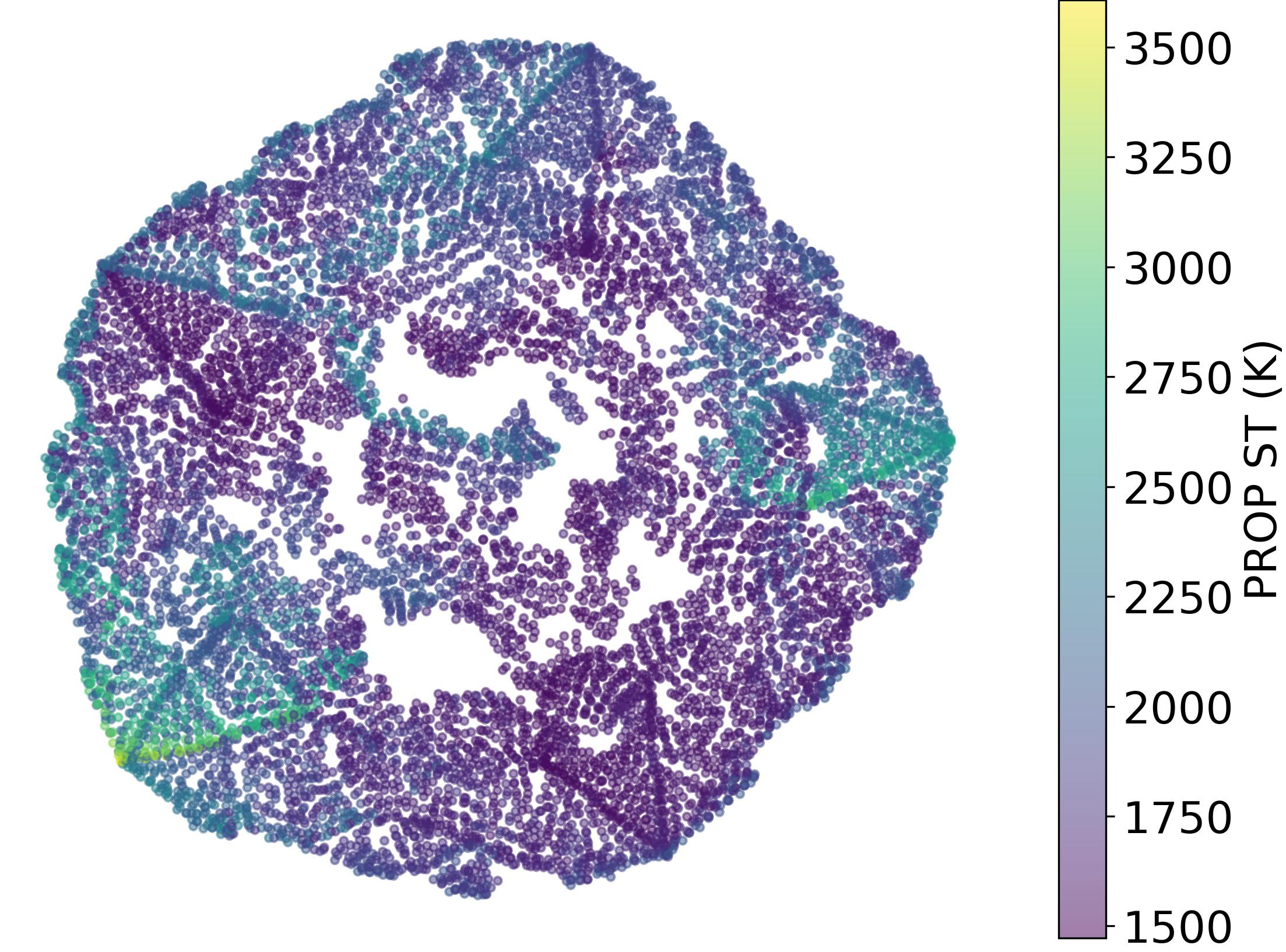}
    \includegraphics[width=0.19\linewidth]{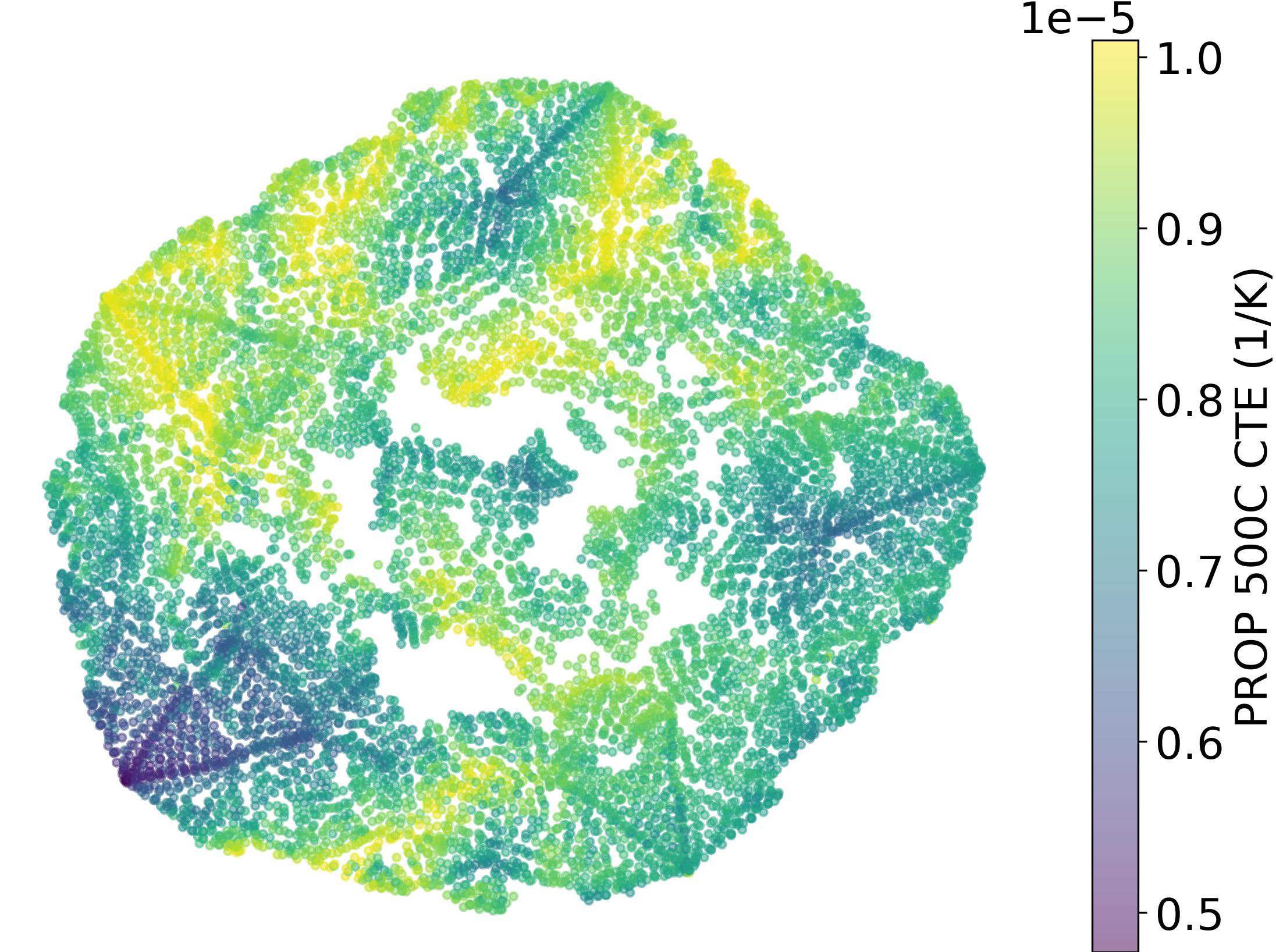}
    \includegraphics[width=0.19\linewidth]{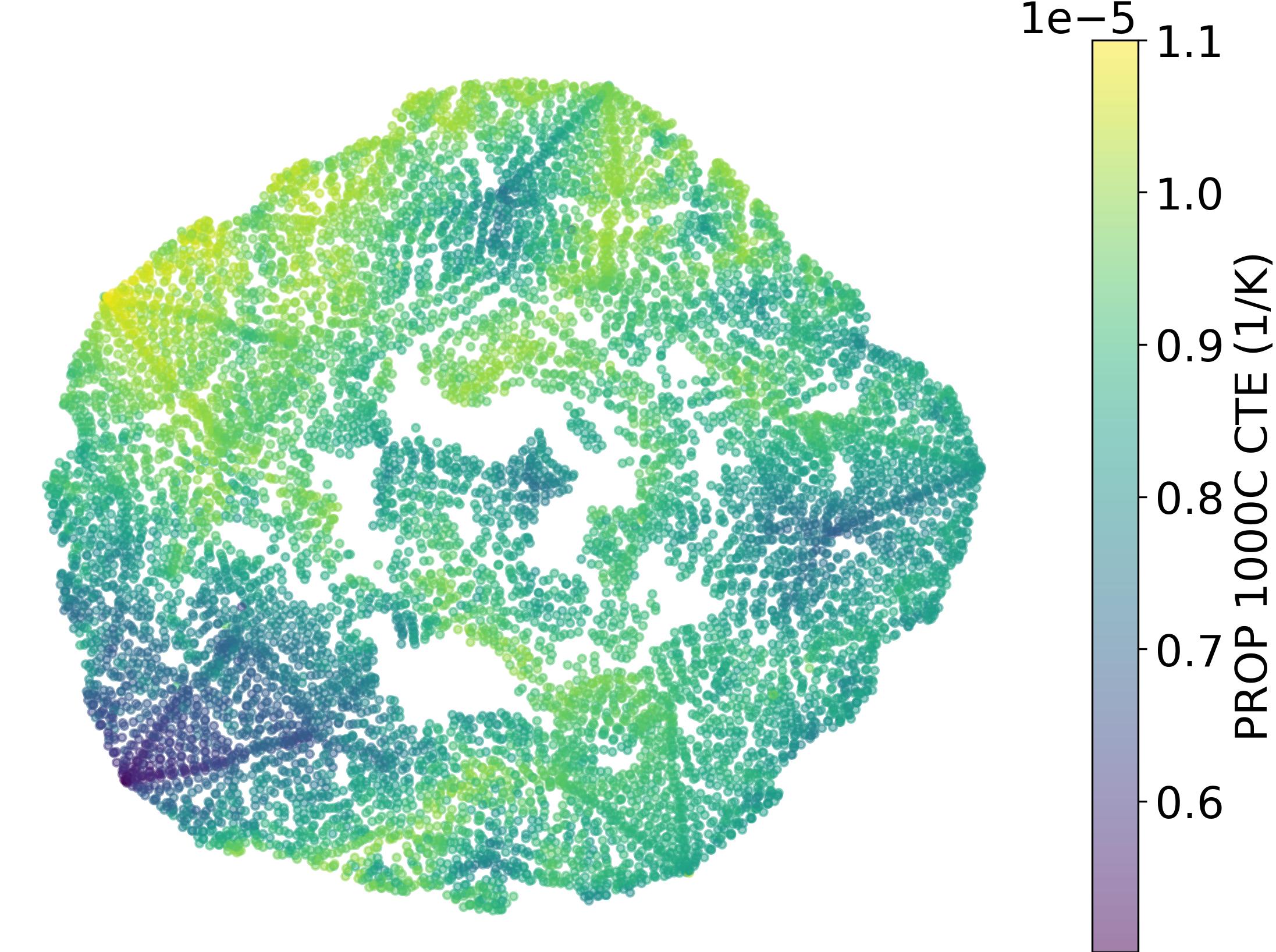}
    \includegraphics[width=0.19\linewidth]{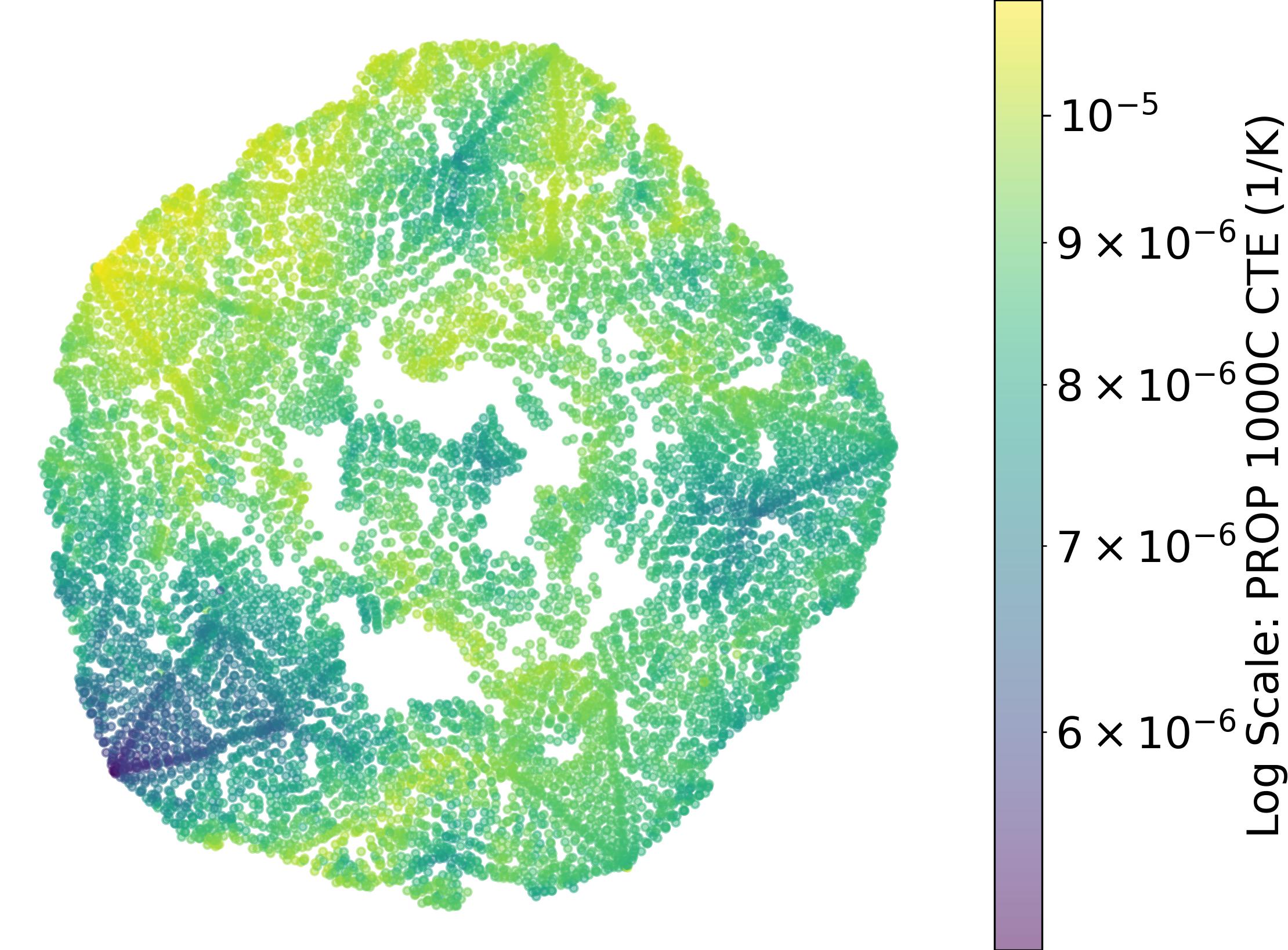}
    %%%%%
    \includegraphics[width=0.19\linewidth]{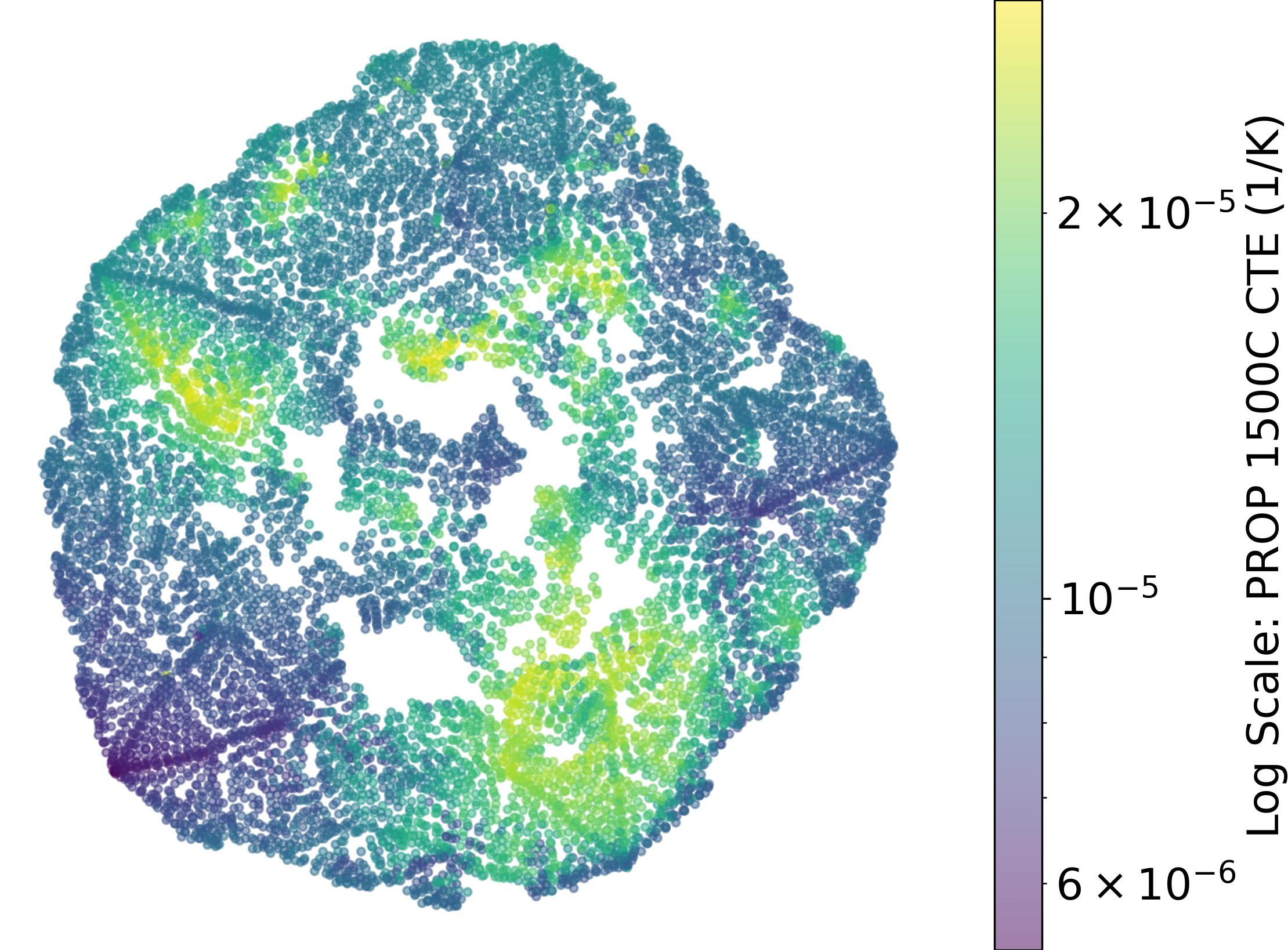}
    \includegraphics[width=0.19\linewidth]{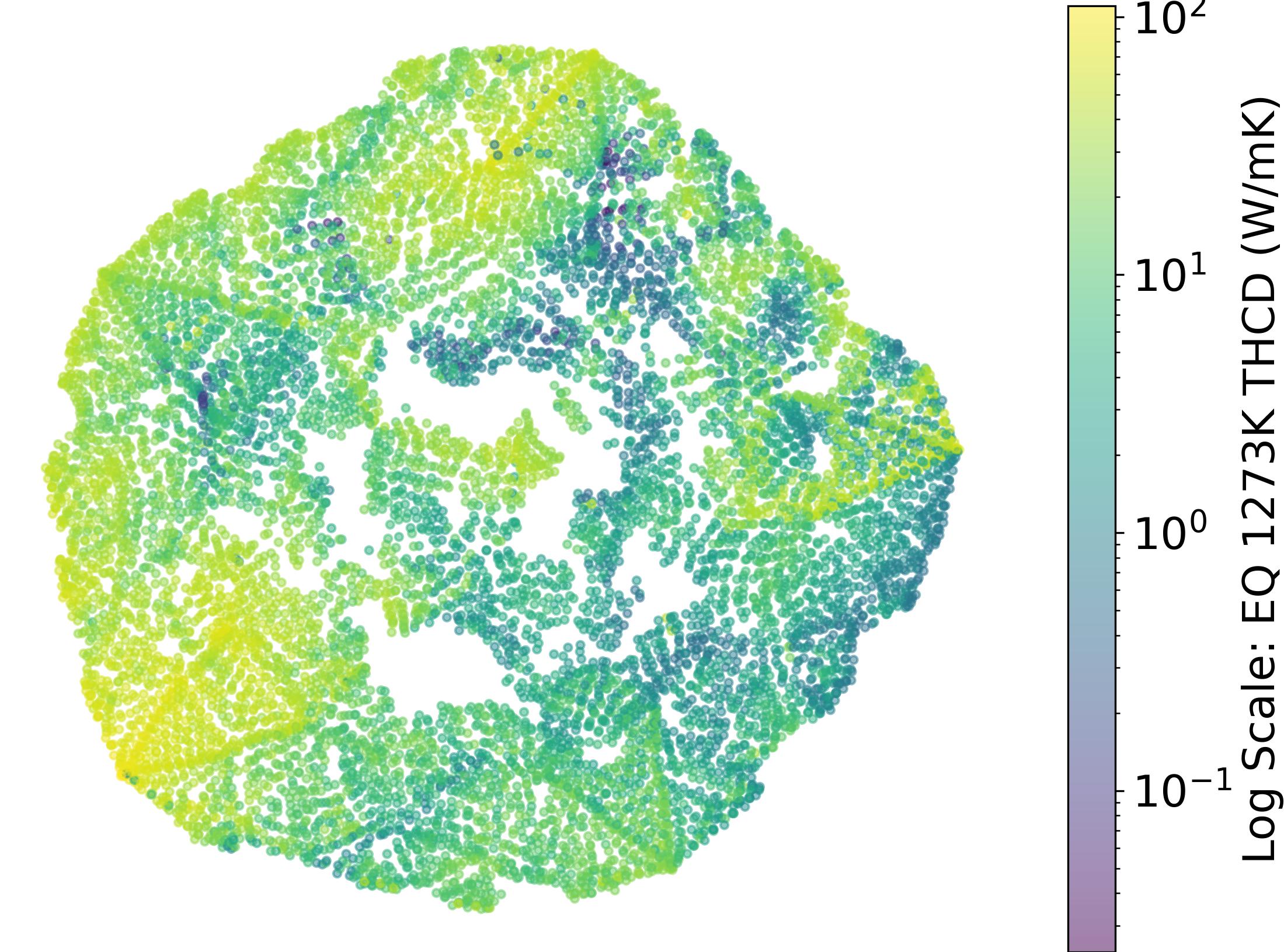}
    \includegraphics[width=0.19\linewidth]{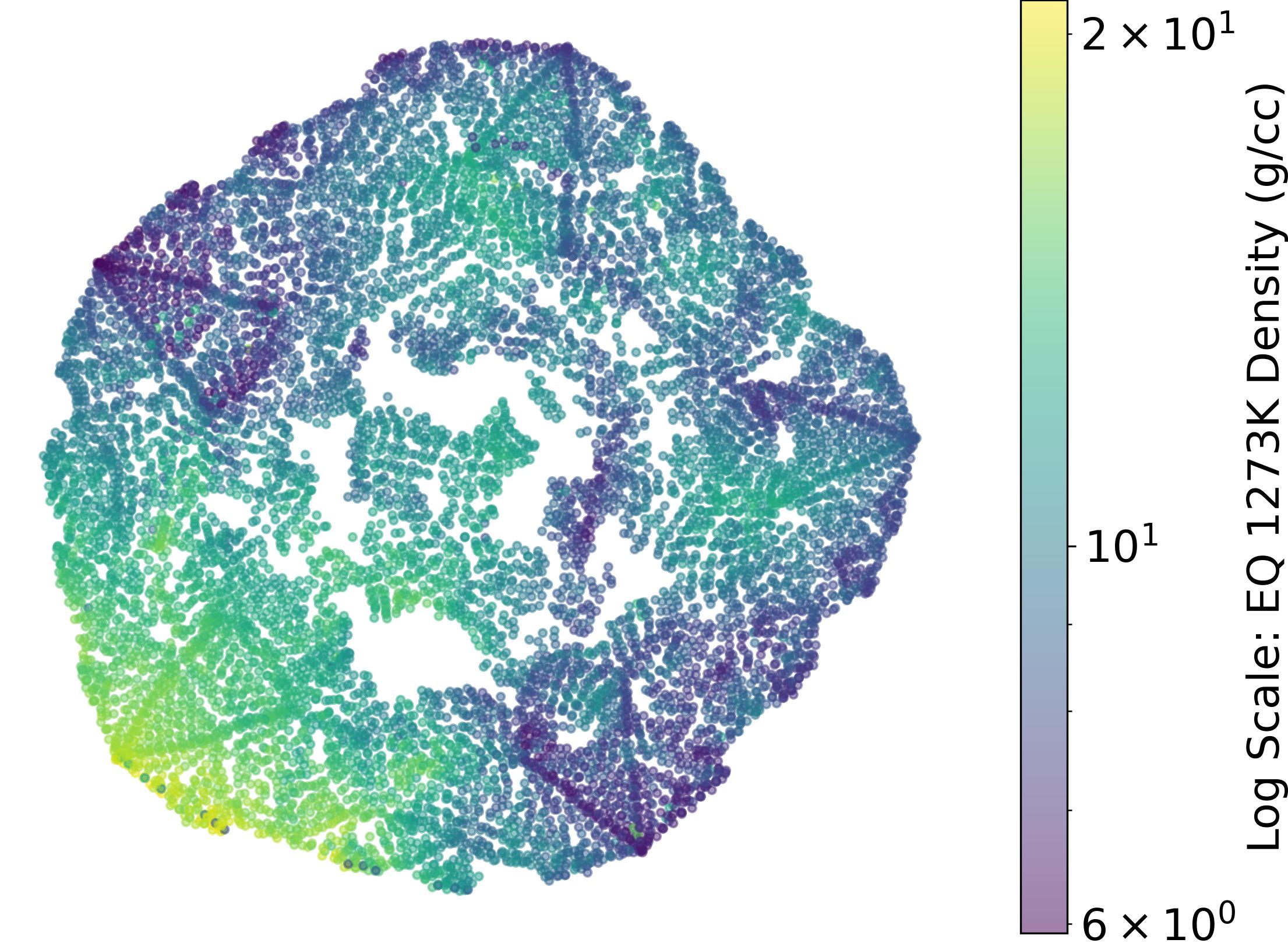}
    \includegraphics[width=0.19\linewidth]{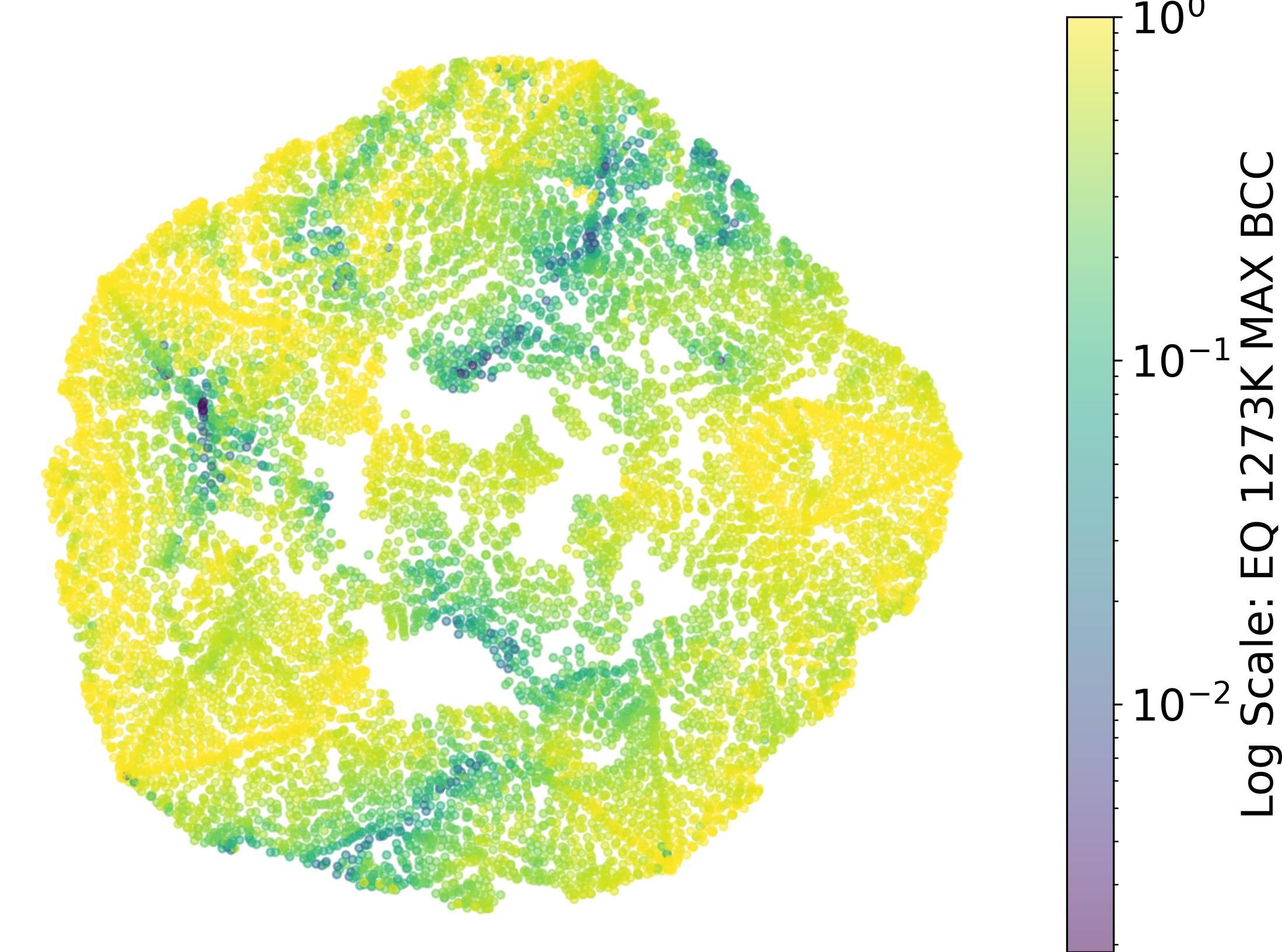}
    \includegraphics[width=0.19\linewidth]{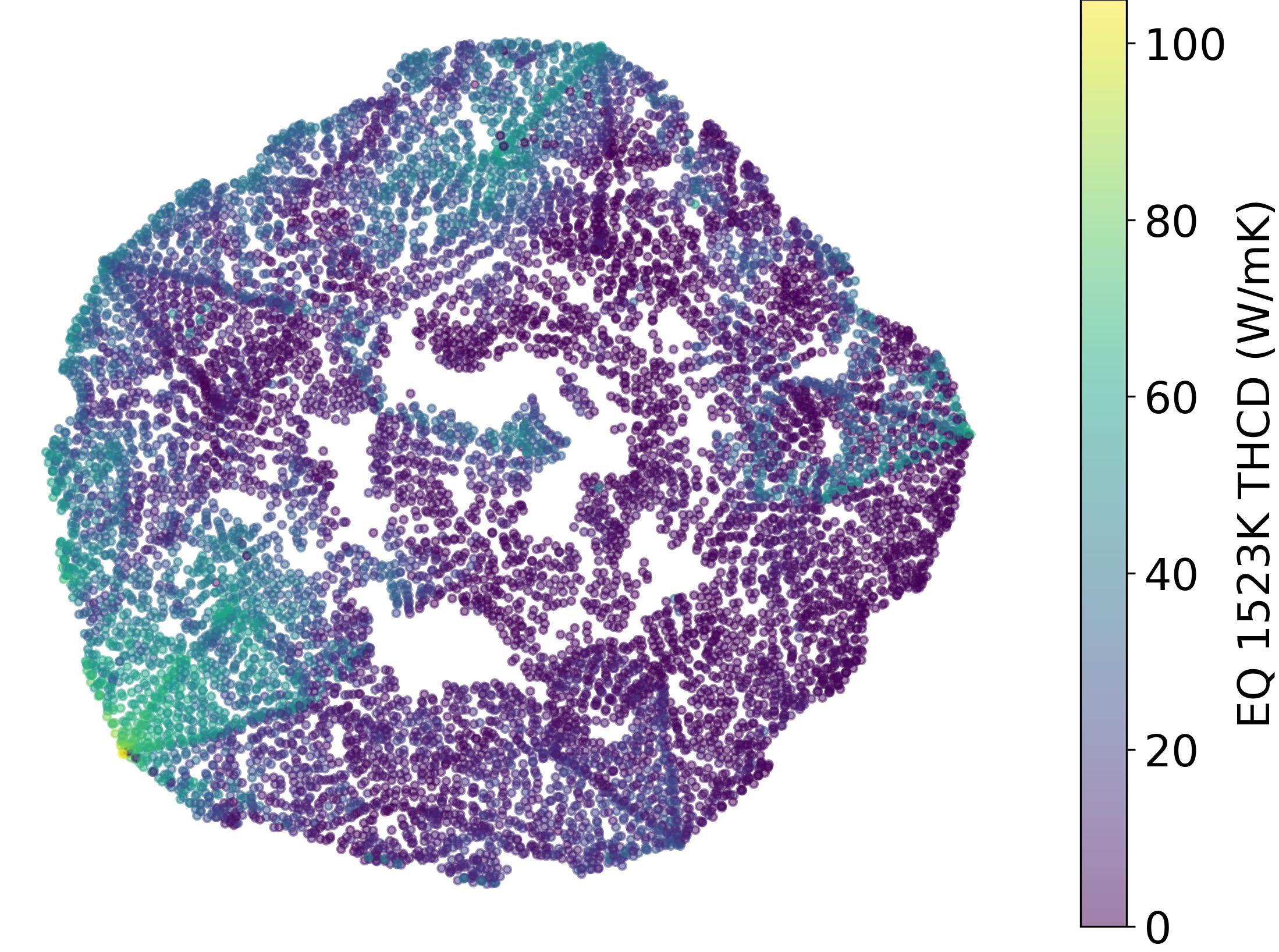}
    \includegraphics[width=0.19\linewidth]{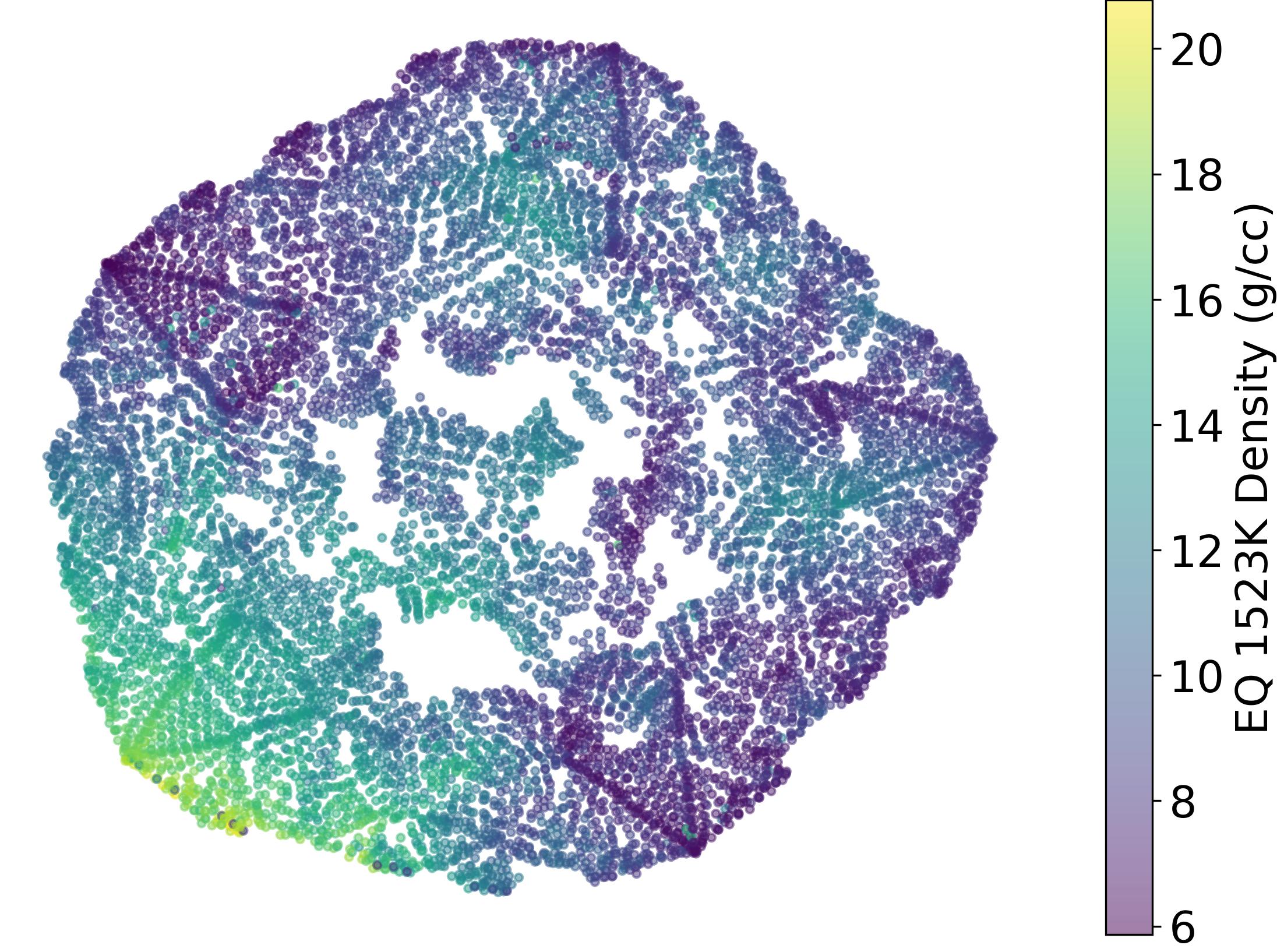}
    \includegraphics[width=0.19\linewidth]{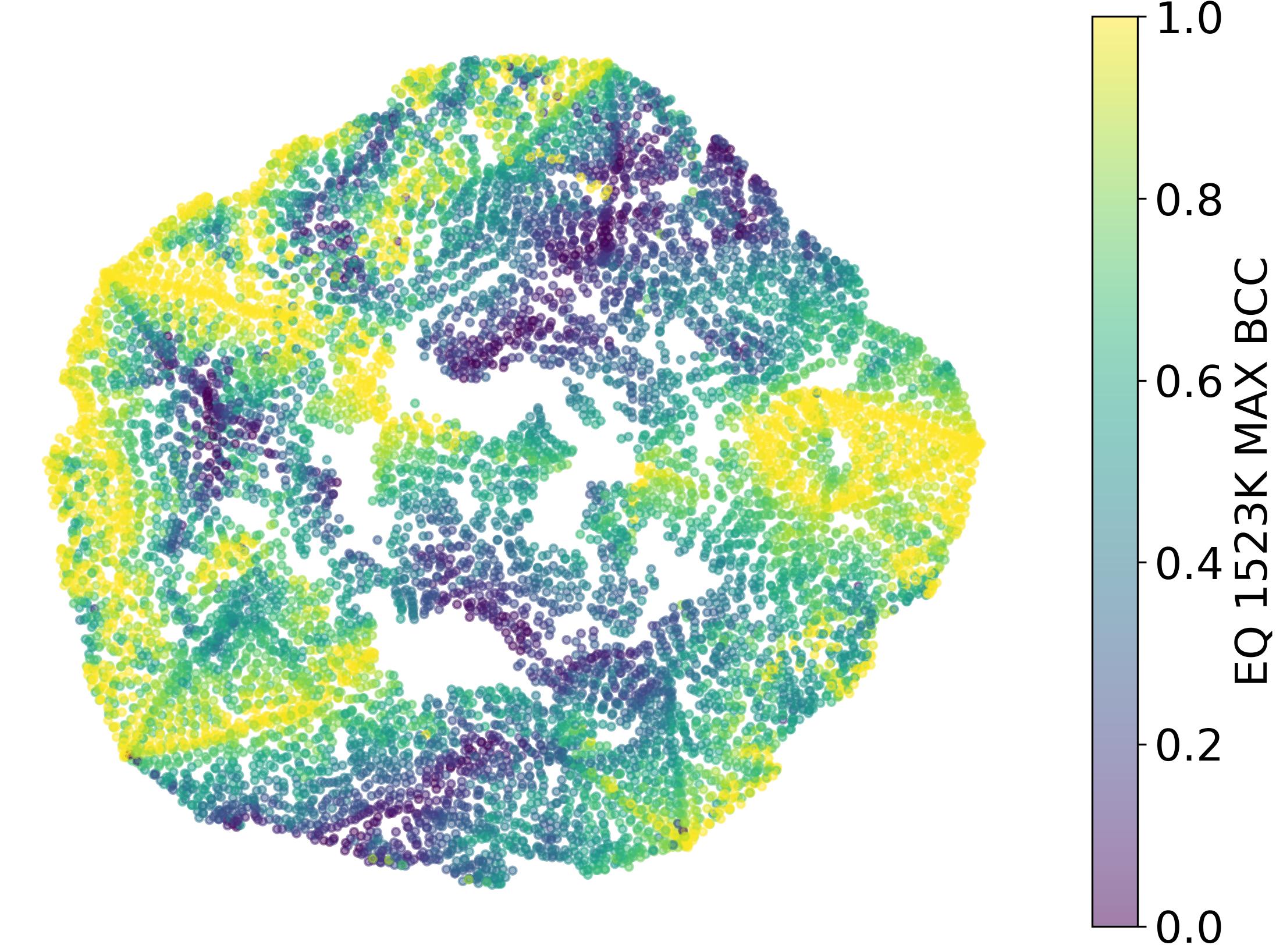}
    \includegraphics[width=0.19\linewidth]{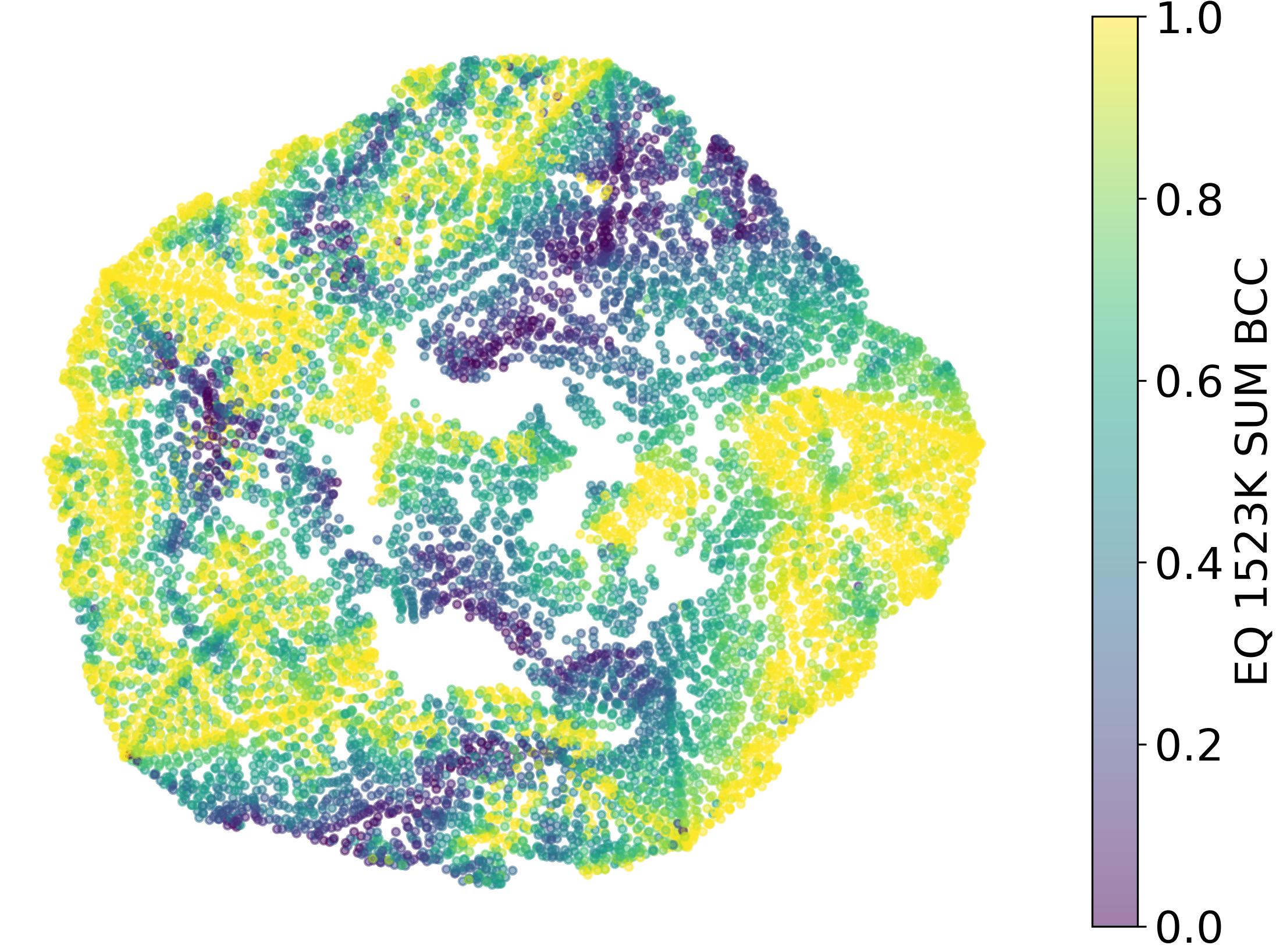}
    \includegraphics[width=0.19\linewidth]{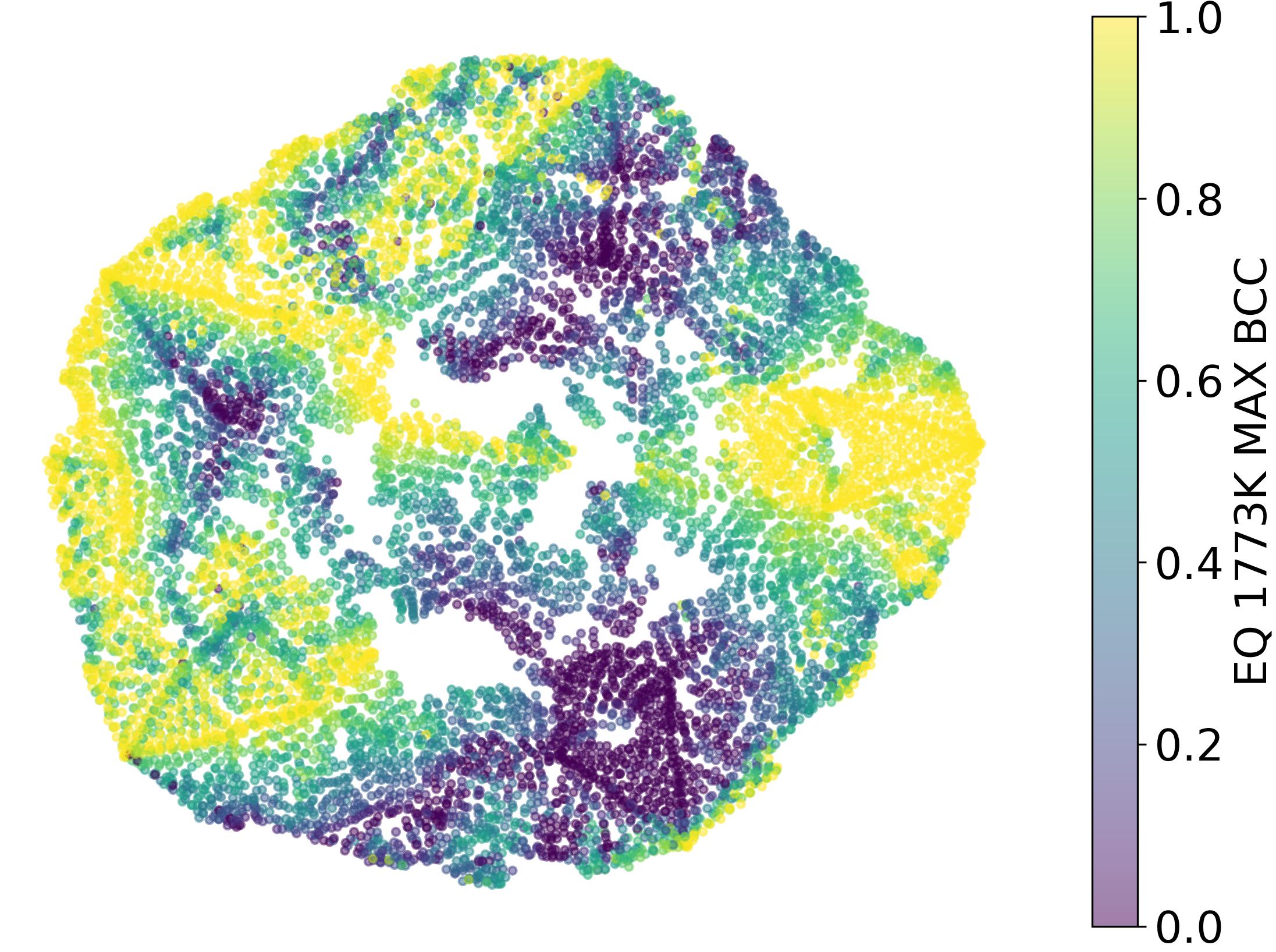}
    \includegraphics[width=0.19\linewidth]{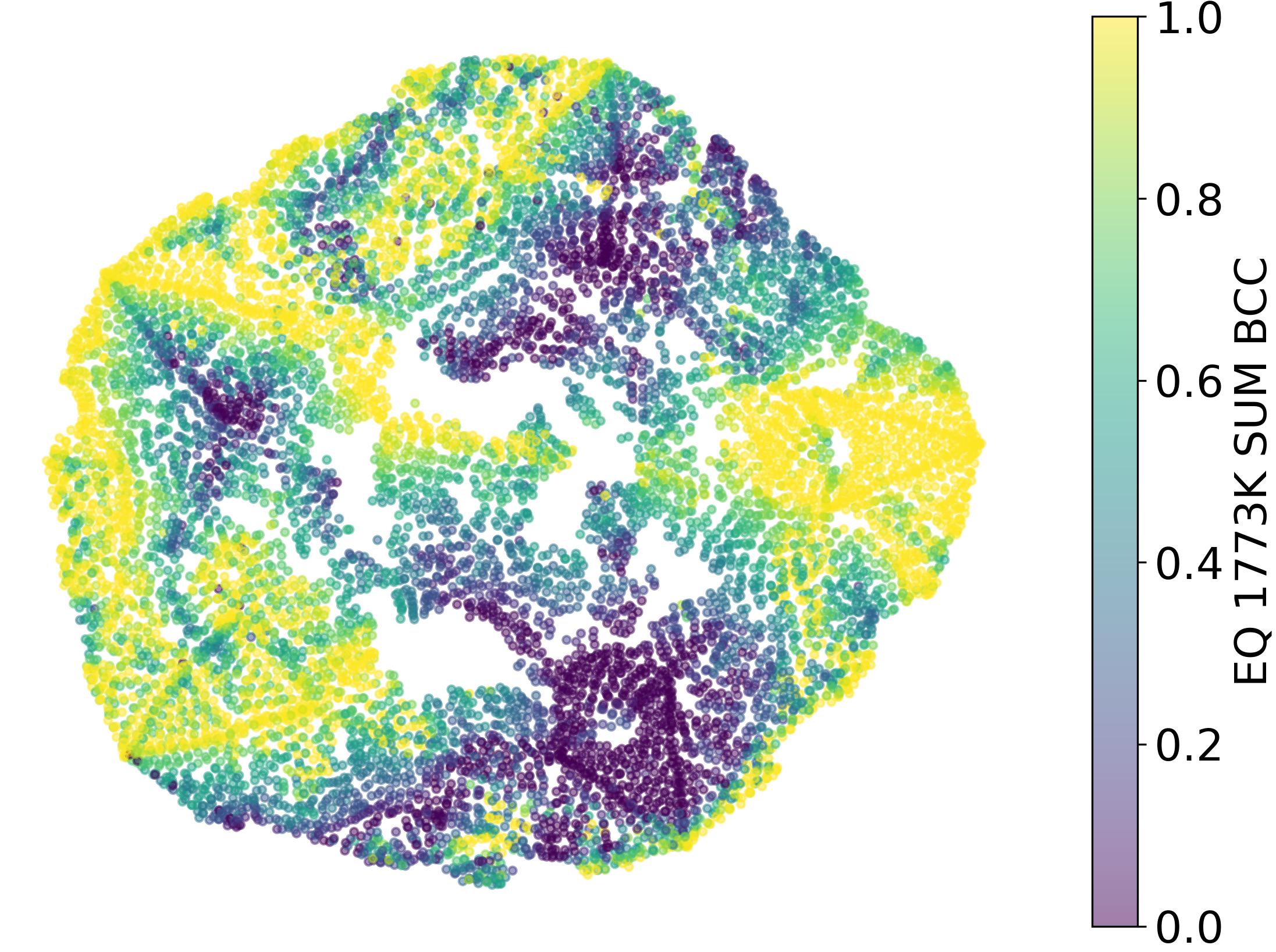}
    \includegraphics[width=0.19\linewidth]{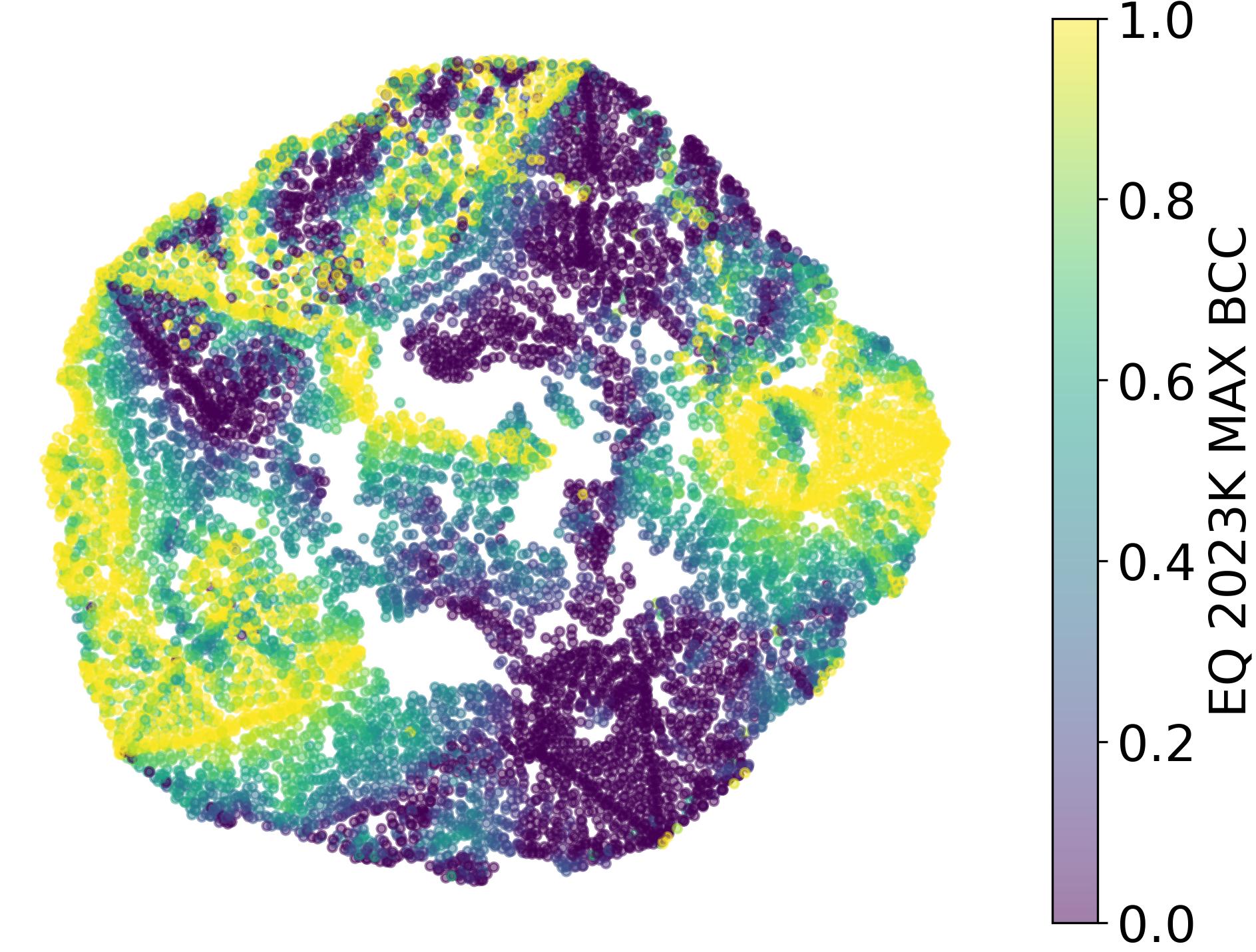}
    \includegraphics[width=0.19\linewidth]{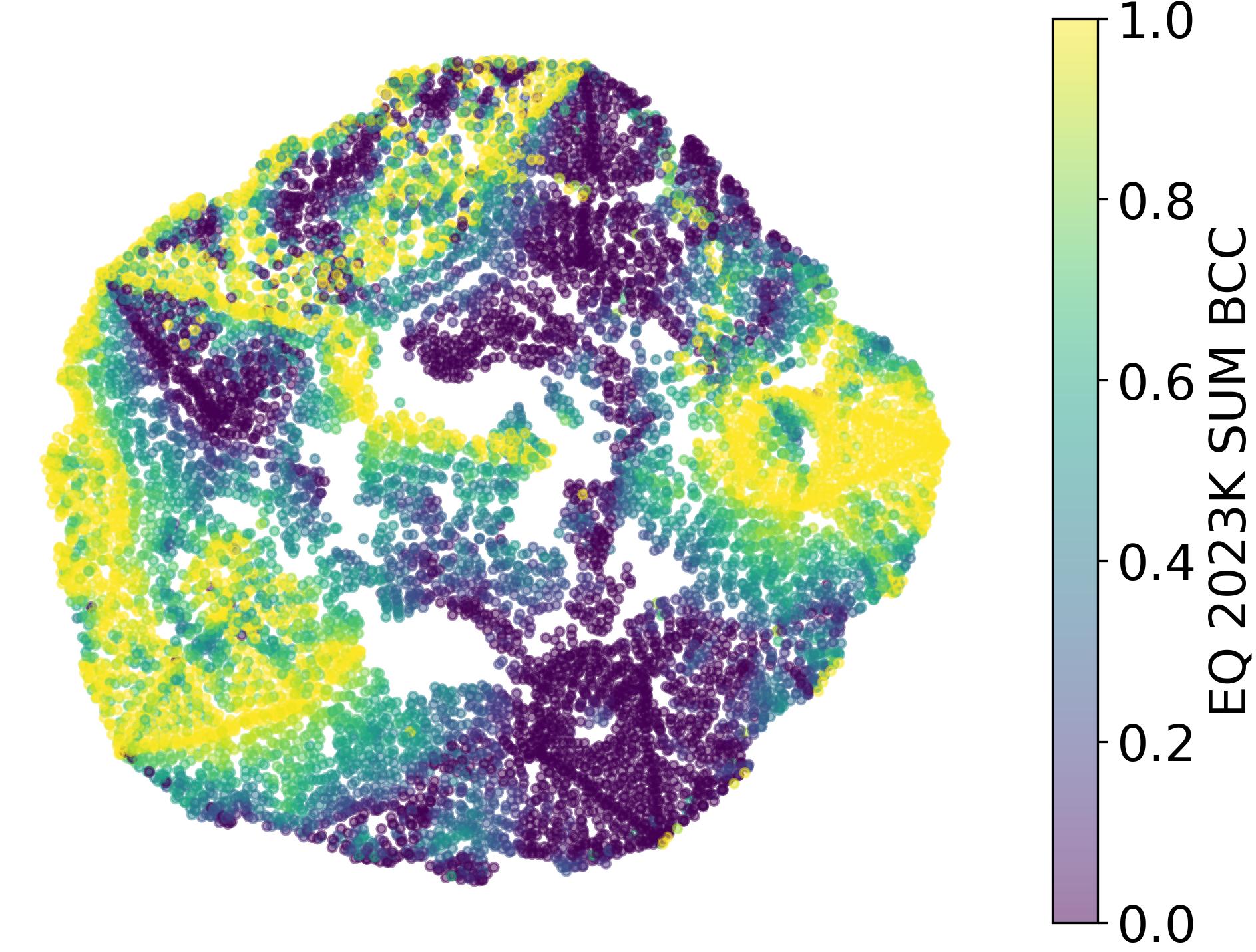}
    \includegraphics[width=0.19\linewidth]{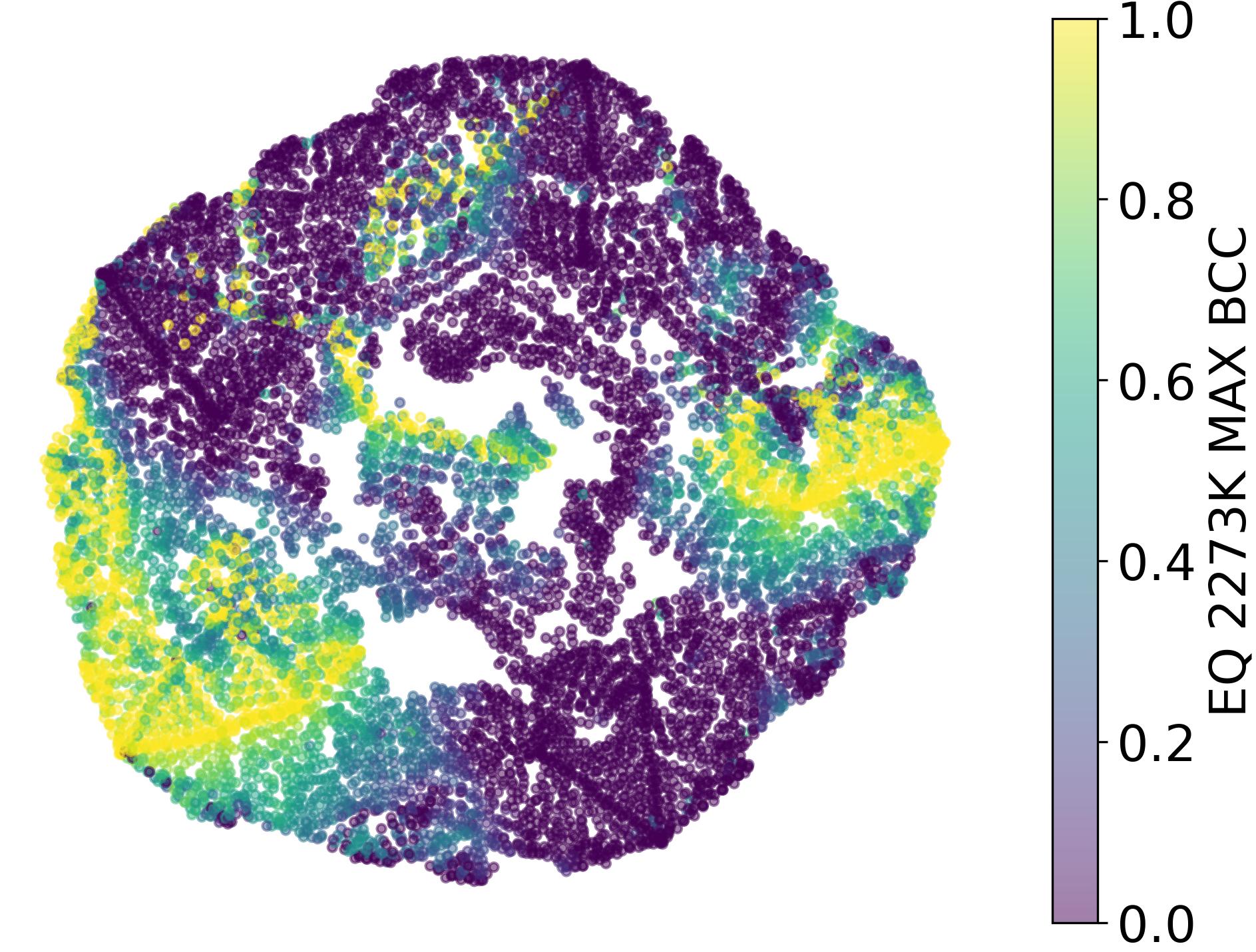}
    \includegraphics[width=0.19\linewidth]{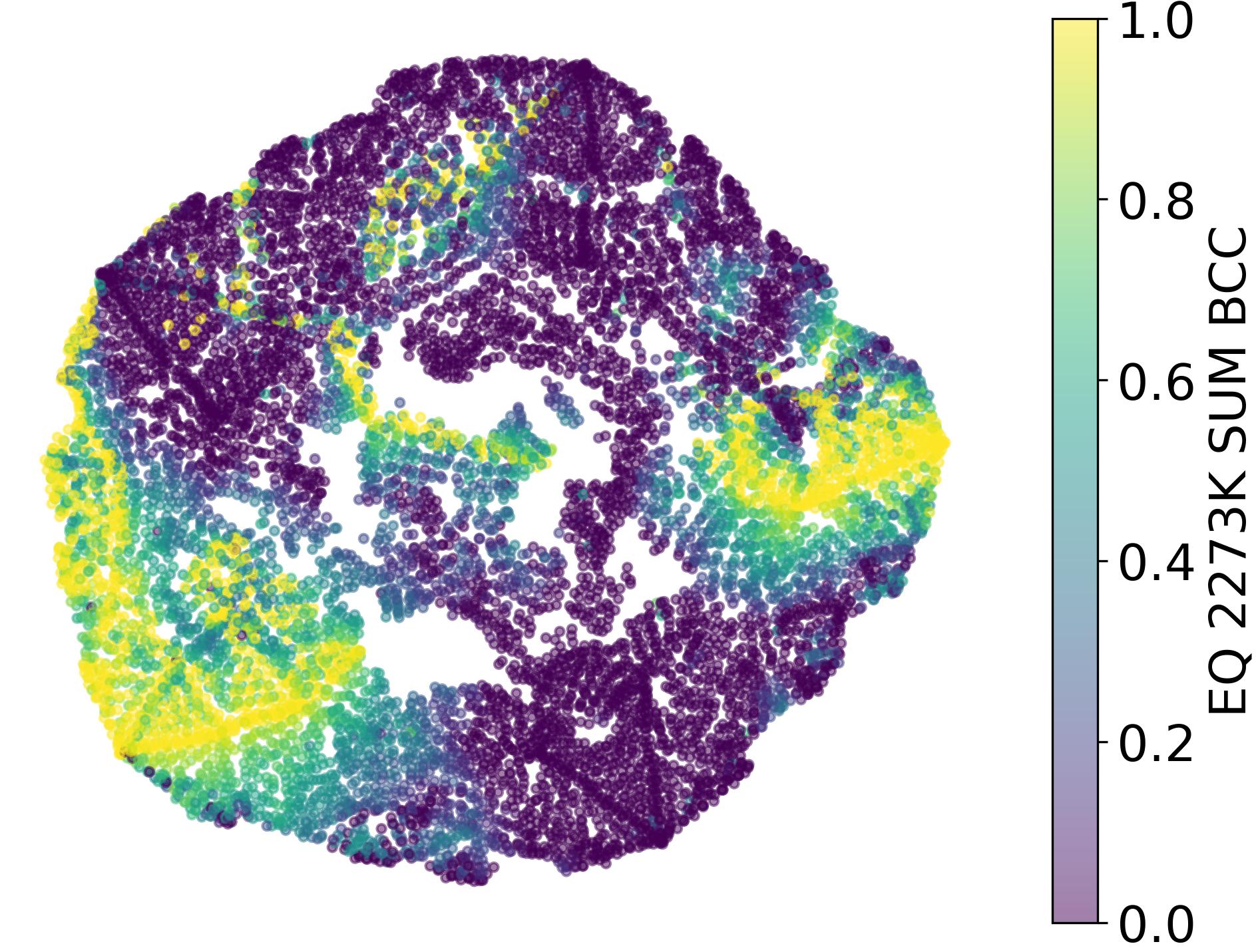}
    \includegraphics[width=0.19\linewidth]{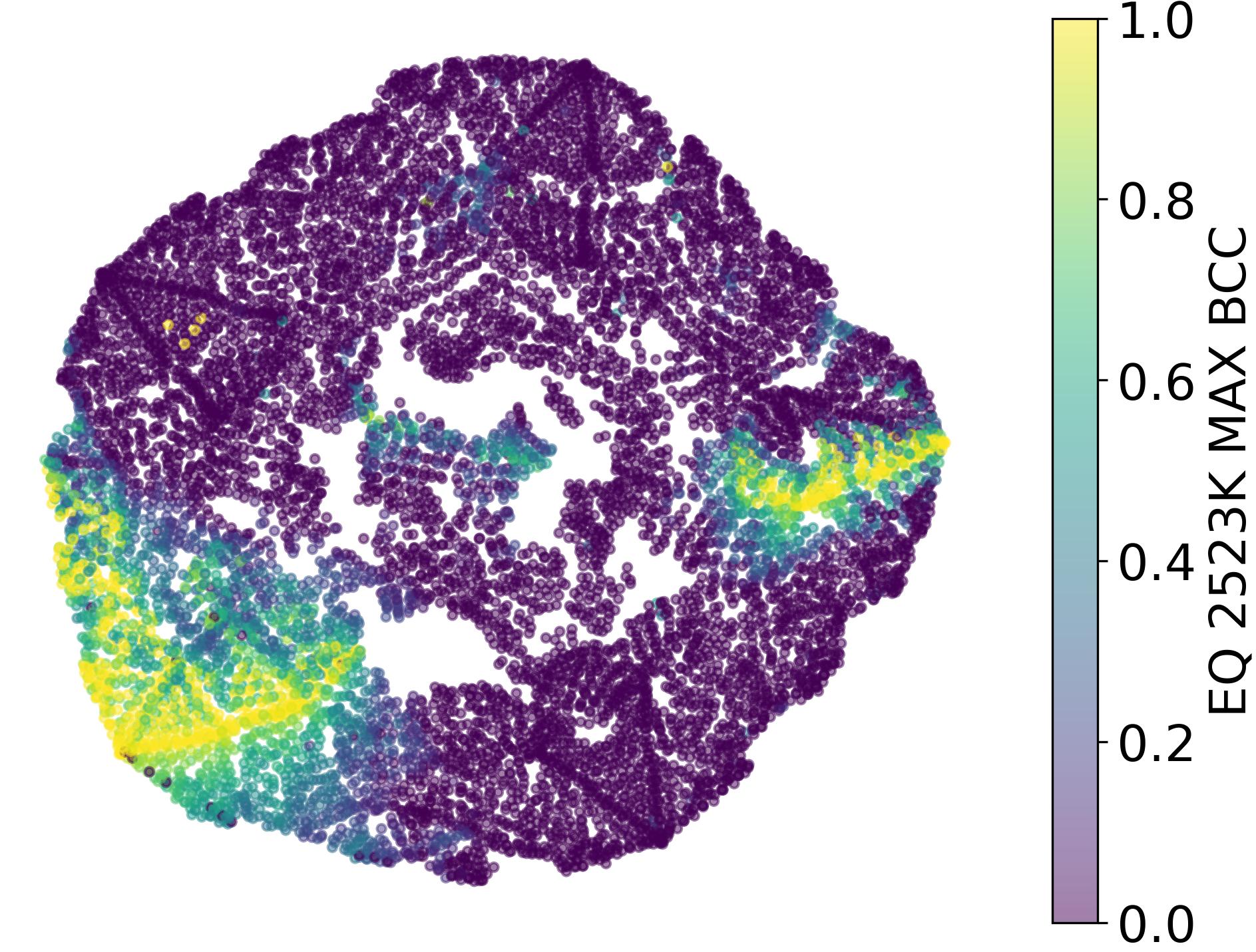}
    \includegraphics[width=0.19\linewidth]{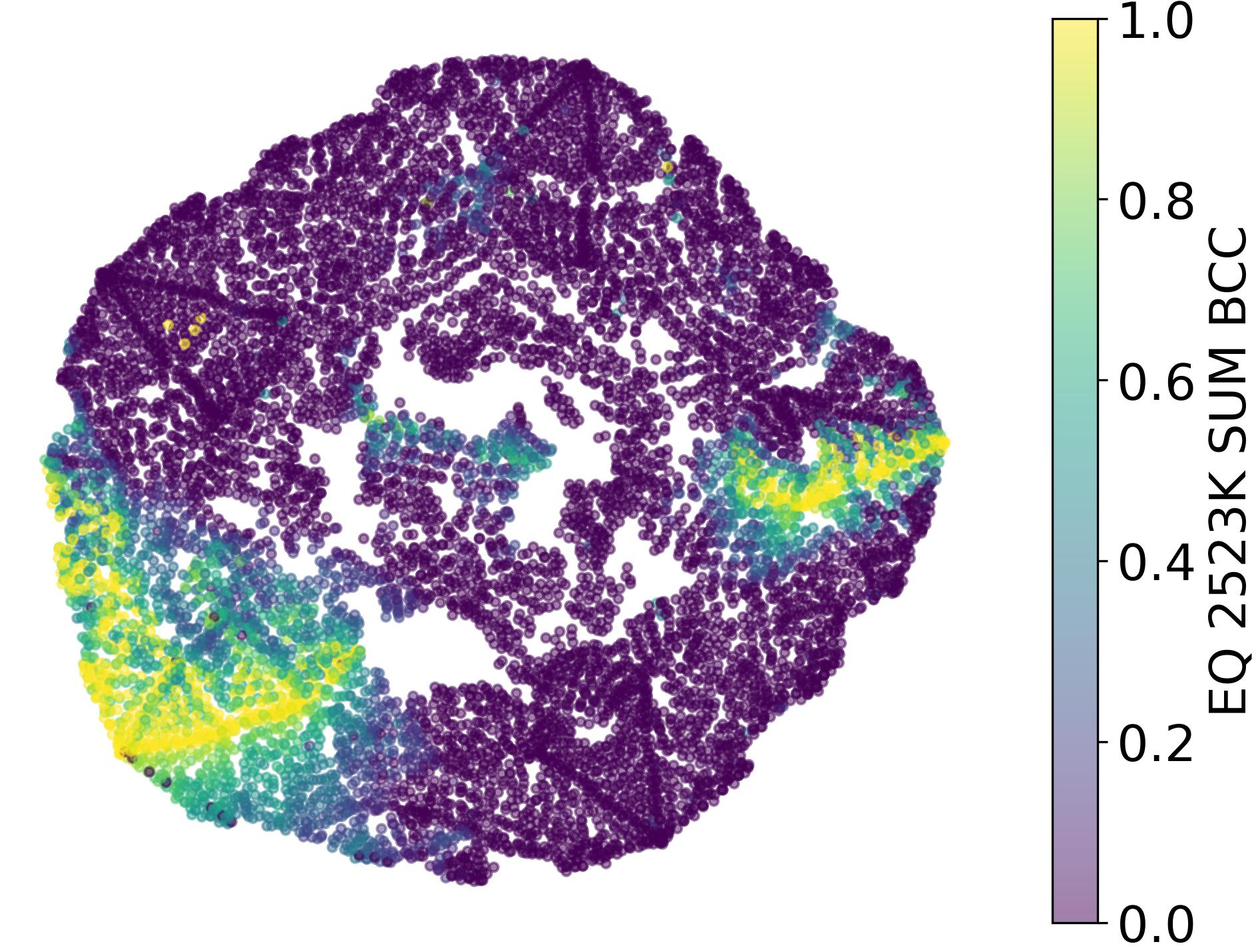}
    \includegraphics[width=0.19\linewidth]{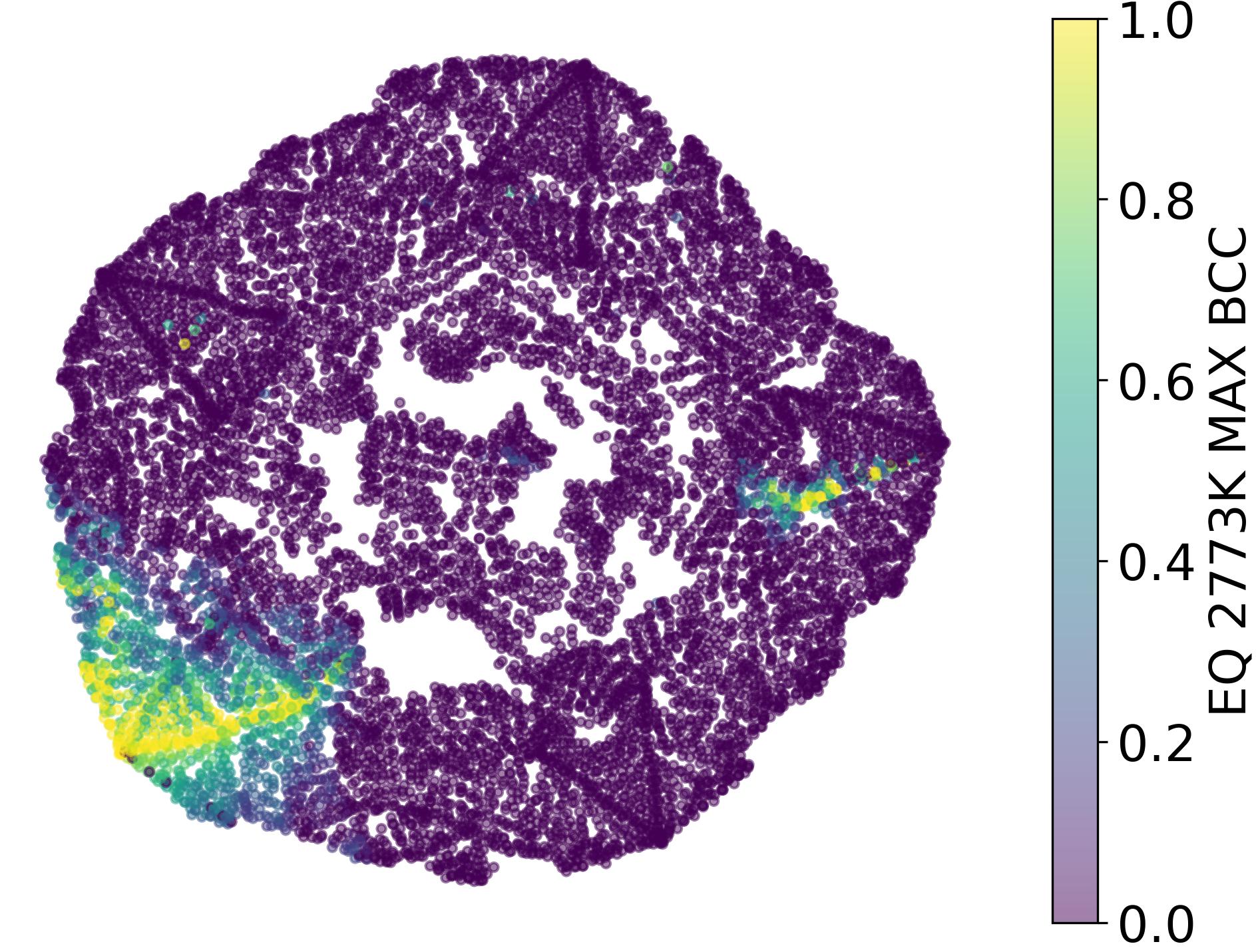}
    \includegraphics[width=0.19\linewidth]{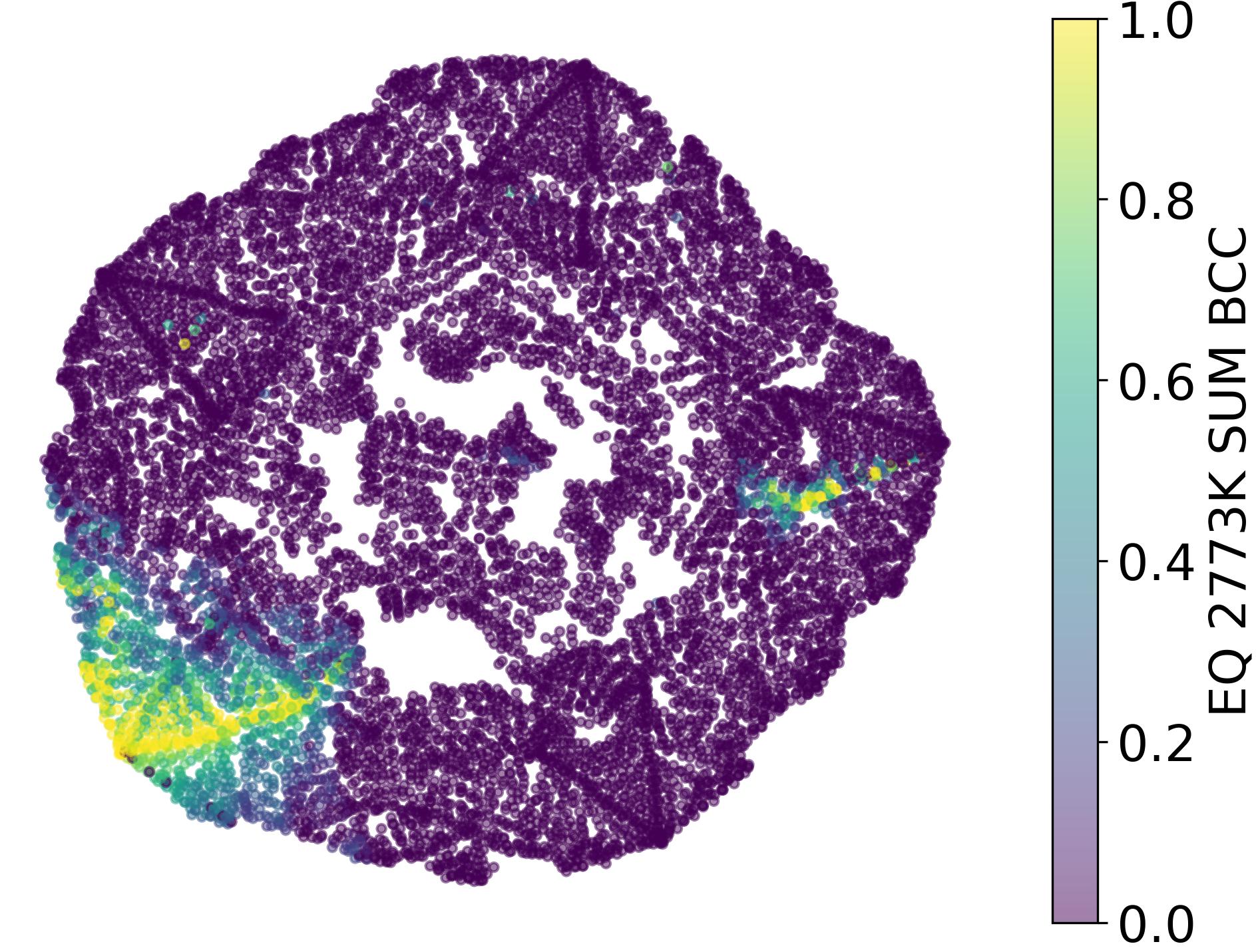}
    \includegraphics[width=0.19\linewidth]{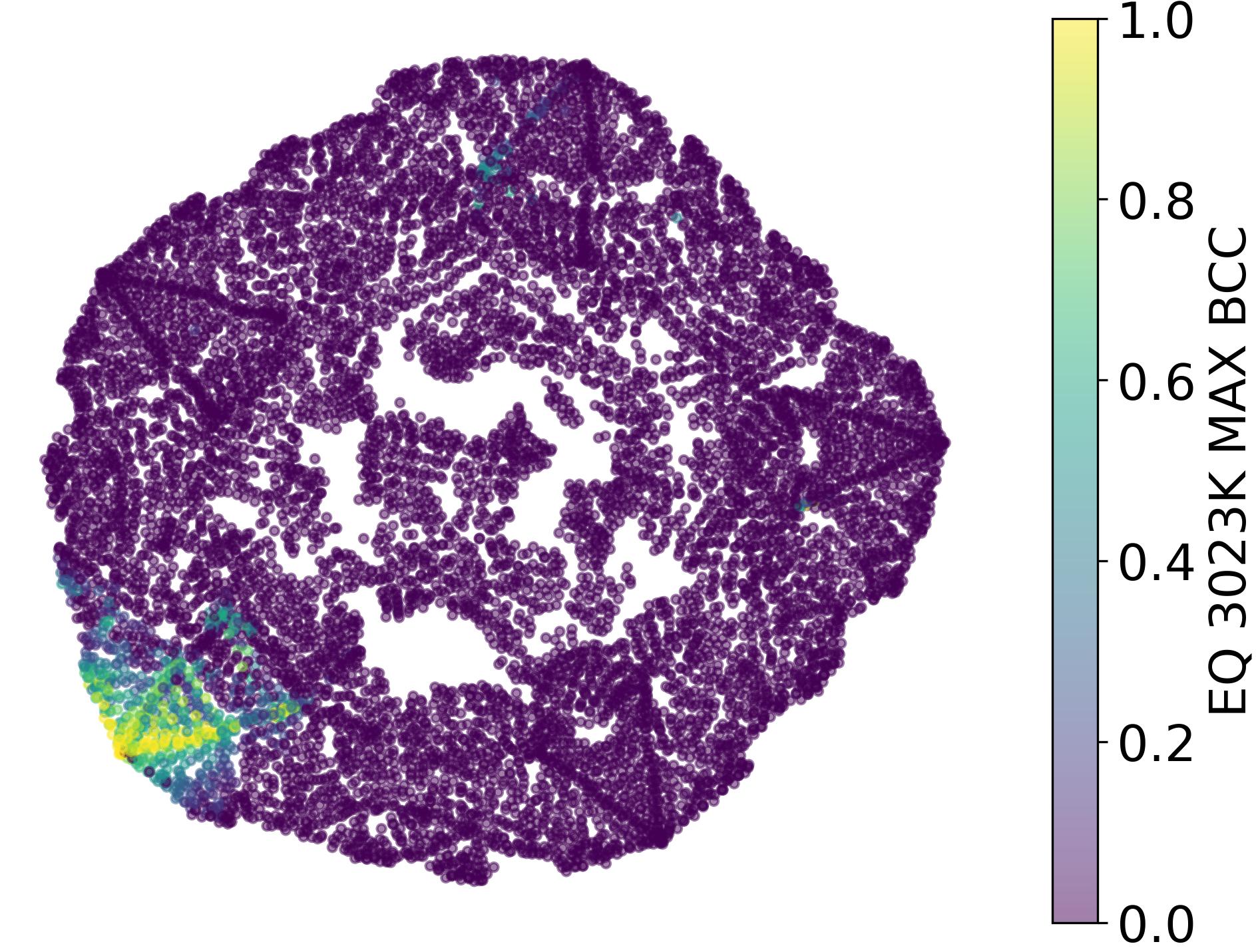}
    \includegraphics[width=0.19\linewidth]{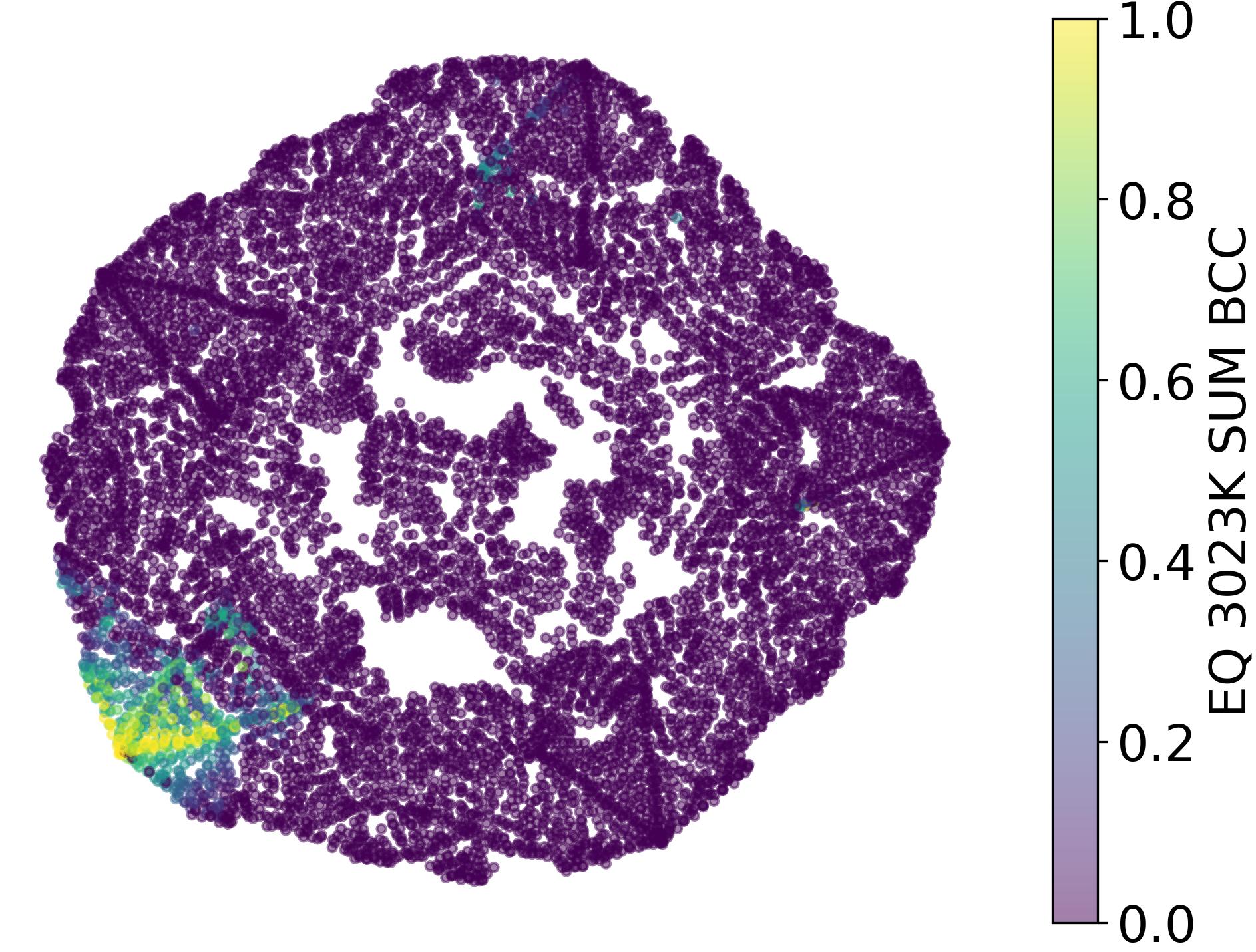}
    \includegraphics[width=0.19\linewidth]{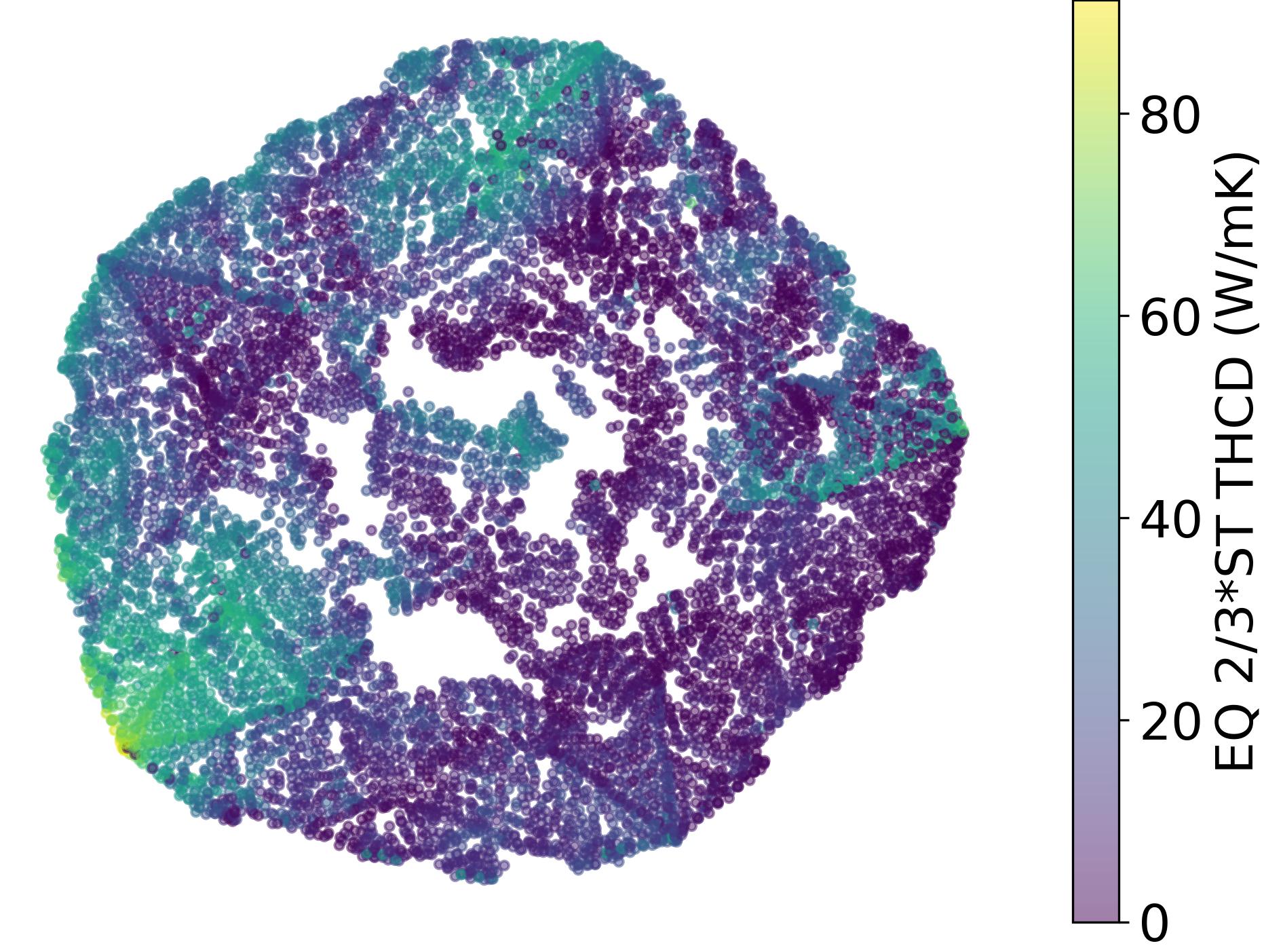}
    \includegraphics[width=0.19\linewidth]{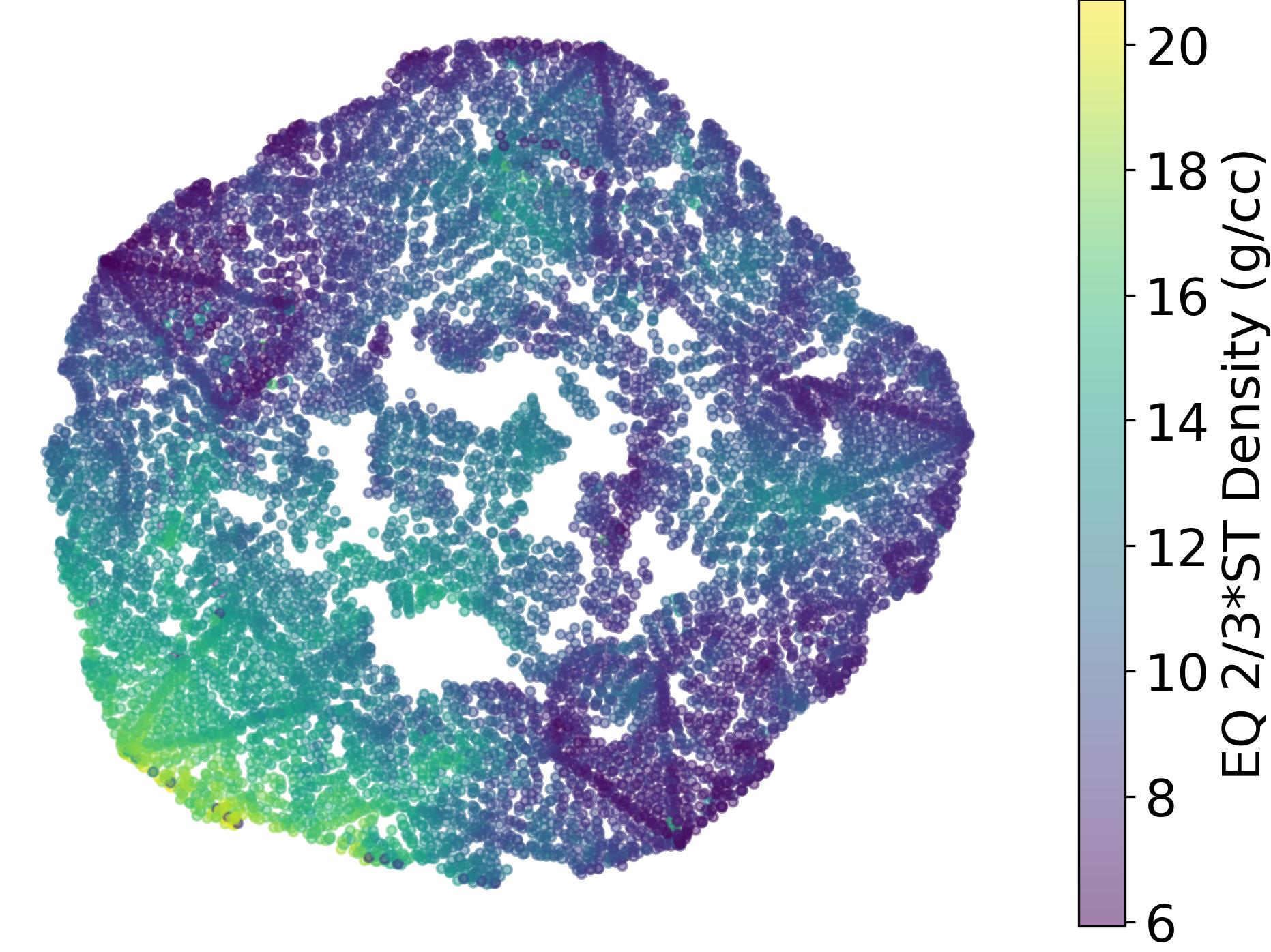}
    \includegraphics[width=0.19\linewidth]{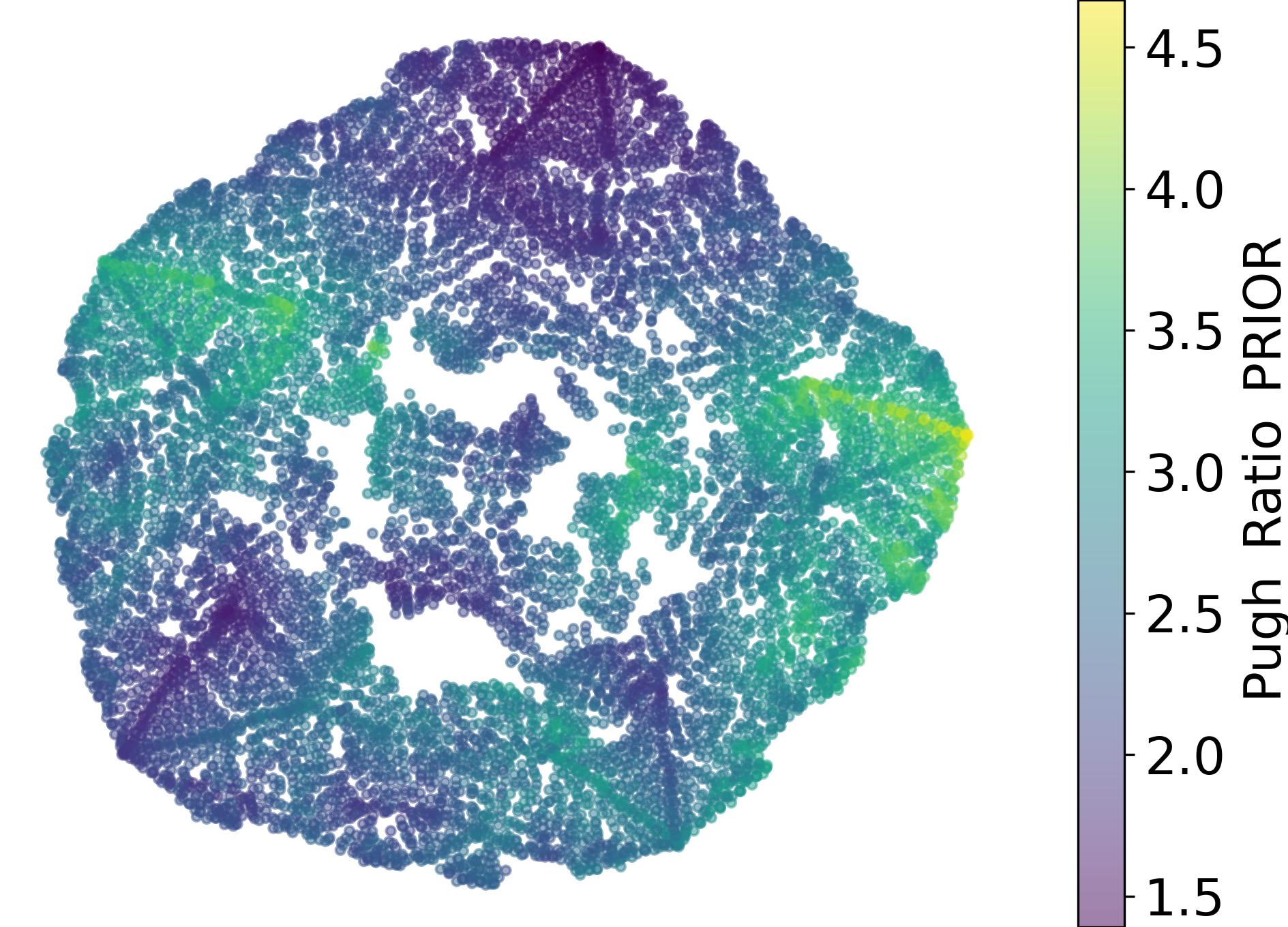}
    \includegraphics[width=0.19\linewidth]{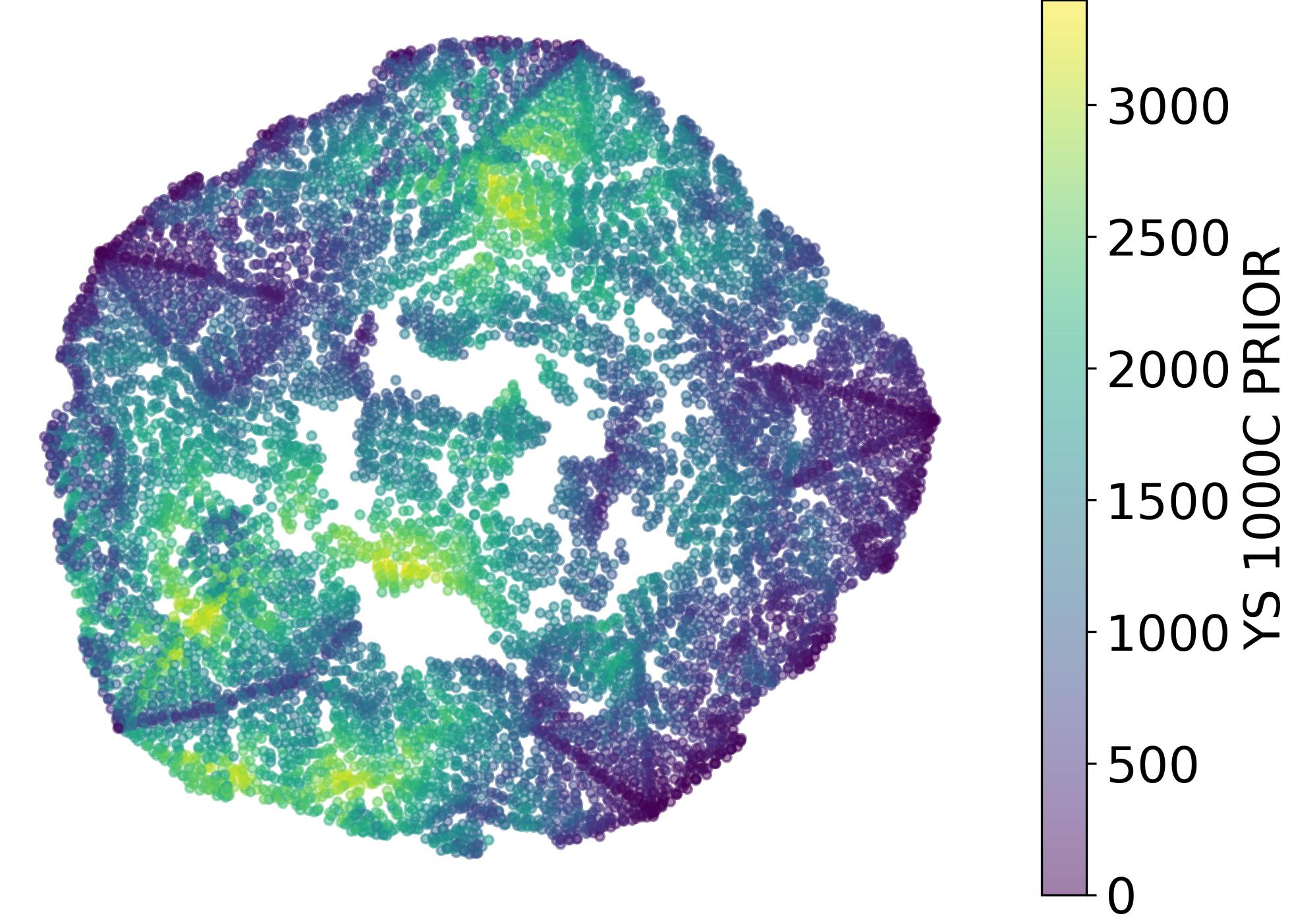}
    \includegraphics[width=0.19\linewidth]{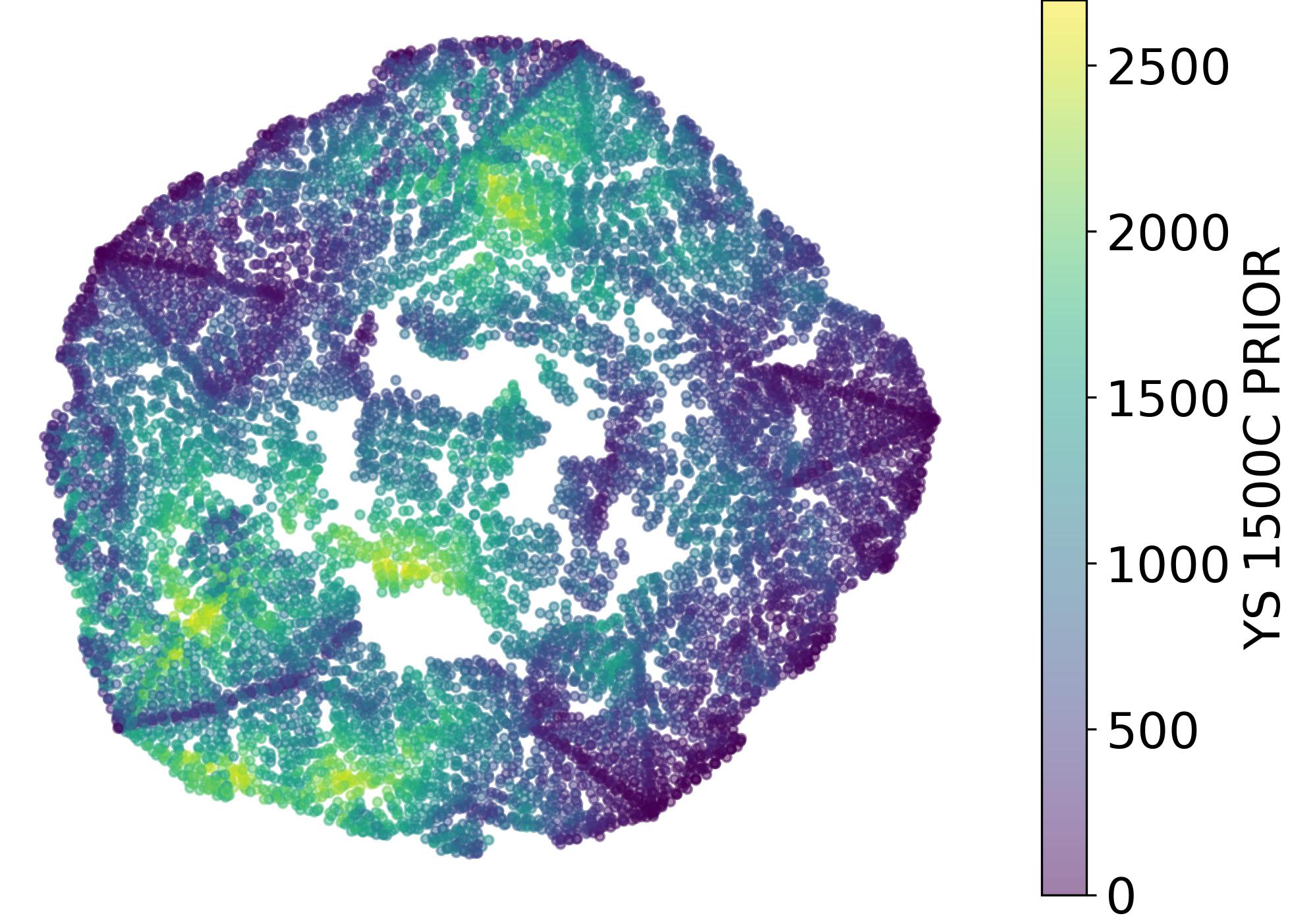}
    \\
    \caption{t-SNE plots for output features}
    \label{fig:output_features}
    \end{subfigure}
    % Main figure caption
    \caption{t-SNE plots of output Features: first 30 Features in the dataset.}
    \label{fig:materials_features}
\end{figure*}

Figures~\ref{fig:input_features} and~\ref{fig:output_features} presents the t-SNE illustrations of various material properties on a logarithmic scale, illustrating their diverse distributions. 

%%%%%%
%%%%%%
%%%%%%
%%%%%%

\subsection{Features}

The dataset consists of the following 65 key features, categorized into alloy compositions as inputs and material properties as outputs. Table~\ref{tab:all_features} summarizes the names of these features and a brief explanation of some of these features are provided below:

\begin{table*}[ht]
\centering
\caption{List of Features in the dataset}\label{tab:all_features}
\resizebox{\textwidth}{!}{%
\begin{tabular}{c c c c c }
\toprule
Nb & Cr & V & W & Zr \\
\midrule
PROP LT (K) & PROP ST (K) & PROP 500C CTE (1/K) & PROP 1000C CTE (1/K) & PROP 1500C CTE (1/K) \\
EQ 1273K THCD (W/mK) & EQ 1273K Density (g/cc) & EQ 1273K MAX BCC & EQ 1273K SUM BCC & EQ 1523K THCD (W/mK) \\
EQ 1523K Density (g/cc) & EQ 1523K MAX BCC & EQ 1523K SUM BCC & EQ 1773K MAX BCC & EQ 1773K SUM BCC \\
EQ 2023K MAX BCC & EQ 2023K SUM BCC & EQ 2273K MAX BCC & EQ 2273K SUM BCC & EQ 2523K MAX BCC \\
EQ 2523K SUM BCC & EQ 2773K MAX BCC & EQ 2773K SUM BCC & EQ 3023K MAX BCC & EQ 3023K SUM BCC \\
EQ 2/3*ST THCD (W/mK) & EQ 2/3*ST Density (g/cc) & Pugh Ratio PRIOR & YS 1000C PRIOR & YS 1500C PRIOR \\
Single BCC & SCHEIL ST & SCHEIL LT & Kou Criteria & SCHEIL BCC B2 \\
SCHEIL BCC B2\#2 & SCHEIL MAX BCC & Kou Criteria Normalized & composition string & Creep Merit \\
25 Min Creep VG [1/s] & 25 Min Creep NH [1/s] & 25 Min Creep CB [1/s] & 25 Min Creep PL5 [1/s] & 500 Min Creep VG [1/s] \\
500 Min Creep NH [1/s] & 500 Min Creep CB [1/s] & 500 Min Creep PL5 [1/s] & 1000 Min Creep VG [1/s] & 1000 Min Creep NH [1/s]$_y$ \\
1000 Min Creep CB [1/s] & 1000 Min Creep PL5 [1/s] & 1300 Min Creep VG [1/s] & 1300 Min Creep NH [1/s]$_y$ & 1300 Min Creep CB [1/s] \\
1300 Min Creep PL5 [1/s] & 1500 Min Creep VG [1/s] & 1500 Min Creep NH [1/s]$_y$ & 1500 Min Creep CB [1/s] & 1500 Min Creep PL5 [1/s] \\
2000 Min Creep VG [1/s] & 2000 Min Creep NH [1/s] & 2000 Min Creep CB [1/s] & 2000 Min Creep PL5 [1/s] &  \\
\bottomrule
\end{tabular}%
}
\label{tab:columns_table}
\end{table*}

\begin{itemize}
    \item \textbf{Material Properties}:
    \begin{itemize}
        \item \textbf{PROP LT (K)}: Low-temperature property measured in Kelvin (K).
        \item \textbf{PROP ST (K)}: Short-term property measured in Kelvin (K).
        \item \textbf{PROP 500°C CTE (1/K)}: Coefficient of thermal expansion (CTE) at 500°C.
        \item \textbf{PROP 1000°C CTE (1/K)}: Coefficient of thermal expansion (CTE) at 1000°C.
        \item \textbf{PROP 1500°C CTE (1/K)}: Coefficient of thermal expansion (CTE) at 1500°C.
        \item \textbf{EQ 1273K THCD (W/mK)}: Thermal conductivity at 1273 K.
        \item \textbf{EQ 1273K Density (g/cc)}: Density at 1273 K.
        \item \textbf{EQ 1273K MAX BCC}: Maximum BCC phase at 1273 K.
        \item \textbf{EQ 1273K SUM BCC}: Sum of BCC phases at 1273 K.
        \item \textbf{EQ 1523K THCD (W/mK)}: Thermal conductivity at 1523 K.
        \item \textbf{EQ 1523K Density (g/cc)}: Density at 1523 K.
        \item \textbf{Pugh Ratio}: Pugh’s ratio for predicting ductility.
        \item \textbf{YS 1000°C PRIOR (MPa)}: Yield strength at 1000°C based on prior information.
        \item \textbf{YS 1500°C PRIOR (MPa)}: Yield strength at 1500°C based on prior information.
        \item \textbf{Creep Rate (s$^{-1}$)}: Minimum creep rates at 1000°C, 1300°C, and 1500°C.
        \item \textbf{SCHEIL ST}: Scheil solidification temperature.
        \item \textbf{SCHEIL LT}: Scheil liquidus temperature.
        \item \textbf{Kou Criteria}: Solidification criteria based on Kou's model.
        \item \textbf{SCHEIL BCC\_B2}: Amount of BCC\_B2 phase from Scheil calculations.
        \item \textbf{SCHEIL MAX BCC}: Maximum BCC phase from Scheil calculations.
    \end{itemize}
\end{itemize}

\subsection{Data Preprocessing}

The raw dataset includes missing values for certain properties, especially at higher temperature ranges. These missing values were filled using an interpolation method. To ensure consistency across varying scales, features were normalized using uniform quantile transformation.

\subsection{Hyperparameter Definitions and Ranges for Each Model}

The key hyperparameters involved in the optimization process for each model  and their range is defined in Table~\ref{tab:hyp_ranges}. 

\begin{table*}[!ht]
    \centering
    \caption{Hyperparameters Tested for Different Models in the Study}\label{tab:hyp_ranges}
    \scriptsize
    \setlength{\tabcolsep}{3pt} % Reduce the space between columns
    \renewcommand{\arraystretch}{1.1} % Reduce space between rows
    \label{tab:combined_hyperparameters}
    \begin{tabular}{p{4cm}p{2.25cm}p{2.25cm}p{5cm}}
        \toprule
        \textbf{Model (Description)} & \textbf{Hyperparameter} & \textbf{Type} & \textbf{Description and Range/Values} \\
        \midrule
        \textbf{XGBoost (Tree-based ensemble)} & n\_estimators & Integer & Number of boosting rounds [10, 50] \\
                         & max\_depth & Integer & Maximum tree depth [3, 10] \\
                         & learning\_rate & Float & Shrinks feature weights [1e-4, 1e-1] (Log Scale) \\
                         & subsample & Float & Fraction of samples per tree [0.5, 1.0] \\
                         & colsample\_bytree & Float & Fraction of features per tree [0.5, 1.0] \\
        \midrule
        \textbf{CNN (Convolutional network)} & latent\_dim & Integer & Dimensionality of latent space [64, 192] \\
                     & drop\_out\_rate & Float & Fraction of units to drop [0.1, 0.3] \\
                     & kernel\_size & Integer & Size of convolution filter [2, 4] \\
                     & learning\_rate & Float & Learning rate for optimizer [1e-4, 1e-3] \\
                     & optimizer & Categorical & Optimization algorithm \{adam, sgd, adadelta\} \\
                     & batch\_size & Integer & Samples per gradient update [32, 64] \\
                     & epochs & Integer & Number of training iterations [10, 50] \\
        \midrule
        \textbf{FDN (Fully dense network)} & latent\_dim & Integer & Dimensionality of latent space [64, 192] \\
                     & drop\_out\_rate & Float & Fraction of units to drop [0.1, 0.3] \\
                     & learning\_rate & Float & Learning rate for optimizer [1e-4, 1e-3] \\
                     & lambda & Float & Regularization strength [1e-5, 1e-2] \\
                     & alpha & Float & Scaling factor in loss function [1e-3, 0.1] \\
                     & optimizer & Categorical & Optimization algorithm (adam) \\
                     & batch\_size & Integer & Samples per gradient update [32, 128] \\
                     & epochs & Integer & Number of training iterations (50) \\
        \midrule
        \textbf{DNF (Logic-based network)} & latent\_dim & Integer & Dimensionality of latent space [64, 192] \\
                     & drop\_out\_rate & Float & Fraction of units to drop [0.1, 0.3] \\
                     & learning\_rate & Float & Learning rate for optimizer [1e-4, 1e-3] \\
                     & num\_conjunctions & Integer & Number of conjunctions [5, 20] \\
                     & conjunction\_units & Integer & Units per conjunction layer [50, 150] \\
                     & optimizer & Categorical & Optimization algorithm (adam) \\
                     & batch\_size & Integer & Samples per gradient update [32, 128] \\
                     & epochs & Integer & Number of training iterations (50) \\
        \midrule
        \textbf{TabNet (Attention-based network)} & n\_d & Integer & Number of decision steps [8, 32] \\
                        & n\_a & Integer & Attention units [8, 32] \\
                        & n\_steps & Integer & Number of steps in TabNet [3, 10] \\
                        & gamma & Float & Relaxation factor [1.0, 2.0] \\
                        & lambda\_sparse & Float & Sparsity regularization [1e-5, 1e-3] \\
                        & learning\_rate & Float & Learning rate for optimizer [1e-4, 1e-2] \\
                        & batch\_size & Integer & Samples per gradient update [32, 128] \\
                        & virtual\_batch\_size & Integer & Virtual batch size [16, 64] \\
                        & epochs & Integer & Number of training iterations (50) \\
                        & optimizer & Categorical & Optimization algorithm (RMSprop) \\
        \midrule
        \textbf{VAE (Generative model)} & latent\_dim & Integer & Dimensionality of latent space [16, 64] \\
                     & lambda & Float & Regularization strength [1e-5, 1e-2] \\
                     & drop\_out\_rate & Float & Fraction of units to drop [0.1, 0.3] \\
                     & alpha & Float & Scaling factor in loss function [1e-3, 0.1] \\
                     & learning\_rate & Float & Learning rate for optimizer [1e-4, 1e-3] \\
                     & optimizer & Categorical & Optimization algorithm (adam) \\
                     & batch\_size & Integer & Samples per gradient update [32, 128] \\
                     & epochs & Integer & Number of training iterations (50) \\
        \bottomrule
    \end{tabular}
\end{table*}

\end{document}